\documentclass[11pt]{article}
\usepackage{color,amsmath,amssymb,path,amsthm,caption,url,changepage}
\newcommand{\ignore}[1]{}

\newtheorem{theorem}{Theorem}[section]
\newtheorem{lemma}[theorem]{Lemma}

\theoremstyle{definition}
\newtheorem{rule0}{Rule}
\newcommand{\Proof}[1]
        {
        \noindent
        \emph{Proof #1.}~
        }
\newsavebox{\smallProofsym}                     
\savebox{\smallProofsym}                        %
        {
        \begin{picture}(7,7)                    %
        \put(0,0){\framebox(6,6){}}             %
        \put(0,2){\framebox(4,4){}}             %
        \end{picture}                           %
        }                                       %
\newcommand{\smalleop}[1]
        {
        \mbox{} \hfill #1~~\usebox{\smallProofsym}\!\!\!\!\!\!\
        }
\newenvironment{theProof}[1]
        {
        \Proof{#1}}{\smalleop{}
        \medskip

        }
\newenvironment{theProofNoSkip}[1]
        {
        \Proof{#1}}{\smalleop{}
        }
\usepackage{graphicx}

\usepackage{dsfont}
\newcommand{\NN}{\ensuremath{\mathbb N}}
\newcommand{\NNnull}{\NN_{^{_0}\!}}

\newcommand{\pl}[1]{#1^{^{_+}}\!}
\newcommand{\Trp}[1]{\pl{{\cal T}}(#1)}
\newcommand{\ex}[1]     {{\,{\ensuremath{\mathbb E}}\!\left(#1\right)}}
\newcommand{\trip}[1]{{\pl{\mathsf{tr}\!}}(#1)}
\newcommand{\supp}[1]{\mathrm{supp}(#1)}
\newcommand{\ch}[2]{\mathrm{ch}_{#1}(#2)}

\newcommand{\contr}{\mathrm{contr}}
\newcommand{\abc}[2]{
  \overbrace{#1}^{\hspace{-0.2ex} \placefig{#2}{3ex}}
}

\DeclareCaptionFormat{myempty}{#1 \hspace{25 mm}}
\DeclareCaptionLabelFormat{fullparens}{#1~#2:}
\newcommand{\marfigsc}[2]{
	\marginpar{
		\captionof{figure}{}

		\label{fi:#1}
		\includegraphics[width=#2 in]{#1.eps}
	}
}

\newcommand{\marfig}[1]{\marfigsc{#1}{1.6}}

\newcommand{\mymark}[1]{\textbf{($\boldsymbol{#1}$)} }

\newcommand{\placefig}[2]
        {\includegraphics[width=#2]{#1.eps}}

\begin{document}

\changetext{}{1.23in}{}{}{}

\pagenumbering{arabic}
\date{}
\title{Counting Triangulations of Planar Point Sets\thanks{%
Work on this paper was partially supported by Grants 155/05 and 338/09 from
the Israel Science Fund. Work by Micha Sharir was also
supported by NSF Grants CCF-05-14079 and CCF-08-30272,
by Grant 2006/194 from the U.S.-Israel Binational Science Foundation,
and by the Hermann
Minkowski--MINERVA Center for Geometry at Tel Aviv University.} }

\author{
 Micha Sharir \\
Blavatnik School of Computer Science \\
Tel Aviv University, Tel Aviv 69978, Israel, and \\
Courant Institute of Mathematical Sciences \\
New York University, New York, NY 10012, USA \\
{\sl michas@tau.ac.il}
\and Adam Sheffer\\
Blavatnik School of Computer Science \\
Tel Aviv University, Tel Aviv 69978, Israel \\
{\sl sheffera@tau.ac.il}
}

\maketitle

\begin{abstract}
We study the maximal number of triangulations that a planar set of $n$ points can have, and show that it is at most $30^n$. This new bound is achieved by a careful optimization of the charging scheme of Sharir and Welzl (2006), which has led to the previous best upper bound of $43^n$ for the problem.

Moreover, this new bound is useful for bounding the number of other types of planar (i.e., crossing-free) straight-line graphs on a given point set. Specifically, we derive new upper bounds for the number of planar graphs ($O^*\left(239.4^n\right)$), spanning cycles ($O^*(70.21^n)$), spanning trees ($160^n$), and cycle-free graphs ($O^*(202.5^n)$).

\end{abstract}


\section{Introduction} \label{sec:Intro}

A {\em planar graph} is a graph that can be drawn on the plane in such a way that its edges intersect only at their endpoints. A {\em planar straight-line graph} is an embedding of a planar graph in the plane such that its edges are mapped into straight line segments. In this paper, we only consider planar straight-line graphs, but refer to them as planar graphs for simplicity.

Given a set $S$ of points in the plane, a \emph{triangulation} of $S$ is a maximal planar graph on $S$. When $S$ is of cardinality at least 5, and is in general position (no three points are collinear), it has at least two different triangulations. Let $\overline{tr}(n)$ ($\underline{tr}(n)$) denote the maximal (minimal) number of triangulations for a planar point set of $n$ points in general position. In this paper, we study the asymptotic behavior of $\overline{tr}(n)$, and focus on its upper bound.

\paragraph{Previous work.}
Variants of this problem have been studied for over 250 years. The first to consider such a variant was probably Euler, who studied the case of $n$ points in convex position. Euler produced a recursion for the number of triangulations of such sets and guessed its solution, but could not prove its validity. In the 19th century, the problem was studied independently by several mathematicians, which were able to produce some findings, including a proof of Euler's guessed solution. That is, the number of triangulations for the convex case is $C_{n-2}$, where
$C_m := \frac{1}{m+1}\binom{2m}{m} = \Theta(m^{-3/2}4^m) = \Theta^*(4^m)$,
$m \in \NNnull$, is the $m$th \emph{Catalan number}\footnote{In the notations $O^*()$, $\Theta^*()$, and $\Omega^*()$, we neglect polynomial factors and just give the dominating exponential term.} (see \cite[page 212]{St99} for a discussion).

During the mid-20th century, Tutte studied several variants of this problem, which did consider points in general position, but had other distinctions from the problem we study (see \cite{Tut62}, and \cite[pages 114-120]{Tut98}).
Avis was perhaps one of the first to ask whether the
maximum number of triangulations of $n$ points in the plane
is bounded by $c^n$ for some $c>0$; see \cite[page 9]{ACNS82}.
This fact was established in 1982
by Ajtai, Chv\'atal, Newborn, and Szemer\'edi \cite{ACNS82},
who show
that there are at most $10^{13n}$ crossing-free graphs on $n$
points---in particular, this bound holds for triangulations.

Further developments have yielded progressively better
upper bounds for the number of triangulations\footnote{Interest
  was also motivated by the obviously related practical question
  (from geometric modeling \cite{Smi89}) of how many bits it takes
  to encode a triangulation of a point set.}
\cite{Smi89,DeSo97,Sei98}, so far culminating in the
previously mentioned $43^n$ bound \cite{ShWe06b} in 2006.
This compares to $\Omega^*(8.48^n)$, the largest known number of
triangulations for a set of $n$ points, derived by
Aichholzer et al.~\cite{AHHHKV05}.

The value of $\underline{tr}(n)$ has also been studied. In a companion paper \cite{SSA}, we derive the bound $\underline{tr}(n) = \Omega(2.43^n)$ (which improves a previous bound by Aichholzer, Hurtado, and  Noy \cite{AHN04}). McCabe and Seidel \cite{MCSe04} showed that when the convex hull has only $O(1)$ vertices, there are $\Omega(2.63^n)$ triangulations.

Hurtado and Noy \cite{HuNo97} presented a configuration of $n$ points in general position and $\Theta^*(\sqrt{12}^n)\approx\Theta^*(3.464^n)$ triangulations, implying $\underline{tr}(n) \approx O^*(3.464^n)$.

\paragraph{Related problems.}
Besides the intrinsic interest in obtaining bounds on the number of triangulations, they are useful for bounding the number of other kinds of planar graphs on a given point set, exploiting the fact that any such graph is a subgraph of some triangulation. We shortly review some of these bounds.

Let $\overline{pg}(n)$ denote the maximal number of planar graphs for a planar point set of cardinality $n$ in general position. A bound of $\overline{pg}(n) = o\left(\overline{tr}(n)\cdot7.98^n\right)$ is derived in \cite{RSW08}, which, combined with Sharir and Welzl's bound of $43^n$ on $\overline{tr}(n)$, yields $\overline{pg}(n) = o\left(43^n\cdot7.98^n\right)=o\left(343.14^n\right)$, which was the best upper bound discovered so far.

Let $\overline{sc}(n)$ denote the maximal number of crossing-free spanning cycles (sometimes referred to as simple polygonizations) for a planar point set of cardinality $n$ in general position. Buchin et al. \cite{BKKSS07} showed that a single triangulation has $O(\sqrt[4]{30}^n)\approx O(2.3403^n)$ spanning cycles as subgraphs, which implies $\overline{sc}(n)=O(\overline{tr}(n)\cdot2.3403^n)$. By using Sharir and Welzl's bound of $43^n$, we get $\overline{sc}(n)\approx O(100.635^n)$, which still falls short of the bound $\overline{sc}(n)\approx O(86.81^n)$, given in \cite{ShWe06} (which is derived in a diffeent manner from an upper bound on the maximal number of crossing-free perfect matchings).

Let $\overline{st}(n)$ denote the maximal number of crossing-free spanning trees for a planar point set of cardinality $n$ in general position. Rib\'o \cite{Rib05} (see also \cite{Rot05}) showed that any planar straight-line graph has at most $\left(5\frac{1}{3}\right)^n$ spanning trees as subgraphs. By using Sharir and Welzl's bound of $43^n$, we get $\overline{st}(n) \leq 43^n\cdot\left(5\frac{1}{3}\right)^n=\left(229\frac{1}{3}\right)^n$, which was the best upper bound discovered so far.

Let $\overline{cf}(n)$ denote the maximal number of cycle-free graphs (i.e., forests) for a planar point set of cardinality $n$ in general position. Such a graph can contain at most $n-1$ edges, which implies that a single triangulation of the point set contains $O^*\left(\binom{3n-6}{n-1}\right)=O^*\left(6.75^n\right)$ cycle-free graphs. Since any cycle-free graph is contained in at least one triangulation, we have $\overline{cf}(n)=O^*\left(6.75^n \cdot \overline{tr}(n)\right)=O^*\left(290.25^n\right)$, using the bound of \cite{ShWe06b}.

Finally, let $\overline{cg}(n)$ denote the maximal number of connected crossing-free graphs for a point set of cardinality $n$ in general position. It can be easily noticed that $\overline{cg}(n) = O(\overline{pg}(n))$.

\paragraph{Our results.}
In this paper, we further decrease the existing gap on $\overline{tr}(n)$ by establishing the new upper bound $\overline{tr}(n) < 30^n$. By using the above relationships, we get improved bounds for all five problems mentioned above (improving also upon bounds obtainable by the alternative technique of \cite{ShWe06}, which is based on crossing-free matchings). Table \ref{ta:OurResults} presents the previous results and their new improvements.

\begin{table}[ht]
\caption{Summary of the new bounds.} \label{ta:OurResults}
\centering
\begin{tabular}{c | c | c}
\hline\hline
An Upper & Previous & Improved \\ [0.5ex]
Bound on & Bound    & Bound \\
\hline
$\overline{tr}(n)$ & $43^n$ \cite{ShWe06b} & $30^n$ \\ [0.5ex]
$\overline{sc}(n)$ & $O^*(86.81)$ \cite{ShWe06}  & $O^*(70.21^n)$ \\ [0.5ex]
$\overline{pg}(n), \overline{cg}(n)$ & $O^*\left(343.14^n\right)$ \cite{RSW08,ShWe06b} & $O^*\left(239.4^n\right)$ \\ [0.5ex]
$\overline{st}(n)$ & $229.\overline{3}^n$ \cite{ShWe06b,Rib05} & $160^n$ \\ [0.5ex]
$\overline{cf}(n)$ & $O^*(290.25)$ \cite{AHHHKV05}  & $O^*(202.5^n)$ \\ [1ex]
\hline
\end{tabular}
\end{table}

\changetext{}{-1.23in}{}{}{}
\section{Degrees in Random Triangulations}

This section, together with the following one, present the basic technique we need in order to derive our bound on $\overline{tr}(n)$. These methods were used in \cite{ShWe06b}, to get the bound $43^n$, and therefore, most of these two sections will repeat the analysis in \cite{ShWe06b}. The ``heart" of this technique is perhaps its charging scheme, which is somewhat similar to Heesch's idea of discharging (Entladung, \cite{Hee69}) employed by the proofs of the Four-Color-Theorem (see \cite{AH77} and \cite{RSST97}). In the next sections, we extend this technique in order to get an upper bound of $30^n$. Moreover, we show that the technique, as presented, cannot achieve a bound of $o\left(\left(28\frac{17}{28}\right)^n\right)$, although the true bound is probably much smaller.

\paragraph{Assumptions and notations.}
We use the general position assumption that no three points are collinear. When there are three (or more) points on the same line, it is easily checked that slightly perturbing the middle point can only increase the number of triangulations. In Section \ref{sec:Intro} we mentioned that for each point set in general position there is an exponential number of triangulations. Interestingly, when there are no restrictions on the number of collinear points, there might be a constant number of triangulations. Figure \ref{fi:assumption1} depicts a set of many points with a single triangulation.\marfigsc{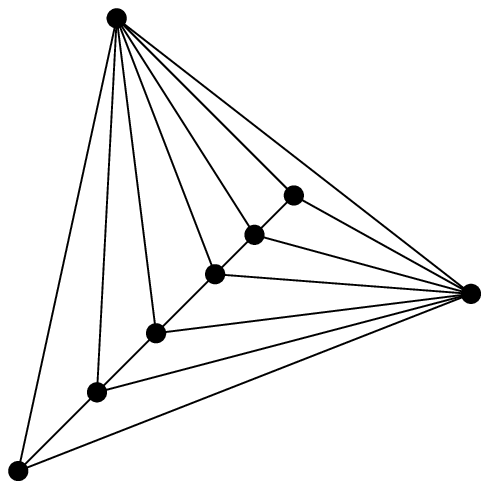}{1.3}Therefore, this assumption is essential for the bounds on $\underline{tr}(n)$, and does not involve any loss of generality for upper bounding $\overline{tr}(n)$.

For a set $S$ of $n$ points in general position, let $S^+$ denote a set of $n+3$ points with a triangular convex hull (i.e., a convex hull of cardinality 3), constructed by taking a triangle that contains $S$ in its interior, and adding the three vertices of the triangle to $S$. Notice that every triangulation of $S$ is contained in at least one triangulation of $S^+$, and thus, an upper bound on the number of triangulations of $S^+$ is also an upper bound on the number of triangulations of $S$.

Notice that every face of any triangulation of $S^+$ has exactly three edges (including the outer face). Using Euler's formula, we find that every triangulation of $S^+$ has exactly $3(n+3)-6=3n+3$ edges and $2(n+3)-5=2n+1$ inner faces. \marfigsc{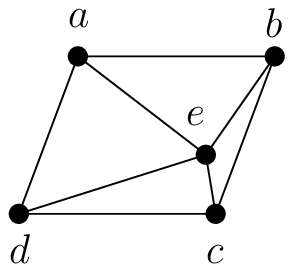}{1.2}

We say that an edge in a triangulation is {\em flippable}, if its two incident triangles form a convex quadrilateral $Q$. A flippable edge can be {\em flipped}, that is, removed from the graph of the triangulation and replaced by the other diagonal of $Q$. Figure \ref{fi:assumption4} depicts a triangulation with exactly two flippable edges --- $ae$ (that can be flipped into $bd$) and $de$ (that can be flipped into $ac$).

\paragraph{Degrees in triangulations.}
Let $\Trp{S}$ denote the set of all triangulations of $\pl{S}$. For $i \in \NN$ and a triangulation $T \in \Trp{S}$, we let $v_i = v_i(T)$ denote the number of points in $S$ (not $\pl{S}$) that have degree $i$ in $T$. Obviously, $v_i \in \NNnull,~~v_1 = v_2 = 0 \mbox{,~~and~~} \sum_i v_i = n$. Let $d_1$, $d_2$ and $d_3$ be the degrees in $T$ of the three vertices of the bounding triangle, then
\begin{equation}
\label{eq:PreDegreesSum}
\mbox{$d_1 + d_2 + d_3 +\sum_i i\, v_i = 2(3n+3) = 6n + 6$}.
\end{equation}

\noindent It is easily seen that for $n \geq 1$, each of the three vertices of the bounding triangle has degree $\geq 3$, and thus, $d_1 + d_2 + d_3 \geq 9$. Hence, (\ref{eq:PreDegreesSum}) implies \marfigsc{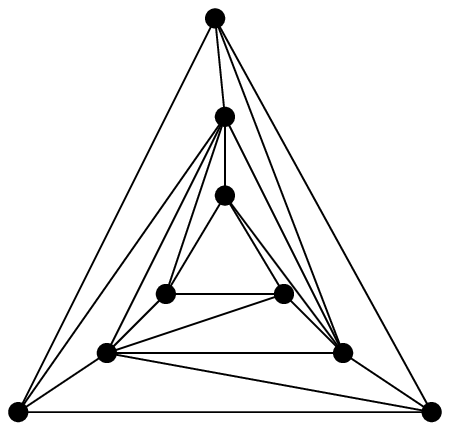}{1.4}

\begin{equation}
\label{eq:DegreesSum}
\mbox{$\sum_i i\, v_i \leq (6n+6) - 9 = 6n - 3 \mbox{,~~~if~} n\geq 1$}.
\end{equation}

\noindent Figure \ref{fi:degrees3} depicts a triangulation of nine points with $v_3=0$. Since we can easily generalize it to a triangulation of $3m$ points, for arbitrarily large values of $m$, we cannot find a better lower bound than $v_3 \geq 0$.

For $i \in \NN$, $i \geq 3$, let
$$
\hat{v}_i = \hat{v}_i(S) := \ex{v_i(T)}
$$
for $T$ uniformly at random in $\Trp{S}$. That is, $\displaystyle \hat{v}_i(S)=\frac{1}{|\Trp{S}|}\sum_{T \in \Trp{S}}v_i(T)$. Due to linearity of expectation, any linear identity or
inequality in the $v_i$'s (such as (\ref{eq:DegreesSum})) will also be satisfied by the
$\hat{v}_i$'s. However, as we will show, the $\hat{v}_i$'s
are more constrained than the $v_i$'s. Some notes concerning these expected degrees are given in \cite{ShWe06b}; they will be extended and improved in a forthcoming companion paper \cite{SSA}. In particular,
there is a constant $\delta > 0$ such that $\hat{v}_3 \ge \delta n$
if $n >0$ and the point set is in general position; recall
Figure \ref{fi:assumption1} to see that general position is indeed necessary here.
Before we establish this bound, let us relate it to the
question about the number of triangulations.
For that, let $\trip{S} := |\Trp{S}|$ and
$\trip{n}:= \max_{|S|=n} \trip{S}$.
\begin{lemma}
\label{le:NumberOfTri}
{\rm (i)} Let $\delta>0$ be a real constant such that, for all $n \in \NN$,
$\hat{v}_3 \ge \delta n$ for any set of $n$ points in general position.
Then, for all $n \in \NNnull$,
\[\trip{n} \leq \left(\frac{1}{\delta}\right)^n ~.\]

\noindent {\rm (ii)} Let $\delta_1 > 0$ be a real constant and $n_0 \in \NN$ such that,
for all $n$, $n_0 \leq n \in \NN$,
$\hat{v}_3 \leq \delta_1\,n$ for any set of $n$ points in general position.
Then for any set $S$ of $n \in \NN$ points in general position, $\trip{S} = \Omega((1 /\delta_{1})^{n})$.
\end{lemma}
\begin{theProof}{\!\!}
(i) Let $S$ be a set of $n>0$ points that maximizes $\trip{S}$
among all sets of $n$ points; without loss of generality,
let $S$ be in general position (a small perturbation of
a point set cannot decrease the number of triangulations).

Note that we can get some triangulations of $\pl{S}$ by
choosing a triangulation of $\pl{S} \setminus \{q\}$
for some $q \in S$, and then inserting $q$ as a vertex
of degree $3$ in the unique face it lands in.
In fact, a triangulation $T \in \Trp{S}$ can be obtained
in exactly $v_3(T)$ ways in this manner (in particular,
if $v_3(T) = 0$, $T$ cannot be obtained at all in this fashion).
This is easily seen to imply that
\begin{equation}
\mbox{$\sum_{T \in \Trp{S}} v_3(T) =
\sum_{q \in S} \trip{S \setminus\{q\}}$} ~.
\label{eq:v3vstr}
\end{equation}\marfigsc{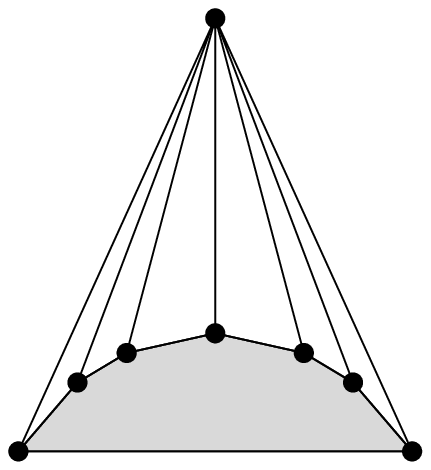}{1.2}
The left hand side of this identity equals $\hat{v}_3\cdot \trip{S}$,
and its right hand side is upper bounded by $n\cdot \trip{n-1}$. Hence,
$$
\mbox{$
\trip{S} \leq \dfrac{n}{\hat{v}_3} \cdot
\trip{n-1} \leq \dfrac{1}{\delta}
\cdot \trip{n-1}
$}
$$
(since we assume that $\hat{v}_3 \geq \delta \, n$),
and thus $\trip{n} \leq \frac{1}{\delta} \cdot \trip{n-1}$ for
all $n \in \NN$. Since $\trip{0} =1$, the lemma follows.

(ii) Along the same lines---omitted.
\end{theProof}

\noindent Recall that $\overline{tr}(n) \leq \trip{n}$, as mentioned above. Therefore, our problem is reduced to finding a large value of $\delta>0$ which satisfies $\hat{v}_3 \geq \delta \, n$ for every $n$-element point set in the plane. Our approach for this problem is explained in Section \ref{sec:charging}, but first, we present an example for analyzing $\hat{v}_3$.

\paragraph{An example.} \marfigsc{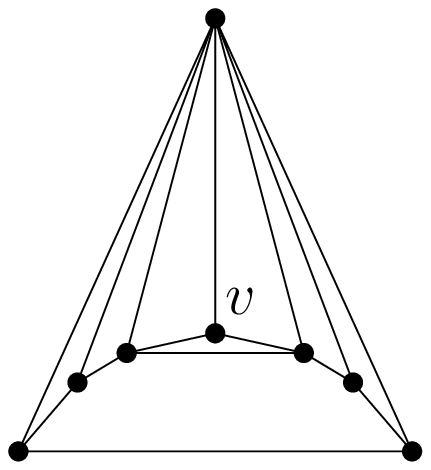}{1.2}

Consider a point set $S^+$ such that $S$ lies on a convex arc that shares its endpoints with an edge of the bounding triangle (as depicted in Figure \ref{fi:degrees4}). Notice that each of the edges depicted in this figure must be present in every triangulation of $S^+$ (since no other edge can cross it). Therefore, the number of triangulations of $S^+$ equals to the number of triangulations of the shaded area. Since this is a convex polygon with $n+2$ vertices, it has $C_n=\Theta^*(4^n)$ triangulations.

For a point in $S$ to have degree 3, its two adjacent vertices in the convex polygon have to be connected to each other, which leaves an $(n+1)$-gon to be triangulated in $C_{n-1}$ ways (as depicted in Figure \ref{fi:degrees5}, where $v$ has degree 3). Therefore, the probabilty that this point has degree 3 is exactly $\frac{C_{n-1}}{C_n}=\frac{n+1}{2(2n-1))}=\frac{1}{4}+O\left(\frac{1}{n}\right)$, and thus, $\hat{v}_3=\frac{n}{4}+O(1)$.

\section{A Lower Bound on $\hat{v}_3$} \label{sec:charging}

The material in this and the following sections is largely borrowed from the earlier paper \cite{ShWe06b}, with the kind permission of Emo Welzl. It is presented here for the sake of completion.

In this section we show how to get a lower bound on $\hat{v}_3$ by using a charging scheme. The basic idea of our analysis is to have each vertex of any triangulation of $S$ charge to vertices of degree $3$. If every vertex charges at least $1$ and each vertex of degree $3$ is charged at most $c$, then we know that $\hat{v}_3 \ge \frac{n}{c}$, so that, by Lemma \ref{le:NumberOfTri}, $\trip{n} \leq c^n$. The actual charging scheme is more involved, for several reasons. First, since there are triangulations that have no degree $3$ vertices, the charging has to go across triangulations. Moreover, we will let vertices charge amounts different from $1$ (even negative charges will occur). However, on average, each vertex will charge at least $1$. The difficulty in the analysis
will be to bound the maximum charge $c$ to a vertex of degree $3$.

\paragraph{A simplified charging scheme.}
We consider the set $S \times \Trp{S}$ and call its elements {\em vints} \
({\em v}ertex {\em in} {\em t}riangulation).
The degree of a vint $(p,T)$ is the
degree (number of neighbors) of $p$ in $T$;
a vint of degree $i$ is called an \emph{$i$-vint}. The overall
number of vints is obviously $n \cdot \trip{S}$, and the
number of $i$-vints is $\hat{v}_i \cdot \trip{S}$. (Note that the three vertices of the enclosing triangle do not participate in this definition.)\marfigsc{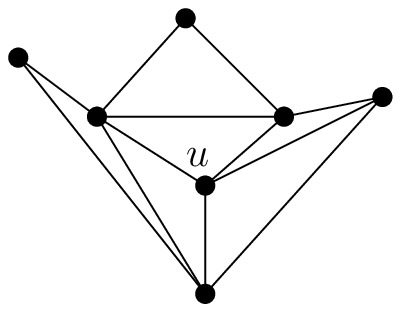}{1.4}

We define a relation on the set of vints. If $u$ and $v$ are vints, then we say that $u \rightarrow v$ if $v$ can be
obtained by flipping one edge incident to $u$ in its triangulation. That is, $u$ and $v$ are associated with the same
point but in different triangulations, and $u$ has to be an $(i+1)$-vint and $v$ an $i$-vint, for some $i\ge 3$.
We denote by $\rightarrow^*$ the transitive reflexive closure of $\rightarrow$, and if $u \rightarrow^* v$, we say that \emph{$u$ can be flipped down to $v$}. Charges will go from vints to $3$-vints they can be flipped down to. For example, the 4-vint $u$ depicted in Figure \ref{fi:charging1} can be flipped down to the 3-vint $v$ in Figure \ref{fi:charging2}.

The \emph{support of a vint $u$} is the number of $3$-vints
it can be flipped down to, i.e.,
$$
\mathrm{supp}(u) :=
\big|\{v \mid \mbox{$v$ is $3$-vint with $u \rightarrow^* v$}\}\big| ~.
$$ \marfigsc{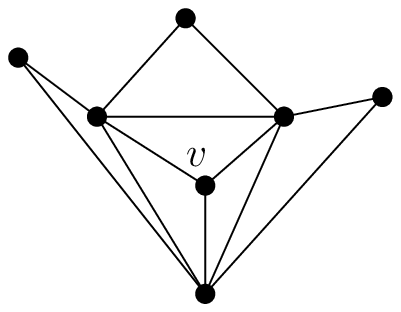}{1.4}

\noindent Out of the four edges incident to the 4-vint $u$ in Figure \ref{fi:charging1}, only one is flippable, and thus, $u$ can only be flipped down to the 3-vint $v$ in Figure \ref{fi:charging2}, and $\supp{u}=1$. The 4-vint $u'$ in Figure \ref{fi:charging3} can be flipped down both to $v$ and to the 3-vint $v'$ in Figure \ref{fi:charging4}, and thus, $\supp{u'}=2$.

A natural charging scheme would let a vint $u$ charge $\frac{1}{\supp{u}}$ to each $3$-vint
it can be flipped down to---in this way, it will charge a total of $1$. In the case depicted in Figures \ref{fi:charging1}--\ref{fi:charging4}, $v$ is charged 1 by $u$ and $\frac{1}{2}$ by $u'$, and $v'$ is charged $\frac{1}{2}$ by $u'$.

Let us gain some understanding of the notion of $\supp{u}$. Note that the removal of an interior point $p$ and
its incident edges in a triangulation $T$ creates a star-shaped polygon (with respect to $p$). We call this the \emph{hole} of the vint $(p,T)$. For a vint $u=(p,T)$, we can remove $p$ and its incident edges from $T$, triangulate the hole that was created,\marfigsc{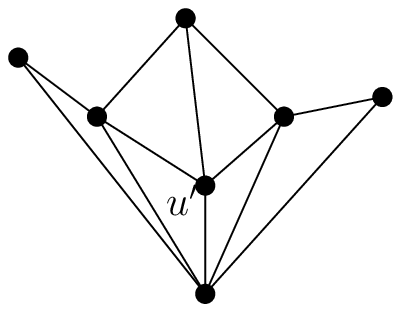}{1.4}and reinsert $p$ as a 3-vint in the unique triangle it lands in. Notice that $u$ flips down to a 3-vint $v$ (and charges it) if and only if $v$ can be obtained as just described. Indeed, each down-flip removes one edge incident to $u$ and the flip cuts off a portion of the hole, until the degree of $u$ becomes 3 and then the removal of $u$ gives a triangulation of its original hole. The converse direction is established similarly. Therefore, $\supp{u}$ equals the number of triangulations of the hole of $u$.

\begin{lemma} For an i-vint $u=(p,T)$:\\[0.2em]
(i) $1 \leq \supp{u} \leq C_{i-2}$, where the upper bound is attained\marfigsc{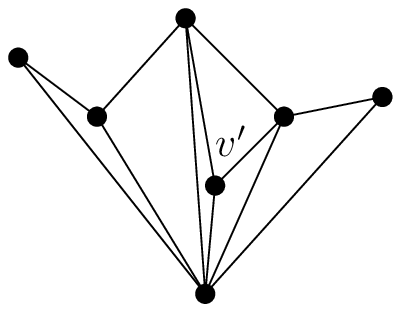}{1.4}
if and only if the hole is
convex.\\[0.2em]
(ii) For a vint $u'$, if $u \rightarrow^* u'$, then $\supp{u} \geq \supp{u'}$.
\end{lemma}
\begin{theProof}{\!\!} (i) This follows from the fact that a convex $i$-gon
has $C_{i-2}$ triangulations, which is the maximum for all $i$-gons. The support is at least 1 since each simple polygon has at least one triangulation.

\noindent (ii) If $u \rightarrow u'$ then the hole of $u'$ is contained in the hole of $u$, with the vertices of the former a
subset of the vertices of the latter. Therefore, every triangulation of the hole of $u'$ can be extended to at least one triangulation of the hole of $u$.
\end{theProof} \marfig{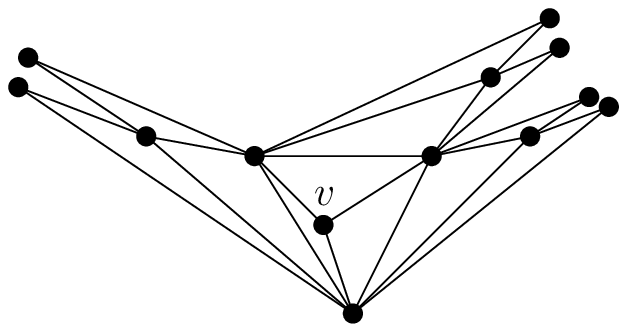}

\begin{lemma} \label{lem:ivints} The number of $i$-vints ($i \geq 3$) that charge a fixed 3-vint is at most $C_{i-1}-C_{i-2}$, and this bound is tight in the worst case.
\end{lemma}

\noindent The general outline of a proof of this lemma can be found in \cite[Lemma 4]{SaSe03}.
\paragraph{The actual charging scheme.}
By Lemma \ref{lem:ivints}, the maximal number of 4-vints that can charge a certain 3-vint is $C_3-C_2=5-2=3$, and the maximal number of 5-vints is $14-5=9$.\marfig{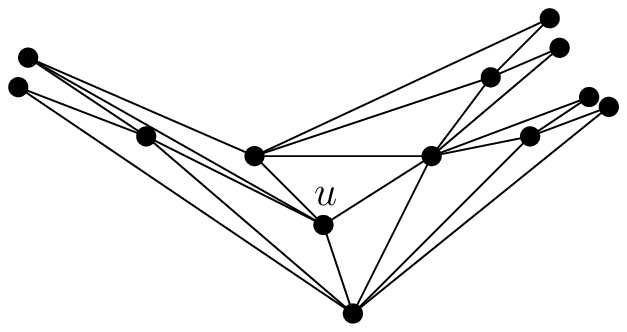}Figure \ref{fi:charging5} depicts a 3-vint $v$ that is charged by three 4-vints and nine 5-vints, and moreover, each of these vints has a support of 1 (i.e., charges 1 to $v$). Figures \ref{fi:charging6} and \ref{fi:charging7} depict two of the 5-vints that charge $v$ (and have a support of 1). This case can easily be extended into a 3-vint charged 1 by $C_{i-1}-C_{i-2}$ $i$-vints, for every $3 \leq i \leq j$. Such a 3-vint is charged at least
\[\overbrace{(C_2-C_1)}^{3-vint}+\overbrace{(C_3-C_2)}^{4-vints}+ \cdots +\overbrace{(C_{j-1}-C_{j-2})}^{j-vints}=C_{j-1}-1=\Theta^*(4^j).\]

\noindent Therefore, in the simplified charging scheme there is no uniform upper bound on the amount charged to individual $3$-vints. \marfig{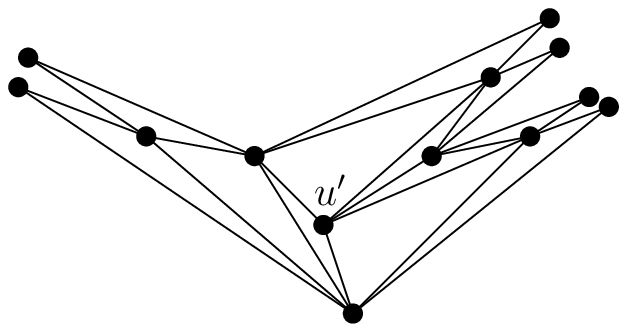}

For that reason, we switch to a charging where
\begin{center}
\emph{\large an $i$-vint $u$ charges $\frac{7-i}{\supp{u}}$ to each
$3$-vint $v$ with $u \rightarrow^* v$.}
\end{center}

\noindent Note that in this scheme, a $3$-vint charges $4$ to itself
(which sounds like bad news), but $7$-vints do not charge at all,
and all $i$-vints with $i \geq 8$ charge a {\em negative} amount,
so that is good news for the $3$-vints (which want to be charged as little as possible).

The overall charge that an $i$-vint can make is $7-i$, so the
overall charge accumulated for all vints associated with
a triangulation $T$ is exactly
$$
\mbox{$
\sum_i (7-i) v_i(T) = \sum_i 7 v_i(T) - \sum_i i\, v_i(T) > 7n-6n=n ,
$}
$$
where we have used (\ref{eq:DegreesSum}) for the inequality. Therefore, on average, each vint gets to charge
at least $1$.

For a $3$-vint $v$ and $i \in \NN$, let $\ch{i}{v}$ be the number of $i$-vints that charge $v$. For an initial upper bound, we can ignore the zero and negative chargings and therefore consider only charges from vints of degree at most 6. Thus, a 3-vint cannot be charged more than
\[4\, \ch{3}{v} + 3\, \ch{4}{v} + 2\, \ch{5}{v} + \ch{6}{v}. \]

\noindent By Lemma \ref{lem:ivints}, $\ch{3}{v} = 1$, $\ch{4}{v} \leq C_3-C_2 = 5-2 = 3$, $\ch{5}{v} \leq 14-5=9$, and $\ch{6}{v} \leq 42-14=28$. Therefore, a 3-vint cannot be charged more than \[4\cdot1 + 3 \cdot 3 + 2 \cdot 9 + 1 \cdot 28 = 59 ,\]

\noindent which implies $\hat{v}_3 \geq \frac{n}{59}$. By
Lemma~\ref{le:NumberOfTri}, this gives an upper bound
of $59^n$ for the number of triangulations of any set of $n$ points.
This bound was established by Santos and Seidel \cite{SaSe03}, which we have derived now
with ideas similar to theirs but in a different setting.

\section{First Improvements} \marfigsc{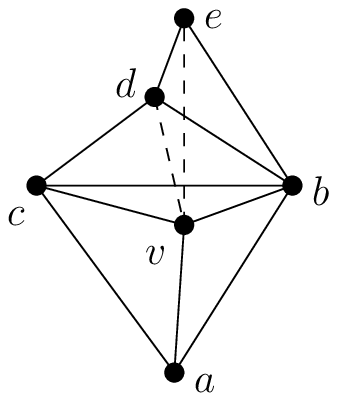}{1.3}

In the current section, we improve the bound $\hat{v}_3 \geq \frac{n}{59}$, presented in the previous section, to the bound $\hat{v}_3 \geq \frac{n}{43}$, repeating the analysis of Sharir and Welzl \cite{ShWe06b}. This improvement is achieved by considering vints with a negative charge (i.e., vints of degree at least 8), and also by taking into account the supports of the positively charging vints (both of which have been ignored in the derivation of the Santos-Seidel bound). We observe that when there is a large positive charge (from vints of degree at most 6), there is also a large negative charge. For example, if indeed $v$ is charged 28 from the 6-vints, it is also charged less than -10164 from 18-vints (the analysis below will clarify this statement).

\paragraph{Flip-trees.}
How do we find the vints that flip down to a given $3$-vint $v=(p_v,T_v)$? Clearly, there is $v$ itself. Consider a flippable edge $e$ (in $T_v$) that is not incident to $p_v$ but is part of a triangle incident to $p_v$. Flipping $e$ yields a $4$-vint $u=(p_v,T_u)$ that can be flipped down to $v$ (by reversing the preceding flip). Similarly, if in the triangulation $T_u$ there is a flippable edge that is
not incident to $p_v$ but part of a triangle incident to $p_v$, then we can flip this edge to get a $5$-vint that can be flipped down to $v$, etc. Figure \ref{fi:firstimp1} depicts a 3-vint $v$, that, by flipping $bc$ into $dv$, turns into a 4-vint that can be flipped down to $v$ (and by afterwards flipping $bd$ into $ev$, turns into a 5-vint that can be flipped down to $v$).

In order to represent this structure, we associate with a
$3$-vint $v=(p_v,T_v)$ a {\em flip-tree} $\tau(v)$, defined as follows. The root of the tree is labeled by the pair $(t_v,N_v)$, where $t_v$ is the hole of $v$ (a triangle) and
$N_v$ is the set of its three vertex points (the neighbors
of $p_v$ in $T_v$).\marfig{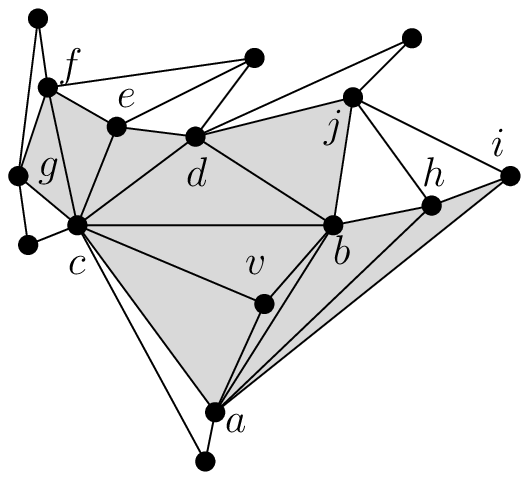}All other nodes of the tree are associated
with a pair $(t,q)$, where $t$ is a face of $T_v$ and
$q$ is a point incident to that face (note that
$t_v$ from the root is not a face of $T_v$---it is the union of the three faces incident to $p_v$). While explaining the structure of the flip-tree in the following paragraphs, we refer to an example depicted in Figures \ref{fi:firstimp2} and \ref{fi:firstimp3}. These figures depict a 3-vint $v$ and its flip-tree, and the nodes of this flip-tree are labeled only by their vertex (and not by their triangle).

(i) Every edge $e$ of $t_v$ gives rise to a child if it can be flipped
in $T_v$. If so, this child is labeled by the triangle
incident to $e$ that is not incident to $p_v$,
and by the point in this triangle which is not
incident to $e$. Therefore, the root has at most three children. In our example, the root has two children---$d$ (since $bc$ is flippable) and $h$ (since $ab$ is flippable). Notice that $\Delta bcd$ is the triangle corresponding to $d$ and $\Delta abh$ is the triangle corresponding to $h$. \marfig{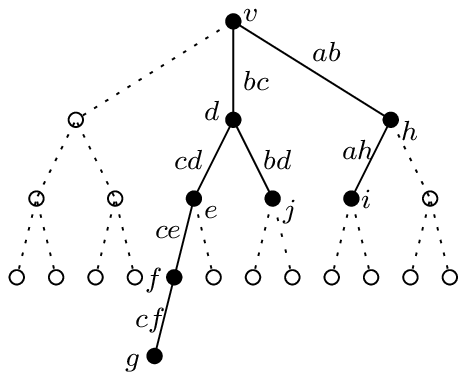}

(ii) Consider now a non-root node of the tree
labeled by $(t,q)$ and an edge $e$ of $t$ incident to $q$. If $e$ is a boundary edge, no child will be obtained via $e$. Otherwise, let $t'$ be the other triangle incident to $e$. If
$t'$ together with the triangle formed by $e$ and
$p_v$ is a convex quadrilateral (where $e$ can be flipped), then
this gives rise to a child of $(t,q)$ labeled by $(t',q')$ where $q'$ is the
vertex of $t'$ that is not incident to $e$. Therefore, a non-root node has at most two children. In our example, the node corresponding to $h$ has a single child, since the quadrilateral $vhia$ is convex, but the quadrilateral $vbjh$ is not.

Note that the union of all triangles of the nodes of
any subtree of $\tau(v)$ (containing the root)
form a polygon that is
star-shaped with respect to $p_v$; this follows easily
by the inductive definition of $\tau(v)$. The triangles (in the triangulation of $v$)
form a triangulation of the polygon, and the subtree is
actually the dual tree of this triangulation. The shaded area in Figure \ref{fi:firstimp2} is the portion of the triangulation dual to the entire flip-tree of $v$. Also, an edge in the flip-tree incident to two nodes that are dual to (i.e., labeled by) the triangles $\Delta_1 , \Delta_2$ in $T_v$, can be regarded as dual to the edge in $T_v$ incident to both $\Delta_1$ and $\Delta_2$. If we retriangulate this polygon in $T_v$ by connecting $p_v$
to all vertices of the polygon, we get a vint that flips
down to $v$. Moreover, every vint $u$ that flips down to $v$ can be obtained in this way (by taking the subtree dual to the hole of $u$). That is:

\begin{lemma}
The subtrees of $\tau(v)$ containing its root
are in bijective correspondence with the vints that
flip down to $v$.
\end{lemma}

\paragraph{Rigid cores.}
In the above, we identified the vints that charge a 3-vint $v=(p_v,T_v)$. The next step is to determine how much these vints charge to $v$. This depends on the support of these vints (i.e., the number of triangulations of their holes)---the smaller the support, the more $v$ is charged. The following analysis only discriminates between vints that have a support of 1, and all other vints.

Consider an edge $e$ of the flip-tree $\tau(v)$, and let us denote the two triangles of $T_v$ that are dual to the nodes adjacent to $e$ as $\Delta_1$ and $\Delta_2$. $e$ is dual to the edge $e'$ of $T_v$, which is adjacent to both $\Delta_1$ and $\Delta_2$. If $e'$ cannot be flipped in the union of these two triangles, then
we say that $e$ is a {\em rigid edge} (with respect to $\tau(v)$). Notice that if one of the two triangles corresponds to the root of $\tau(v)$, $e'$ may be flippable in $T_v$ but not in $\Delta_1 \cup \Delta_2$. For an example, we return to the case depicted in Figures \ref{fi:firstimp2} and \ref{fi:firstimp3}, where the edge $ab$ is flippable in the triangulation, but not in $\Delta abc \cup \Delta abh$. Figure \ref{fi:firstimp4} depicts (again) the flip-tree of $v$ from Figure \ref{fi:firstimp2}, with the distinction that the solid lines represent rigid edges and the dashed lines represent non-rigid edges.  \marfig{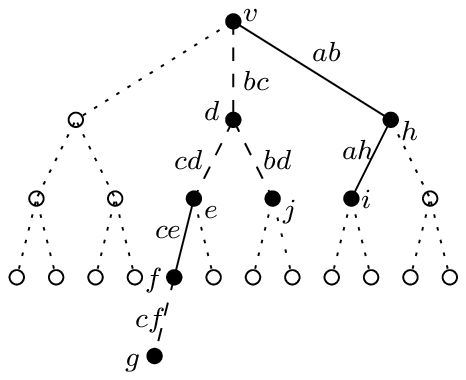}

The \emph{rigid core}, $\tau^*(v)$, of $\tau(v)$ is defined to be
the maximal subtree of $\tau(v)$ that includes the root and
consists exclusively of rigid edges. $\tau^*(v)$ is non-empty, since
it always contains the root of $\tau(v)$. In Figure \ref{fi:firstimp4}, the rigid core consists of the edges dual to $ab$ and $ah$, and of the nodes incident to these edges.

\begin{lemma}
The subtrees of the rigid core $\tau^*(v)$ containing the
root are in bijective correspondence with the vints $u$ that flip
down to $v$ and have a support of $1$.
\end{lemma}

\begin{theProofNoSkip}{\!\!}
Consider a vint $u$ that flips down to $v$. We recall that
$\supp{u} = 1$ if and only if the hole of $u$
has exactly one triangulation. Note that one
triangulation of this polygon can be obtained by
taking the set of triangles in the subtree corresponding
to $u$.\begin{list}{\labelitemi}{\leftmargin=1em}
 \item If all edges in this subtree are rigid, then none
of the dual edges in the triangulation can be flipped.
That is, there is only one triangulation of the hole, since the
set of triangulations of a polygon is connected via edge-flips (as shown by Hurtado et al.\ \cite{HNU99}).
 \item If any of the edges is not rigid, then its dual
edge can be flipped, and so obviously there are at least two
triangulations.
\end{list}
\end{theProofNoSkip} \marfig{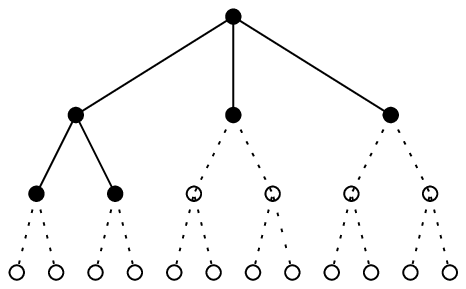}

We next analyze the contribution of a rigid core $R$ to the charging of its 3-vint $v$. Each $j$-edge subtree of $R$ (containing the root) corresponds to a ($j+3$)-vint, and therefore, charges $7-(j+3)=4-j$. Let contr$^+(R)$ (contr$^-(R)$) denote the sum of positive (negative) charges coming from subtrees of $R$. That is, contr$^+(R)$ (resp., contr$^-(R)$) is the sum of the charges coming from subtrees with $j \leq 3$ (resp., $j \geq 5$) edges.

Given a tree, we let the level of an edge denote the level of the node at its bottom (where the root is of level 0). Given a rigid core, we let $\lambda_i$, $i \in \{1,2,3\}$, denote the number of level-$i$ edges it contains. Moreover, we denote the number of nodes at level 1 with two child-edges by $\nu_2$. There are several restrictions on these parameters: $\lambda_1 \leq 3$, $\lambda_2 \leq 2\lambda_1$, $\lambda_3 \leq 2\lambda_2$, and $\nu_2 \leq \lambda_2/2$. For example, for the rigid core depicted in Figure \ref{fi:firstimp5}, we have $\lambda_1=3$, $\lambda_2=2$, $\lambda_3=0$, and $\nu_2=1$.

We can express contr$^+(R)$ by using the above parameters:
\begin{eqnarray}
&&
4 \cdot
     \overbrace{1}^{\hspace{-0.2ex} \placefig{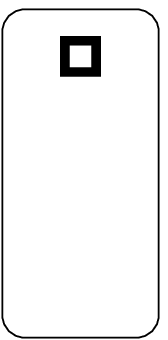}{3ex}} +
3 \cdot
     \abc{\lambda_1}{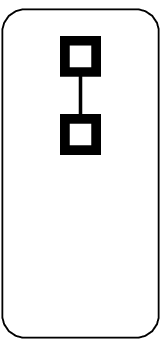} +
2 \cdot (
     \abc{\mbox{$\binom{\lambda_1}{2}$}}{type5-2} +
     \abc{\lambda_2}{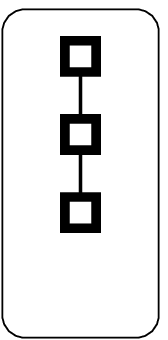}
) \nonumber \\
&+&
1 \cdot (
     \abc{\mbox{$\binom{\lambda_1}{3}$}}{type6-3} +
     \abc{\lambda_2(\lambda_1-1)}{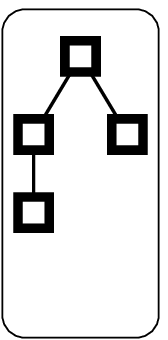} +
     \abc{\nu_2}{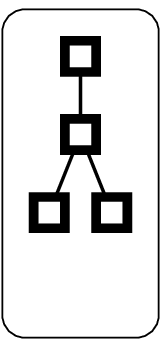} +
     \abc{\lambda_3}{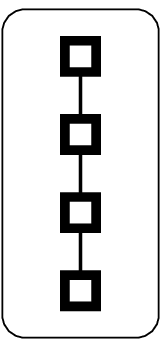}
)\nonumber \\
&=&
4 + \mbox{$\binom{\lambda_1}{3}$} + \lambda_1^2 + 2\lambda_1
  + (\lambda_1 + 1)\lambda_2 + \lambda_3 + \nu_2
\nonumber \\
&=&
\left\{
\begin{array}{ll}
20 + 4 \lambda_2 + \lambda_3 + \nu_2 & \mbox{if $\lambda_1=3$,}\\
12 + 3 \lambda_2 + \lambda_3 + \nu_2 & \mbox{if $\lambda_1=2$, and}\\
7 + 2 \lambda_2 + \lambda_3 + \nu_2 & \mbox{if $\lambda_1=1$.}
\end{array}
\right.\label{eq:contrpl}
\end{eqnarray}

\noindent For example, if $R$ is a complete tree of height 3, then $\lambda_1 = 3$, $\lambda_2 = 6$, $\lambda_3 = 12$, and $\nu_2=3$. Therefore, contr$^+(R)= 20+4\cdot6+12+3=59$.

\begin{lemma}
Let $R$ be a rigid core with $m$ edges and without any level-4 edges, then
\begin{flalign*}
 (i) & \quad \contr^+(R) \leq \frac{13+9m}{2}. \\
 (ii) & \quad \contr^-(R) \leq min\{0,14-3m\}.
\end{flalign*}
\end{lemma}

\begin{theProof}{\!\!}
(i) Note that $\nu_2 \leq \frac{\lambda_2}{2}$ and $\lambda_2+\lambda_3 = m - \lambda_1$. If $\lambda_1=3$, then by using (\ref{eq:contrpl}) we get
\[ \contr{}^+(R) \leq 20 + \frac{9}{2}\lambda_2+\lambda_3 \leq 20 + \frac{9}{2}(\lambda_2+\lambda_3) = 20 + \frac{9}{2}(m-\lambda_1) = \frac{13+9m}{2}.\]

\noindent In a similar manner, we get a bound of $\contr^+(R) \leq \frac{10+7m}{2}$ when $\lambda_1=2$, and a bound of $\contr^+(R) \leq \frac{9+5m}{2}$ when $\lambda_1=1$. Obviously, these latter bounds are dominated by the bound of $\lambda_1=3$.

(ii) If $m \leq 4$, then $R$ does not contain any $i$-vints with $i \geq 8$, and thus, $\contr^-(R) = 0$. If $m = 5$,  there is a single 8-vint that consists of the entire rigid core, and thus, $\contr^-(R) = 7 - 8 = -1$. Notice that the above bound holds for both of these cases. \marfig{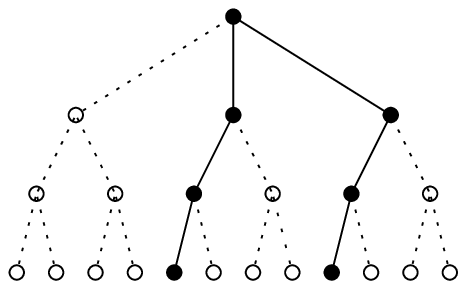}

For $m \geq 6$, the vint that consists of all the edges of the rigid core is an ($m+3$)-vint that charges $7-(m+3)=4-m <0$. By removing a single leaf from the rigid core, we get an ($m+2$)-vint that charges $7-(m+2)=5-m <0$. A rigid-core of size at least 6 that has no level-4 edges must have at least two leaves, and therefore, contains at least two ($m+2$)-vints. (Figure \ref{fi:firstimp6} depicts a rigid core with $m=6$ and exactly two leaves.) By summing up the above, we get $\contr^-(R) \leq (4-m) + 2(5-m) = 14-3m <0$, which implies that the bound holds for this case too.
\end{theProof}

\paragraph{The maximal charge of a flip-tree.}
We are now ready to analyze how much can a 3-vint $v$ get charged by the vints of its flip-tree (which are the only vints that charge it, as shown above).

First, for $j \geq 4$, we ignore $j$-level edges of the flip-tree. Since such edges cannot participate in 4-, 5-, or 6-vints, this can only increase the charge of the flip-tree. Moreover, we assume that every 4-, 5-, or 6-vint that is not entirely in the rigid core has a support of 2. Since such a vint has a support of at least 2, this also can only increase the charge of the flip-tree. Finally, we consider $i$-vints with $i \geq 8$, only if they have a support of 1 (i.e., contained in the rigid core). Since such vints with a larger support have a negative charge, ignoring them can only increase the charge of the flip-tree.

We further simplify the analysis, by assuming that the flip-tree is complete up to level 3 (i.e., the root has three child edges, and every level-1 or level-2 node has two child edges). If an edge is missing in the flip-tree, we can add it as a non-rigid edge. Since we only consider vints with a non-rigid edge if they have a positive charge, this can only increase the charge of the flip-tree.

By using all of the above assumptions, we notice that $v$ cannot be charged by more than (the second term represents the charge from vints not entirely in the rigid core)
\begin{flalign*}
& \contr{}^+(R)+\frac{1}{2}(59-\contr{}^+(R))+\contr{}^-(R) \\
= \quad & \frac{59}{2} + \frac{\contr{}^+(R)}{2}+\contr{}^-(R) \\
\leq \quad & \frac{118+(13+9m)}{4}+\contr{}^-(R) \\
= \quad & \frac{131+9m}{4}+\contr{}^-(R),
\end{flalign*}

\noindent where $R$ is the rigid core of the flip-tree, and $m$ is the number of its edges. If $m \leq 4$, then $\contr^-(R) = 0$, and the expression is bounded by $\frac{131+36}{4}=41\frac{3}{4}$. If $m \geq 5$, then the expression is bounded by \[\frac{131+9m}{4}+(14-3m)=\frac{187-3m}{4} \leq \frac{187-3\cdot5}{4} = 43.\]

\noindent Therefore, we get a bound of $\hat{v}_3 \geq \frac{n}{43}$ for any set of $n$ points. Figure \ref{fi:firstimp7} depicts a flip-tree that achieves this bound by using our pessimistic and simplified analysis (as before, the solid lines represent rigid edges and the dashed lines represent non-rigid edges).\marfig{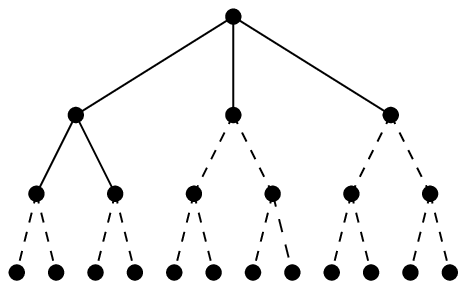}In this flip-tree, the rigid core generates a 3-vint (which is the root), three 4-vints, five 5-vints (out of the possible 9), six 6-vints (out of possible 28), and one 8-vint. This implies that the charge of this flip-tree (again, using our pessimistic form of analysis) is \[ 4\cdot1+ 3\cdot3 + 2\left(5+\frac{4}{2}\right)+\left(6+\frac{22}{2}\right)-1\cdot1=43 . \]

\paragraph{Can we do better?}
We now discuss possible improvements for the bound presented above. There are some obvious places where the simplified analysis presented above can potentially be improved --- it considers vints with a negative charge only if they are entirely in the rigid core, and it assumes that every vint with a positive charge has a support of at most 2. For example, we can improve the analysis by noticing that every vint with at least two non-rigid edges has a support of at least 3.

The following sections present a more complex analysis that exploits these issues, and shows that the maximum charge to a 3-vint is smaller than 30, thus yielding the bound of $\hat{v}_3 > \frac{n}{30}$ (and $\overline{tr}(n)<30^n$).\marfig{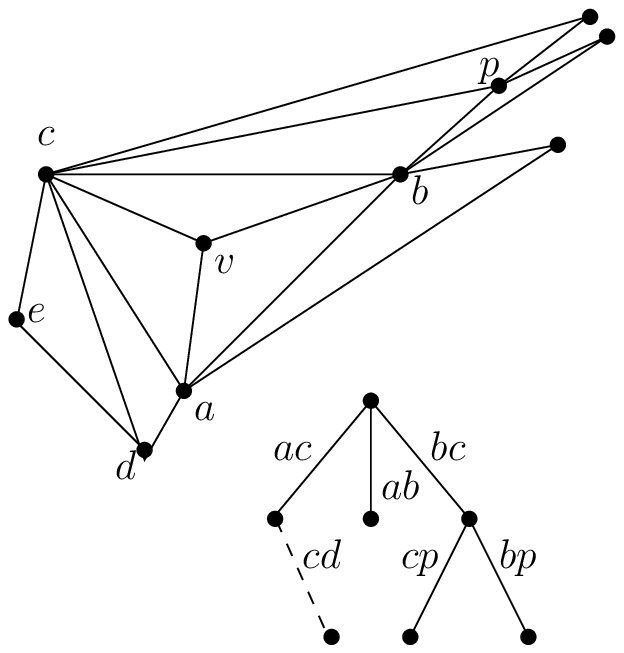}A natural question would be how much further can we improve this bound. To answer this, we consider the 3-vint $v$ depicted in Figure \ref{fi:firstimp8} (together with the respective flip-tree). This is exactly the flip-tree in Figure \ref{fi:firstimp7}, after removing all of its non-rigid edges, except for $cd$. For the charge coming from the rigid core, we can repeat the above analysis, and get $ 4\cdot1 + 3\cdot3 + 2\cdot5 + 1\cdot6 - 1\cdot1 = 28$. The non-rigid edge is present in one 5-vint, two 6-vints, three 8-vints, and one 9-vint. In the following sections, we explain how to analyze the supports of such vints (i.e., count the number of triangulations of their holes). For now, we only state that the 5-vint has a support of 3, both 6-vints have a support of 4, two 8-vints have a support of 8, the third 8-vint has a support of 7, and the 9-vint has a support of 12. (All of these statements can be verified directly, though tediously, from Figure \ref{fi:firstimp8}.) Therefore, $v$ gets charged \[28 + 2\cdot1\cdot\frac{1}{3} + 1\cdot2\cdot\frac{1}{4} - 1\left(2\cdot\frac{1}{8}+1\cdot\frac{1}{7}\right )-2\cdot1\cdot\frac{1}{12}=28\frac{17}{28}.\]
This implies that even an optimal analysis of the flip-tree will not achieve a better bound than $28\frac{17}{28}$. We believe that this is indeed the flip-tree with the largest charge possible. However, recall that our technique gives a bound for the worst-case 3-vint, when we  actually need a bound for the average 3-vint. Therefore, it might be possible to achieve a much smaller bound than $28\frac{17}{28}$, by using methods that consider the average charge to a 3-vint. It seems likely that the actual value of $\overline{tr}(n)$ is much closer to the current lower bound of $8.4853^n$ than to our upper bound of $<30^n$.

\section{Infrastructure for an Improved Analysis} \label{sec:foundations}

The three remaining sections of this paper describe an improved analysis, proving that a 3-vint always gets charged less than 30. This extended analysis proceeds by case analysis according to the possible RCs (rigid cores). The current section presents some notations and rules which will be used repeatedly in the analysis of charges of 3-vints. Section \ref{sec:vintExt} provides more advanced rules that are used to bound the supports of vints with negative charges. Finally, Section \ref{sec:analysis} presents the analysis itself. \marfigsc{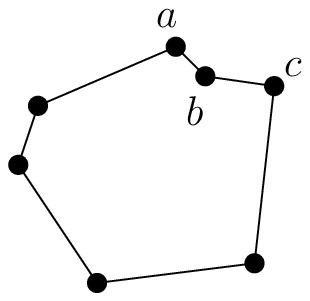}{1}

\paragraph{Catalan numbers --- extensions.} The Catalan numbers were introduced in Section \ref{sec:Intro}, for counting triangulations of point sets in convex position. We will also need the following extension of these numbers, for point sets in ``almost" convex position. Consider a simple polygon with $n+1$ vertices in convex position, and an additional reflex vertex $b$, which blocks the visibility between its two direct neighbors, $a$ and $c$, and not between any other pair of vertices (see Figure \ref{fi:Catalan1}). The number of triangulations of this polygon is equal to the number of triangulations of a convex set of $n+2$ points, which do not contain the edge $ac$ (see Figure \ref{fi:Catalan2}). This number is easily seen to be $C_n-C_{n-1}$, and we denote it by $C'_n$. \marfigsc{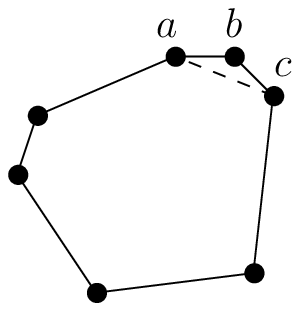}{1}

Consider a simple polygon with $n$ vertices in convex position, and two additional reflex vertices, which are not direct neighbors, so that, as above, each of them only blocks the visibility between its two neighbors (as depicted in Figure \ref{fi:Catalan3}). Similarly to the previous case, the number of triangulations of this polygon is equal to the number of triangulations of a convex set of $n+2$ points, which do not contain the edge $ac$ and the edge $df$. By using the inclusion-exclusion principle, this number is easily seen to be $C_n-2C_{n-1}+C_{n-2}$, and we denote it by $C''_n$.

We can further generalize this notation into a polygon with $r\leq\frac{n}{2}$ reflex vertices, with the above minimal-blocking property, when no two of these vertices are neighbors. By using the inclusion-exclusion principle again, it can be easily seen that the number of triangulations of such a polygon is $C^{(r)}_n=\sum_{i=0}^{r}(-1)^{i}\binom{r}{i}C_{n-i}$. \marfigsc{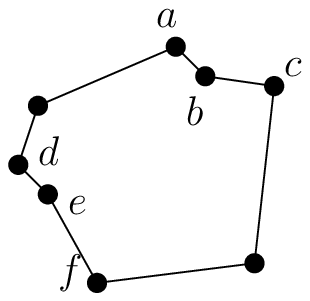}{1}

\paragraph{The $tr(\cdot)$ function.} This function is defined with respect to a simple star-shaped polygon $P$, and its input is an internal chord of $P$. The value of the function is the number of triangulations of $P$ which contain the chord. For example, when referring to the polygon in Figure \ref{fi:trFunc}, we have $tr(bd)=C_2=2$ and $tr(ad)=C'_2=1$. When we wish to refer to the number of triangulations which contain more than one chord, we put a plus sign between the chords. For example, using the same polygon, we have $tr(bd+be)=1$.

We usually use this notation when each triangulation must contain exactly one out of two specific chords, $A$ and $B$.\marfigsc{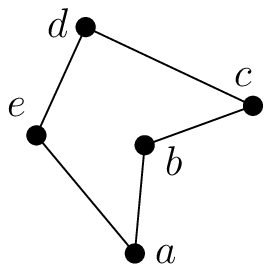}{1}In such a case, the number of triangulations of the polygon is $tr(A)+tr(B)$. For example, the polygon in Figure \ref{fi:trFunc} has $tr(ad)+tr(be)=1+2=3$ triangulations.

\paragraph{The vertex of an edge.}This term is used with respect to a specific flip-tree. Consider an edge $H$ in the flip-tree, which is dual to an edge $pq$ of the triangulation. Let $pqa$ and $pqb$ be the triangles adjacent to the edge $pq$, so that the node in the flip-tree dual to $pqa$ is the parent of the node dual to $pqb$. In this case, we say that $b$ is {\em the vertex of the edge} $H$, or, equivalently, of the edge $pq$. (Recall that $b$ was used earlier to label the node dual to $pqb$.) The {\em vertices} of a vint $v$ are the vertices of the edges in the flip-tree of $v$, plus the three vertices of the triangle containing the point of the vint. \marfig{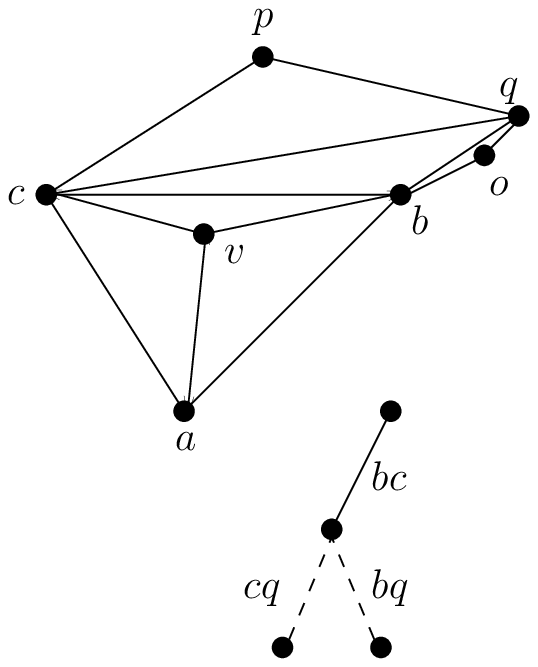}

For example, in Figure \ref{fi:handicapped}, $q$ is the vertex of $bc$ and $p$ is the vertex of $cq$ (in the flip-tree of $v$).

\begin{rule0} \label{rule:SingleFlip}
Let $D$ be a level-1 or level-2 edge which is part of the rigid core (RC). Assume that $D$ has two child-edges in the flip-tree, $E$ and $F$, and that they are not part of the RC. Flipping $E$ or $F$ might cause $D$ to be flippable, but it is not possible for both of them to have this property. \end{rule0}

\paragraph{Explanation.}For an example of the assumptions in the rule, see Figure \ref{fi:handicapped}. In this figure, $v$ is the 3-vint, and $bc$, $cq$, and $bq$ are dual to $D$, $E$, and $F$, respectively. In the flip-tree, a dashed line represents a non-rigid edge, and a solid line represents a rigid edge. In the notation of the figure, assume, without loss of generality, that $abq$ forms a right turn. Since $o$ is to the right of the line supporting $\overrightarrow{bq}$ (directed from $b$ to $q$), $abo$ is also a right turn. This means that, after flipping $bq$, $bc$ remains unflippable. \hfill $\square$

We refer to Figure \ref{fi:handicapped} again, and consider the 5-vint which uses $bc$ and $bq$. Such a 5-vint can have a support of at most 2, no matter where $o$ is, since $o$ can never see $a$. We refer to such a 5-vint as a {\em handicapped} 5-vint. In other words, it is a 5-vint which uses a rigid level-1 edge, and the level-2 edge which cannot cause its parent to be flippable, no matter where its vertex is. Rule \ref{rule:SingleFlip} implies that each rigid level-1 edge with two non-rigid child-edges, produces at least one handicapped 5-vint. Note that the level-2 edge of a handicapped 5-vint may or may not be rigid.

\begin{rule0} \label{rule:RemainsRigid}Let $D$ be a level-1 or level-2 edge which is part of the RC, so that it has a non-rigid child $E$ and a rigid child $F$. After flipping $E$, $F$ remains unflippable. \marfig{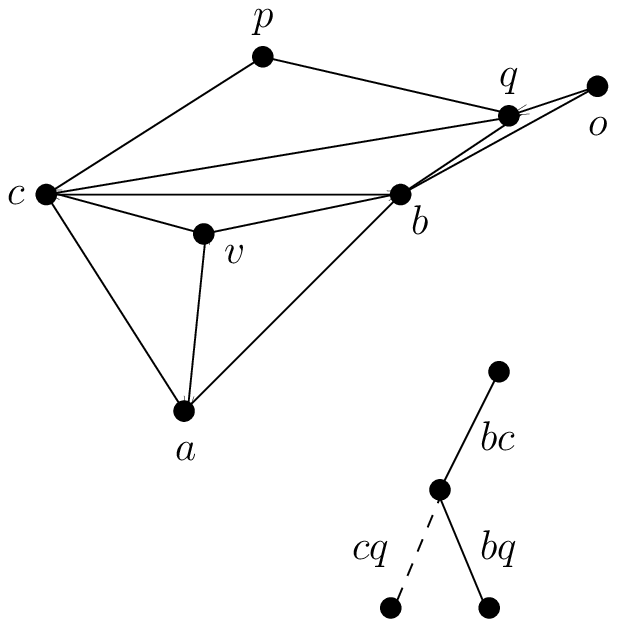} \end{rule0}

\paragraph{Explanation.} Without loss of generality, we assume that $D$ is a level-1 edge. For an example of the assumptions in the rule, see Figure \ref{fi:SiblingFlip}. In the figure, $D$, $E$, and $F$ are dual to $bc$, $cq$, and $bq$, respectively. For $F$ to be rigid, $o$ has to be to the left of the line supporting $\overrightarrow{cq}$ (it cannot be to the right of the line supporting $\overrightarrow{cb}$, or else it would not be visible from $v$). This means that $o$ is also to the left of the line supporting $\overrightarrow{pq}$. This implies that the quadrilateral $pboq$ is non-convex, and that after flipping $cq$, $bq$ remains unflippable. The same argument implies that $bq$ remains unflippable after flipping any child edges of $cq$. The symmetric case, where the flippable edge belongs to the handicapped 5-vint, is depicted in Figure \ref{fi:SiblingFlip2}. Notice that the proof also remains valid if we replace $E$ with one of its child edges. \hfill $\square$

\smallskip

\begin{rule0} \label{rule:LargestRC}\marfig{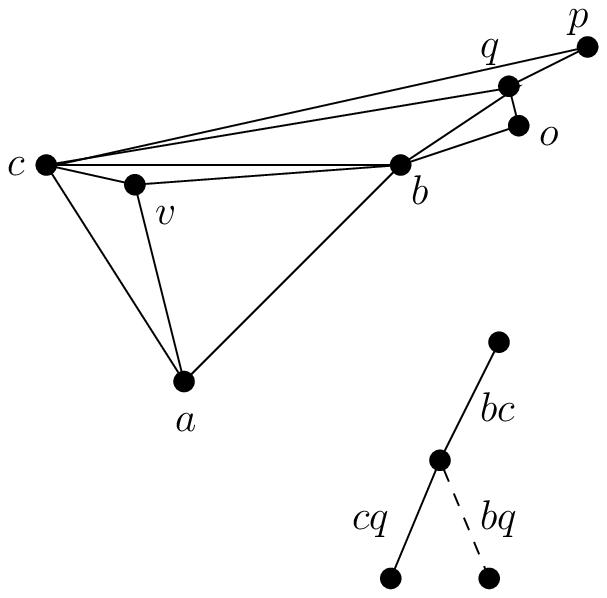}Consider a rigid core with at least four edges. Expanding it by adding a level-3 edge $H$ to the RC, cannot increase the charge (the entire charge of the 3-vint, not only from rigid core edges). \end{rule0}

\paragraph{Explanation.}There is a single vint with a positive charge that uses $H$, which is a 6-vint. Let $m>1$ denote the support of this 6-vint in the original configuration. If the 6-vint did not exist, we define $m=\infty$. In the new layout, in which $H$ becomes rigid, the charge gained from the 6-vint increases by $1-\frac{1}{m}$. There is at least one 8-vint which contains the 6-vint and two additional RC edges. In the original triangulation, this 8-vint had a support of at least $m$. In the new layout, the charge received from the 8-vint is $-1$, which means that it decreased by at least $1-\frac{1}{m}$. Therefore, adding $H$ to the RC cannot increase the charge. \hfill $\square$

\section{Vint extensions} \label{sec:vintExt} One of the techniques that are used in order to bound positive charges of vints, is to extend them into vints with negative charge, such that this charge neutralizes some (or all) of the positive charge (see, for example, Rule \ref{rule:LargestRC} above). Typically, but not exclusively, we add RC edges to the vint, since they have a relatively small influence on the support of the vint.\marfigsc{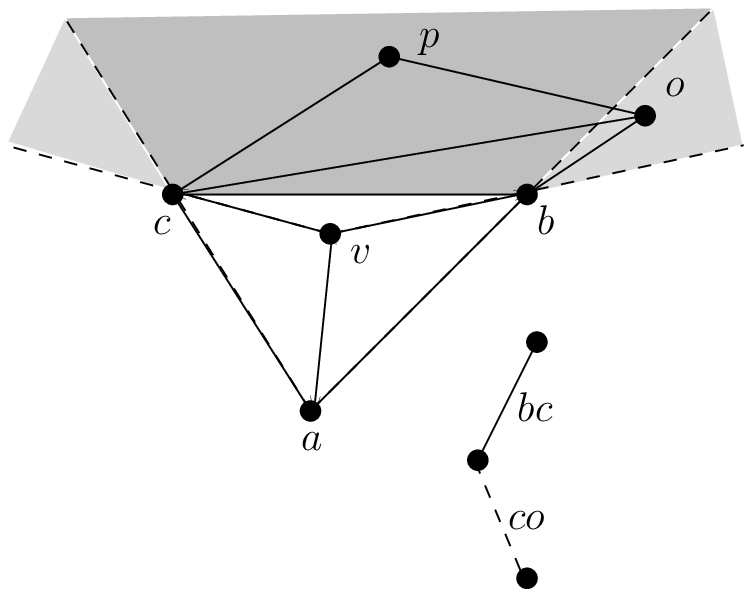}{1.6}This section presents additional rules which calibrate the effect of such extensions on the total charge.

\subsection{Non-visible terrains} \label{sec:non-visib}

Let $v$ be a 3-vint whose vertex has $a$, $b$, and $c$ as neighbors (see Figure \ref{fi:NVter1}). The {\em non-visible terrain} of $bc$ is defined as follows. Draw two half-lines from the vertex $a$, one passing through $b$, and the other through $c$. The truncated unbounded wedge bounded by these two half-lines and by the edge $bc$ is referred to as the {\em visible terrain} of $bc$; it is the darkly shaded area in Figure \ref{fi:NVter1}. The {\em non-visible terrain} of $bc$ consists of two wedges, lightly shaded in Figure \ref{fi:NVter1}, one bounded by the halflines that emanate from $c$, lie on the lines $\overrightarrow{ac}$, $\overrightarrow{cv}$, and do not contain $a$, $v$, and the other bounded by the halflines that emanate from $b$, lie on the lines $\overrightarrow{ab}$, $\overrightarrow{bv}$, and do not contain $a$, $v$. Consider the subtree of the flip-tree of $v$, which is formed by taking the edge dual to $bc$ and its descendants.\marfigsc{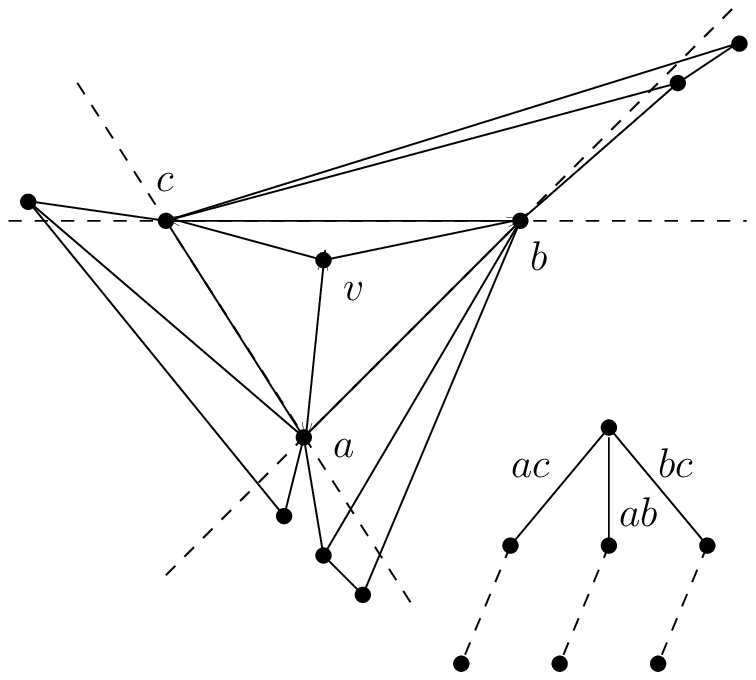}{1.6}When talking about vertices in the non-visible terrain of $bc$, we only refer to vertices of edges dual to edges in this subtree \footnote{Note that any such vertex must lie either in the visible or in the non-visible terrain of $bc$.}. Any vertex in the non-visible terrain of $bc$ cannot see any of the vertices in the non-visible terrains of $ab$ and $ac$ (in the sense that the segment between them is not fully contained in the hole of the respective vint; such a case is depicted in Figure \ref{fi:NVter2}). We say that the vertex of an edge $E$ is in its non-visible terrain, if it is in the non-visible terrain of the level-1 ancestor of $E$ (which may be $E$ itself). By definition, the vertex of an RC edge has to be in its non-visible terrain.

\begin{rule0} \label{rule:non-visib} Consider two (or three) vints without a common level-1 edge, and assume that all of their vertices are in their non-visible terrains. We can create a larger vint by appending the edges of these vints. The support of this larger vint will be the product of the supports of the original vints.
\end{rule0}

\paragraph{Explanation.}By the above definitions, vertices from different vints cannot see each other, which implies the rule. \hfill $\square$ \marfig{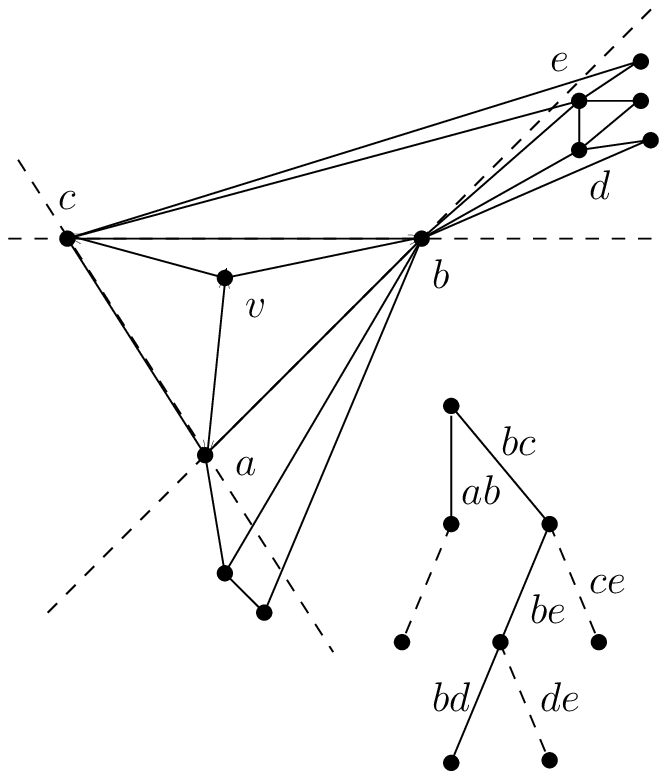} \smallskip

For a typical application of this rule, consider a 5-vint with a rigid level-1 edge, a non-rigid level-2 edge, and a support of 2. The vertex of the level-2 edge must be in its non-visible terrain, for otherwise the 5-vint would have a support of 3 (this is depicted in Figure \ref{fi:NVter1}, where the 5-vint has a support of 3 if and only if $p$ is in its visible terrain.). See Figure \ref{fi:NVter2} for an example, which depicts such a 5-vint in each of the subtrees of the flip-tree. Appending the edges of two such 5-vints results in a 7-vint with a support of four. Appending the edges of three such 5-vints results in a 9-vint with a support of eight. Additional RC edges can also be appended without increasing the support.

The support of more complex vints can be bounded this way. For example, building a vint using all the edges in Figure \ref{fi:NVter3}, results in a 10-vint with a support of 12 (a 5-vints with a support of 2 in the subtree of $ab$;\marfig{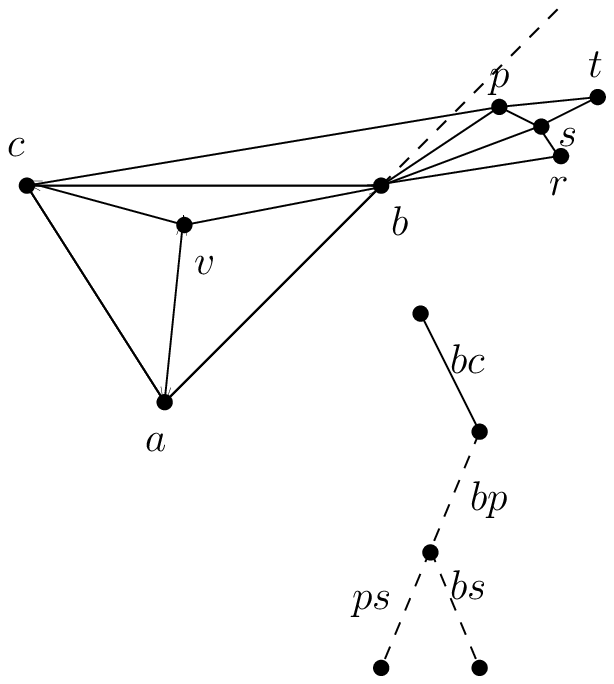}in the subtree of $bc$, $ce$ must be present, and when $cd$ is present there are four possible triangulations, giving a total support of 6).

\begin{rule0} \label{rule:non-visib2} Consider a 5-vint with a rigid level-1 edge, a level-2 edge, and a support of at most 2. (a) At least one of the two 6-vints, which extend the 5-vint with a level-3 edge, is entirely in its non-visible terrain (i.e., the vertices of its three edges are in their non-visible terrain). (b) If the 5-vint, which uses the same level-1 edge with a different level-2 edge (the sibling edge), has a support of 3, both 6-vints (extending the first 5-vint) are entirely in their non-visible terrain. (c) If the 5-vint is handicapped, the two 6-vints are entirely in their non-visible terrain. \end{rule0}

\paragraph{Explanation.} In each of the cases (a)--(c), the vertices of the level-1 and level-2 edges are in their non-visible terrain, as easily follows from the assumptions.\marfig{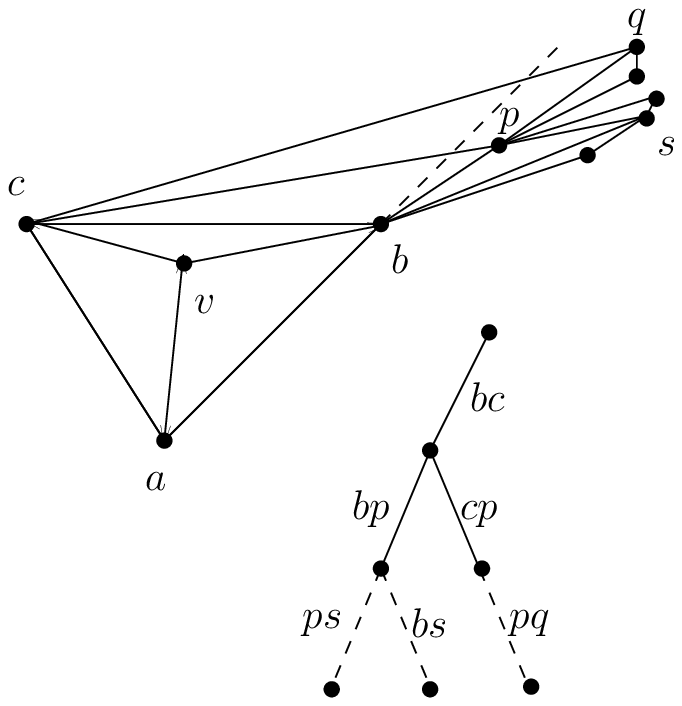}Thus, we only need to show that the vertices of the level-3 edges are in their non-visible terrain. Consider first case (c) of a handicapped 5-vint, as depicted in Figure \ref{fi:NV6vint1} (the 5-vint containing the edges $bc$ and $bp$ is handicapped). Since $p$ must be to the right of $\overrightarrow{ab}$, the halfline which emanates from $b$, lies on the line $\overrightarrow{ab}$, and does not contain $a$, must be counterclockwise to the ray from $a$ through $p$. The vertices of the level-3 edges of the 6-vints which extend the 5-vint ($t$ and $r$ in the figure), must be to the right of the ray from $a$ through $p$, since otherwise they will not be able to see $v$. This implies that these vertices are to the right of $\overrightarrow{ab}$, and thus, in their non-visible terrain.

Consider a 5-vint with a support of 3, a rigid level-1 edge, and a level-2 edge $A$. The 5-vint which uses the sibling edge of $A$ is either a handicapped 5-vint, or entirely in the RC (see Rule \ref{rule:SingleFlip}; the distinction is because we have defined handicapped 5-vints only for non-RC vints). Such a 5-vint is depicted in Figure \ref{fi:NV6vint2} (the 5-vint which contains $bc$ and $bp$).\marfig{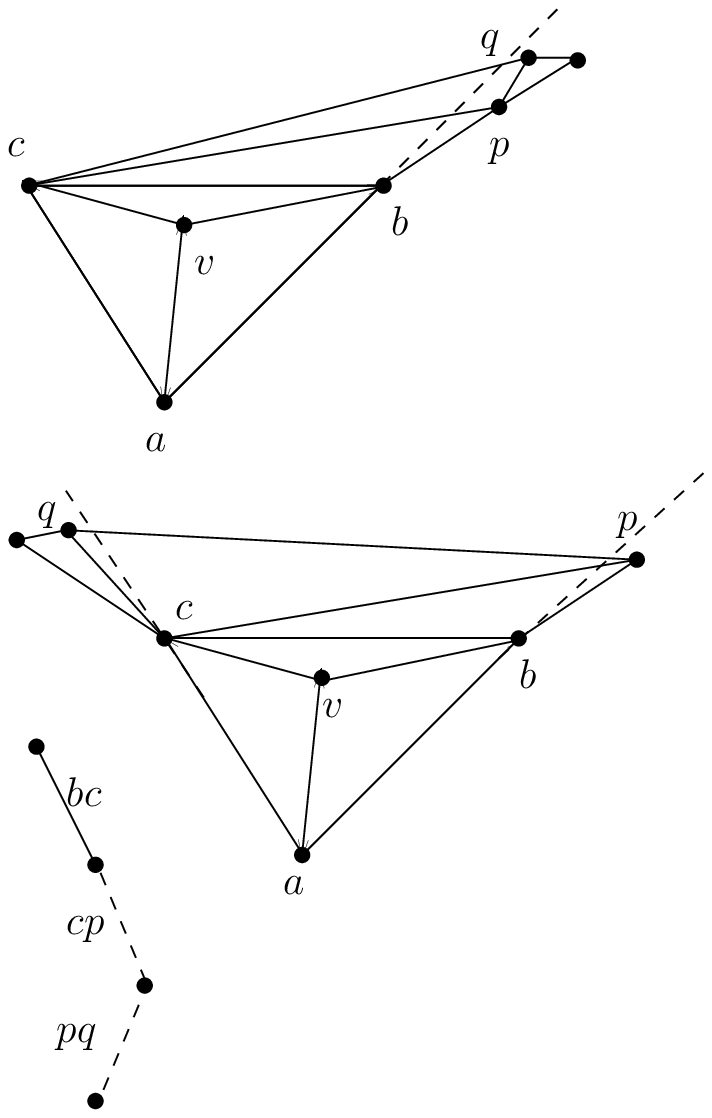}The analysis in the preceding paragraph applies here as well, and implies the claim in (b).

Finally, consider case (a). Let $q$ denote the vertex of the level-2 edge of the 5-vint. It is easily checked that if the 5-vint has a support of at most 2, $q$ cannot see $a$ (see Figures \ref{fi:NV6vint2} and \ref{fi:NV6vint3}), so $q$ must be in its non-visible terrain, and thus, at least one of the level-3 triangles with $q$ as a vertex lies in its non-visible terrain.

\subsection{Non-visible subtrees} \label{sec:non-visib2}

This subsection presents an additional application of non-visible terrains. We do not present it as a rule, since it is a general method, and we will later use several of its variants.

Consider a level-1 edge and one of its child edges, both belonging to the RC. By construction, the vertices of these edges are in the same wedge of their non-visible terrain. This case is depicted in Figure \ref{fi:NVsubt1}, where $o$, $p$, and $q$ are in their non-visible terrain wedge bounded by the ray from $c$ through $b$ and the ray from $v$ through $b$ (the shaded wedge in the figure). These vertices cannot see any of the vertices from the subtree of the flip-tree rooted at $ab$ (such as the vertex $d$ in Figure \ref{fi:NVsubt1}); we refer to this subtree as the {\em non-visible subtree} of the vertices in the wedge.\marfig{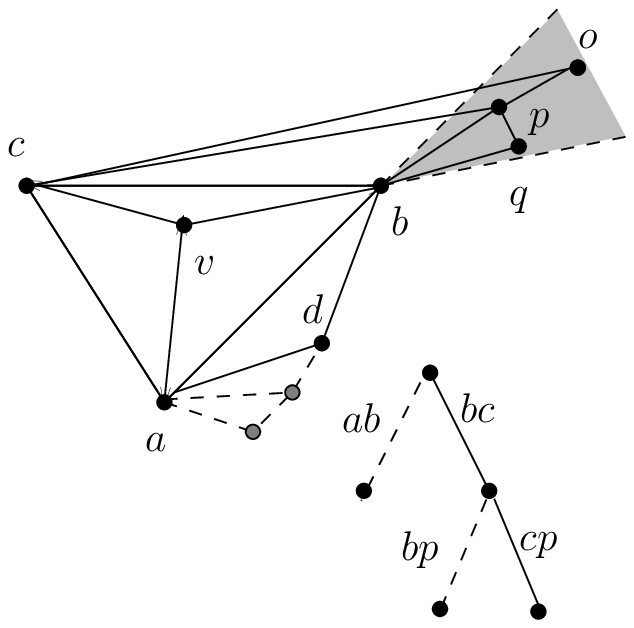}Each non-visible subtree can contain up to five 6-vints, four of which use a level-3 edge, and one of which uses two level-2 edges. The two RC edges assumed above do not participate in any of these 6-vints, since they are in a different subtree. Using these two edges, any of the five 6-vints can be extended into an 8-vint with the same support. In Figure \ref{fi:NVsubt1}, $bc$ and $cp$ are two such RC edges, which can be used to extend possible 6-vints in the subtree of $ab$.

Here are two additional applications of this observation:

(i) Consider a handicapped 5-vint. The vertices of the 5-vint are in the same wedge, and hence have the same non-visible subtree. This time, each 6-vint from the non-visible subtree can be extended into an 8-vint with a double support. In Figure \ref{fi:NVsubt1}, $bc$ and $bp$ create such a 5-vint, which can be used to extend 6-vints in the subtree of $ab$. In this way, half of the charge of these 6-vints is eliminated. \marfig{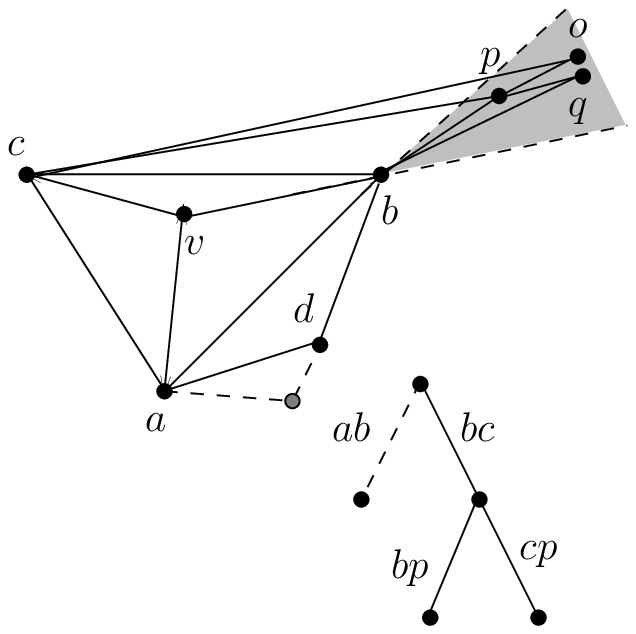}

(ii) Consider a level-1 edge which has two child edges, all contained in the RC. The vertices of the three edges must be in the same wedge. The non-visible subtree of these vertices can contain up to two 5-vints; each can be extended into an 8-vint with the same support, halving the charge of any such 5-vint. Such a case is depicted in Figure \ref{fi:NVsubt2}, where the subtree of $ab$ is the non-visible subtree of $bc$, $bp$, and $cp$.

\subsection{Non-rigid subtrees}
\begin{rule0} \label{rule:subtree}Consider a level-1 edge, $e$, which is not part of the rigid core. We refer to the subtree which is rooted at $e$ as a {\em non-rigid subtree}. There are at most five 6-vints in this subtree, and the overall charge from these 6-vints and their extensions cannot exceed $2$. \end{rule0}

\paragraph{Explanation.}Since the level-1 edge is not rigid, each of the 6-vints has a support of at least 2, which implies a trivial bound of $\frac{5}{2}$ on their overall charge. To improve this bound to the one asserted in the rule, we distinguish between the following cases: \begin{list}{\labelitemi}{\leftmargin=1em}
 \item At most four of the 6-vints exist. Then the bound cannot exceed $\frac{1}{2}\cdot4=2$. In the following cases we may therefore assume that all five 6-vints are present, so the subtree is full up to level 3.
 \item All of the 6-vints have a support of 2. \marfig{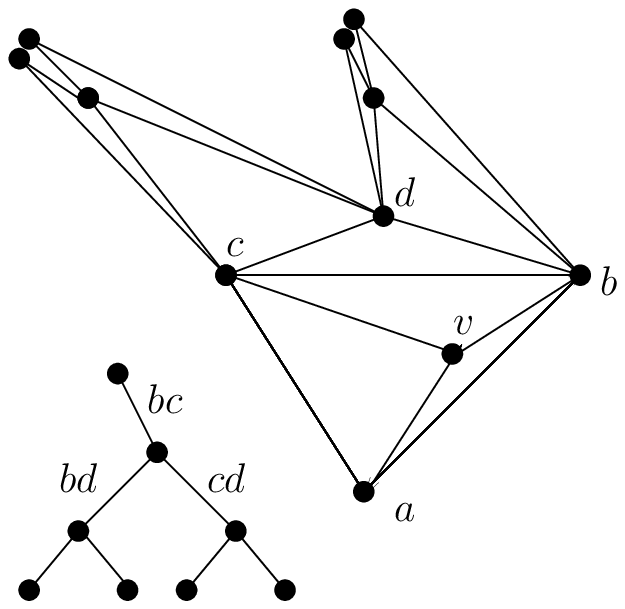} In this case, all the edges in the non-rigid subtree must be rigid, except for the level-1 edge. In addition, these edges must remain unflippable (in the hole of $v$) after the level-1 edge is flipped, since none of the vertices in the subtree, except $d$, can see $a$; see Figure \ref{fi:NRsubt1}. This implies that the 10-vint which contains the entire non-rigid subtree must also have a support of 2. The overall charge in this case is lower than $\frac{1}{2}(1\cdot5-3\cdot1)=1$.
 \item Exactly four 6-vints have a support of 2, and the fifth 6-vint uses a level-3 edge. Appending the edges of the first four 6-vints generates a 9-vint with a support of 2. The charge in that case is at most $\frac{1}{2}(1\cdot4-2\cdot1)+\frac{1}{3}=1\frac{1}{3}$.
 \item Exactly four 6-vints have a support of 2, and the fifth 6-vint uses two level-2 edges. The fact that the support of the fifth 6-vint is at least 3 implies that at least one of the level-2 edges must be flippable (possibly only after flipping the level-1 edge). Two of the other 6-vints also use this level-2 edge, and thus have a support larger than 2. We thus conclude that this case cannot occur.
 \item Exactly three 6-vints have a support of 2. For the same reason as in the previous case, the two 6-vints with the higher support must contain a level-3 edge. Appending the edges of the other three 6-vints generates an 8-vint with a support of 2. The charge is at most $\frac{1}{2}(1\cdot3-1\cdot1)+\frac{1}{3}\cdot2=1\frac{2}{3}$.
 \item There are at most two 6-vints with a support of 2. The charge is at most $\frac{1}{2}\cdot2+\frac{1}{3}\cdot3=2$. \hfill $\square$
\end{list}

\vspace{5mm} \noindent We do not use the following rules in the analysis of $\lambda_1=0$ and $\lambda_1=1$, and thus, it is possible to skip forward to the respective subsections of Section \ref{sec:analysis} before reading them.

\subsection{Eliminating 6-vints with two level-2 edges} \label{sec:2level2}

\begin{rule0} \label{rule:6-vint1} Consider a 6-vint with a rigid level-1 edge and two non-rigid level-2 edges. Assume that there are at least three additional RC edges, not involved in the 6-vint, which are not level-3 edges. Using these edges, it is possible to extend the 6-vint into at least two 8-vints, which neutralize its positive charge. \end{rule0}

\paragraph{Explanation.}\marfig{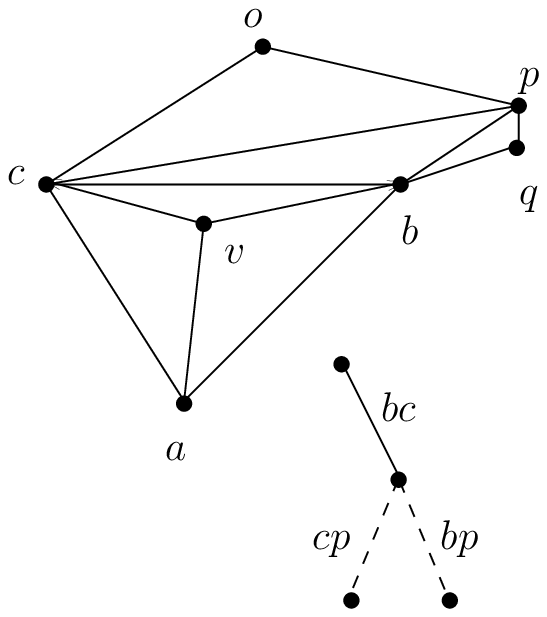}See Figure \ref{fi:2level2} for an example of such a 6-vint (the additional RC edges are not shown). We use the notation depicted in the figure and consider the handicapped 5-vint which uses the edge $bp$. By construction, the two vertices of this 5-vint ($p$ and $q$) are in their non-visible terrain. If the vertex $o$ of the other level-2 edge cannot see $a$, then it is also in its non-visible terrain. In this case, adding two extra RC edges cannot increase the support of the 6-vint, since vertices in their non-visible terrains cannot see vertices of RC edges from other subtrees. Hence, each extended 8-vint has the same support as the original 6-vint, and thus fully neutralizes the charge of the 6-vint. We may therefore ignore this case, and assume that $o$ sees $a$ (as depicted in Figure \ref{fi:2level2}).

In order to show that the 8-vints can neutralize the charge of the 6-vint, we need to count the triangulations of the hole of the 6-vint, and of the holes of the potential 8-vints. We first claim that in any of these triangulations, exactly one of the edges $bc$ and $ao$ must be present. This is obvious for the 6-vint, since $ao$ is the only chord of its hole which crosses $bc$, so when $bc$ is absent, $ao$ must be present. For an 8-vint, its hole is obtained by appending two triangles through the edges $ab$ and/or $ac$. When considering an 8-vint which extends the 6-vint, we refer to the additional RC edges as the {\em added edges}. Since $o$ is the only vertex of the 6-vint which is in its visible terrain, it is the only vertex that might be able to see vertices of the added edges. Hence, the only chords of the hole of such an 8-vint which can cross $bc$ are incident to $o$. Moreover, if any such chord, other than $ao$, is part of the triangulation, then it is obtained after flipping one of the edges $ab$ and $ac$, and the first time such a flip occurs, $ao$ must be present in the triangulation, and remain in it thereafter.\marfig{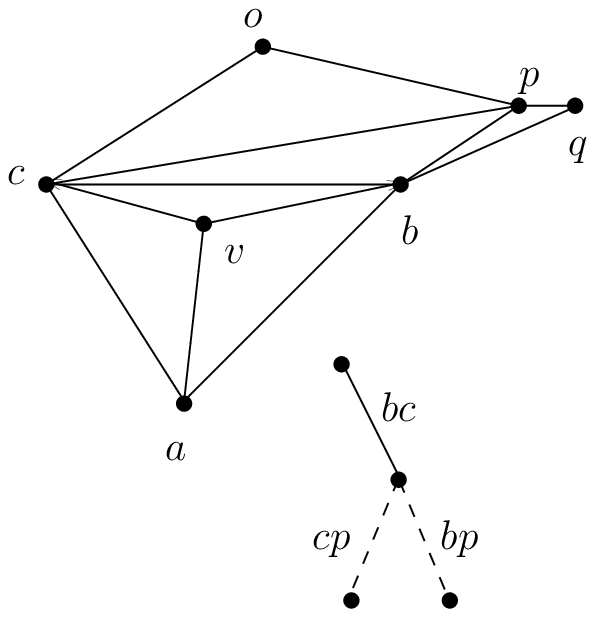}Moreover, in a triangulation of the hole of an 8-vint, an added edge is flippable only if both of its vertices are connected to $o$. This implies that when $bc$ is present, no added edge is flippable; that is, the 6-vint and the 8-vints have the same number of triangulations which contain $bc$.

It remains to count the triangulations which contain $ao$. Notice that when $ao$ is present, $bo$ must also be present. We distinguish between the following cases:

(i) $o$ cannot see $q$, as depicted in Figure \ref{fi:2level2b}. In this case, $bp$ must be present, which implies that each triangulation can be uniquely determined by the set of vertices of added edges which are connected to $o$. The 6-vint has only one triangulation, which corresponds to the empty set. After adding two RC edges, there can be at most $2^2=4$ such sets, including the empty one. This implies that the support of an 8-vint can be higher than that of the 6-vint by at most 3. Since $o$ can see $a$, the 6-vint has a support of at least $tr(bp)=C'_3=3$, which means that two 8-vints are always sufficient to neutralize its charge.

(ii) $o$ can see $q$, as depicted in Figure \ref{fi:2level2}. In this case, when $bc$ is present there are five triangulations, both for the hole of the 6-vint and for any hole of an 8-vint (the number of triangulations of the convex pentagon $bcopq$). In order to count triangulations which contain $ao$, we use the same method as in the previous case. This time, since the quadrilateral $obqp$ has two triangulations, each subset of vertices of added edges corresponds to two triangulations. The 6-vint has only the empty set, so its support is $5+1\cdot2=7$. An 8-vint can have, as above, at most four such subsets, and thus, a support of at most $5+4\cdot2=13$. Thus, two 8-vints are always sufficient to neutralize the charge of the 6-vint. \hfill $\square$

\subsection{Reducing the charge of 6-vints with a level-3 edge} \label{sec:longPr}

\begin{rule0} \label{rule:6-vint2} Consider a 6-vint with a level-3 edge and a rigid level-1 edge.\marfigsc{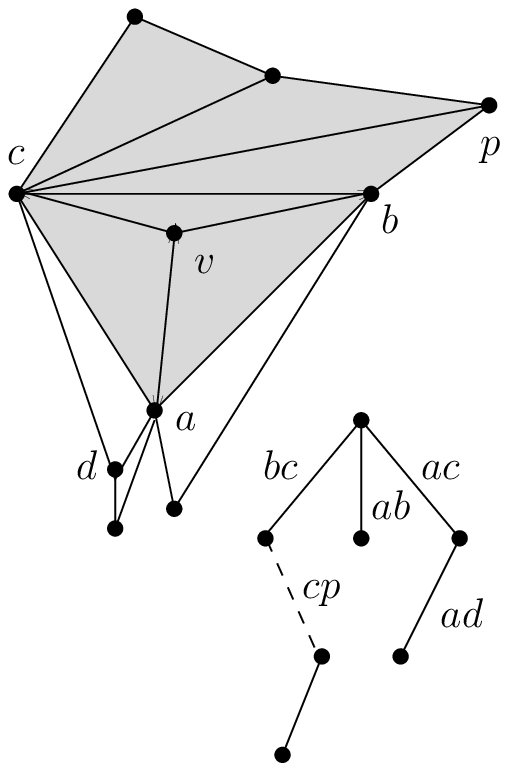}{1.4}Assume that there are at least three RC edges which do not participate in the 6-vint and are not level-3 edges. Using these edges, the 6-vint can be extended into at least two 8-vints and one 9-vint. The overall charge of the 6-vint and its extensions cannot exceed $\frac{1}{1400}$.
\end{rule0}

\paragraph{Explanation.} Establishing this rule is the most complex part of the analysis, and in fact, in what follows, we also provide a more general analysis of how one can reduce the charge of a 6-vint of the form described in the rule, using extensions of the vint into vints with negative charge, even when there are only two RC edges that can be exploited. When encountering a reference to this section in the analysis of an RC, it is best to refer to Table \ref{ta:app}, which presents bounds for the support of 8-vints which extend the possible 6-vints. The table also presents improved bounds for {\em standard} 8-vints, which are 8-vints that extend the 6-vint using an additional level-1 edge and one of its child edges (as opposed to extensions that use two additional level-1 edges, or descendant edges of the level-1 edge of the 6-vint).\marfigsc{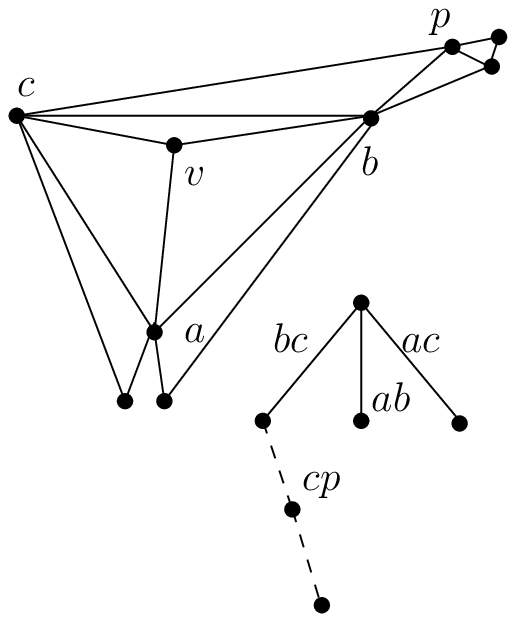}{1.4}Such a case is depicted in Figure \ref{fi:longrule01}, where the shaded area represents the hole of the 6-vint, the 8-vint created using $ac$ and $ad$ is a standard 8-vint, and the 8-vint created using $ab$ and $ac$ is not.

\begin{table}[ht]
\caption{The results of Section \ref{sec:longPr}} \label{ta:app}
\centering
\begin{tabular}{c c c c}
\hline\hline
Case & Support   & Support   & Support of \\
	 & of 6-vint & of 8-vint & standard 8-vint \\[0.5ex]

\hline
Level-2 edge is rigid & $\leq 3$ & same as 6-vint & - \\
Level-2 edge is rigid & 4 & $\leq7$ & - \\
$q$ and $o$ cannot see $a$ & any & same as 6-vint & - \\
only $o$ can see $a$ & 4 & $\leq7$ & $\leq6$ \\
only $o$ can see $a$ & 6 & $\leq9$ & $\leq8$ \\
only $o$ can see $a$ & 7 & $\leq13$ & $\leq11$ \\
only $q$ can see $a$ & 3 & $\leq8$ & - \\
only $q$ can see $a$ & 4 & $\leq7$ & - \\
only $q$ can see $a$ & 5 & $\leq13$ & - \\
only $q$ can see $a$ & 6 & $\leq9$ & - \\
only $q$ can see $a$ & 7 & $\leq13$ & - \\
$q$ and $o$ can see $a$ & 9 & $\leq25$ & $\leq20$ \\
$q$ and $o$ can see $a$ & 6 & $\leq20$ & $\leq15$ \\
$q$ and $o$ can see $a$ & 4 & $\leq14$ & $\leq11$ \\ [1ex]
\hline
\end{tabular}
\end{table}

In the analysis (and in the table), we refer to Figures \ref{fi:longrule03}--\ref{fi:longrule19}. In these figures, $v$ is the vertex of the 3-vint and $bc$ is the rigid level-1 edge of the 6-vint, whose vertex $p$ lies to the right of $\overrightarrow{ab}$. The 6-vint has two additional vertices $q$ and $o$. Unless otherwise stated, $q$ is the vertex of the level-2 edge, and $o$ is the vertex of the level-3 edge. If the level-2 edge of the 6-vint is $bp$ (as depicted in Figure \ref{fi:longrule02}), all the vertices of the 6-vint are in their non-visible terrain, and adding RC edges will not increase the support of the vint (see Section \ref{sec:non-visib}). If there are three additional RC edges, as prescribed in the rule, we can form at least two 8-vints and one 9-vint which extend the 6-vint and (more than) neutralize its charge.\marfigsc{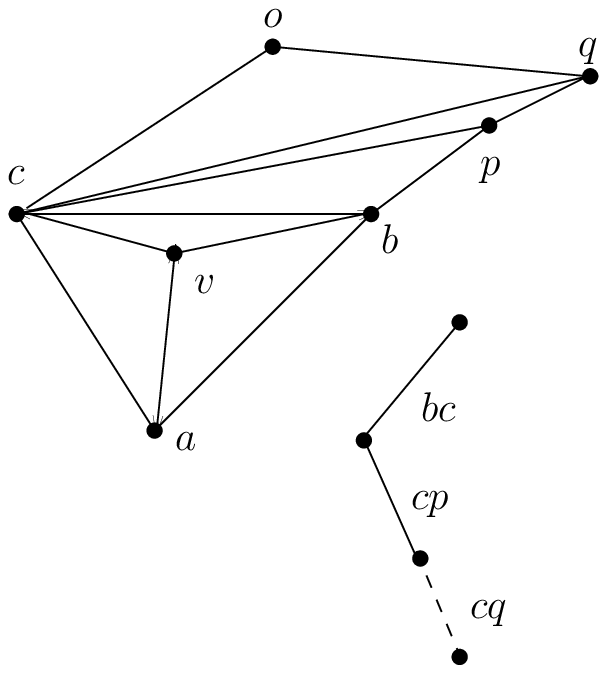}{1.4}If only two RC edges are present, we can neutralize its charge by the resulting 8-vint extension. We may therefore ignore this case and assume that the level-2 edge is always $cp$.

Assume that $bp$ is one of the additional RC edges. By Rule \ref{rule:RemainsRigid}, appending $bp$ (and any of its descendants) to the 6-vint cannot increase its support. This implies that replacing $bp$ with other RC edges can only increase the support of the 9-vint and two 8-vints. That is, it suffices to consider the cases where all the additional RC edges are in the subtrees of $ab$ or of $ac$. In particular, if three RC edges are added, they induce at least one standard extending 8-vint. We may therefore assume that $bp$ is not one of the additional RC edges. For the rest of the analysis, we distinguish between five possible cases:

\setcounter{enumi}{0}
\addtocounter{enumi}{1}
\paragraph{\arabic{enumi}. The level-2 edge is rigid.} An example of this case is depicted in Figure \ref{fi:longrule03}.\marfigsc{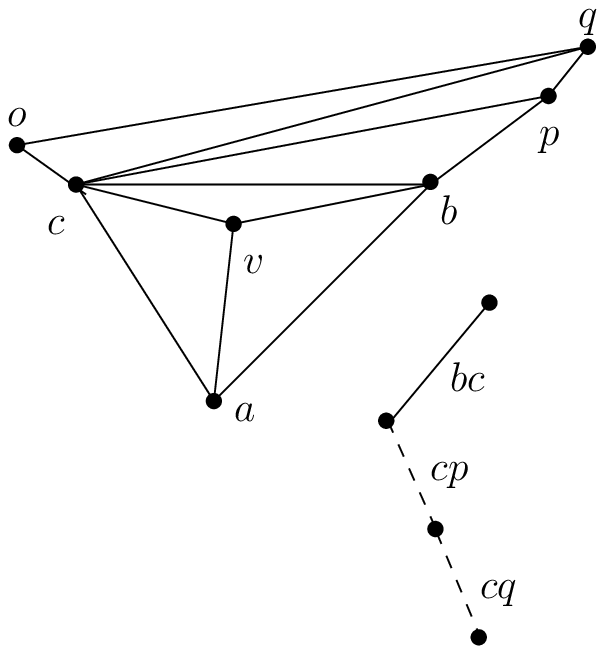}{1.4}If $o$ (the vertex of the level-3 edge) is in its non-visible terrain, all of the vertices of the 6-vint are in their non-visible terrains, and thus, adding RC edges will not increase its support. We may therefore assume that $o$ is in its visible terrain (as in the figure), which also implies that the 6-vint has a support of 4 (see also the argument below).

In every triangulation of the hole of the 6-vint or of an extension 8-vint, exactly one of the edges $bc$ and $ao$ must exist (this is explained in the proof of Rule \ref{rule:6-vint1}). If $bc$ exists, there are exacly $C'_3=3$ triangulations (of the 6-vint). If $ao$ exists, $bo$ and $op$ must also exist, and only $o$ can see the vertices of the added RC edges. Each triangulation which contains $ao$ can be uniquely determined by its set of vertices of added edges which are connected to $o$ (a similar argument can be found in Rule \ref{rule:6-vint1}).\marfigsc{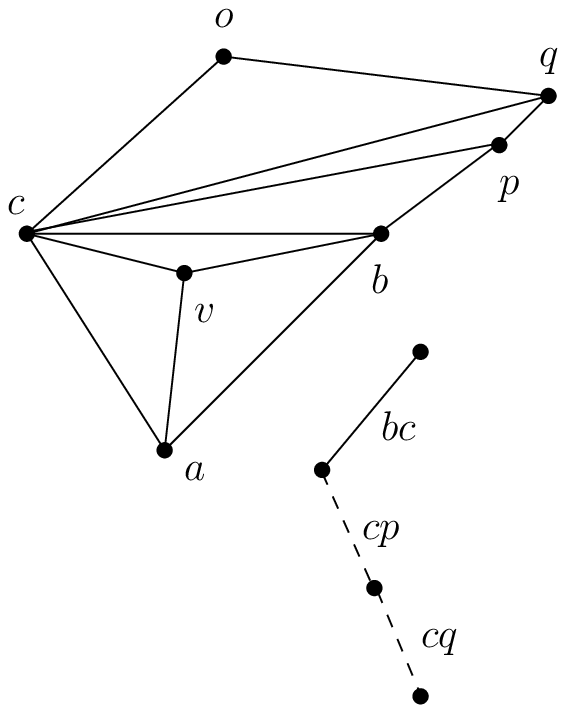}{1.4}This implies that the support of an extension 8-vint is 3 plus the number of these sets. Adding two RC edges yields at most $2\cdot2=4$ sets, and thus, each 8-vint has a support of at most 7. Since we have at least two 8-vints which extend the 6-vint, they always (more than) neutralize its charge.

In the following cases, the level-2 edge is assumed not to be rigid.

\addtocounter{enumi}{1}
\paragraph{\arabic{enumi}. Both $q$ and $o$ cannot see $a$.}An example of this case is depicted in Figure \ref{fi:longrule04}. Since all the vertices of the 6-vint are in their non-visible terrains, the addition of RC edges cannot increase its support, so even a single extension 8-vint will neutralize the charge.

\addtocounter{enumi}{1}
\paragraph{\arabic{enumi}. Only $o$ can see $a$.} Examples of this case are depicted in Figures \ref{fi:longrule05}--\ref{fi:longrule07}. Since we assume that the level-2 edge is not rigid, $q$ can see $b$.\marfigsc{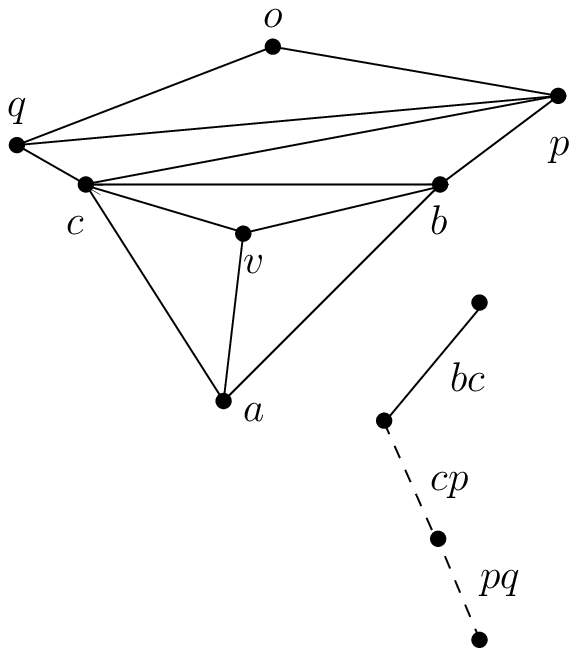}{1.4}$o$ can also see $b$ and $c$, since in order not to see one of them, $o$ has to be in its non-visible terrain, which implies that it cannot see $a$ either. Since $q$ is in its non-visible terrain, it cannot see any of the vertices of the added RC edges. Let $x$ be a binary variable, which is 1 if and only if $o$ can see $p$ (see Figure \ref{fi:longrule07} for a case where $x=0$). Similarly, let $y$ be a binary variable, which is 1 if and only if $x=1$ and $bq$ does not cross $ao$ ($y=1$ in Figure \ref{fi:longrule05} and 0 in Figures \ref{fi:longrule06} and \ref{fi:longrule07}).

Notice that exactly one of the edges $bc$ and $ao$ must exist in any triangulation of the hole of the 6-vint or of any extension 8-vint.\marfigsc{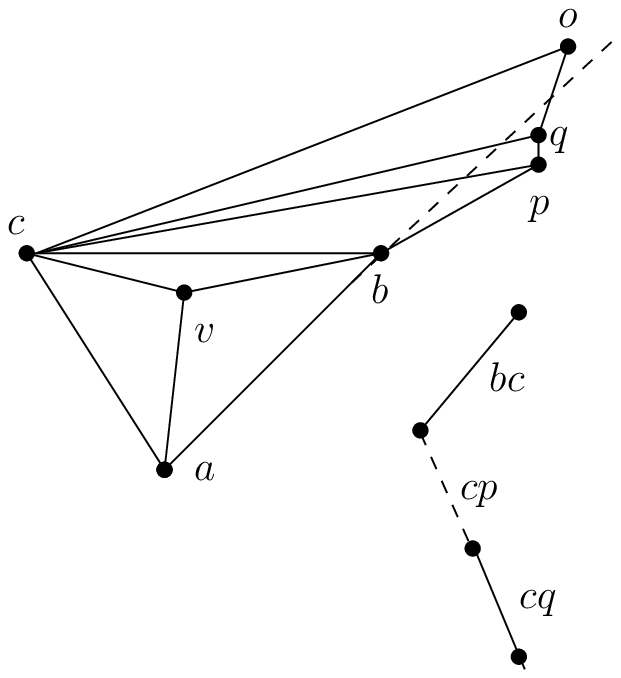}{1.4}The 6-vint has $3+2x$ triangulations which contain $bc$ (it has either $C_3=5$ or $C'_3=3$ triangulations), and $1+y$ triangulations which contain $ao$ (as can be easily checked; see Figures \ref{fi:longrule05}, \ref{fi:longrule06}, and \ref{fi:longrule07}). That is, the 6-vint has $4+2x+y$ triangulations. An extension 8-vint has the same number of triangulations when $bc$ is present (recall that we only consider extensions through $ab$ or $ac$). When $ao$ is present, $bo$ must also be present, and the quadrilateral $oqpb$ has $x+1$ triangulations (or, as in Figure \ref{fi:longrule06}, it might not be present at all in these triangulations). Once again, we count the number of possible sets of vertices of added edges which are connected to $o$. This time, each set represents $y+1$ triangulations (which contain $ao$). The support of the 8-vint is therefore $3+2x$, plus the number of sets multiplied by $y+1$. Adding two RC edges can create at most four possible sets, which implies that an 8-vint has a support of at most $(3+2x)+4(y+1)=7+2x+4y$. For a standard 8-vint, there are at most three sets, and thus, the support is at most $(3+2x)+3(y+1)=6+2x+3y$.\marfigsc{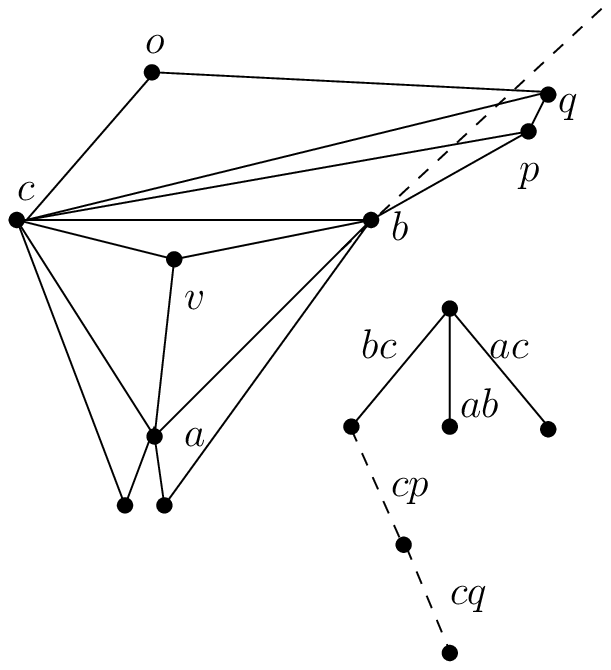}{1.4} From the above, we conclude that two 8-vints always neutralize the charge of the 6-vint, since $\frac{1}{4+2x+y}=\frac{2}{8+4x+2y}\leq\frac{2}{8+2x+4y}<\frac{2}{7+2x+4y}$ (by definition, $x\geq y$).

A case with $x=y=1$ is depicted in Figure \ref{fi:longrule08}, where the 6-vint has a support of 7 (5 when $bc$ is present, and 2 when $ao$ is present) and the 8-vint has a support of 13 (5 when $bc$ is present, and $2\cdot4=8$ when $ao$ is present).

\addtocounter{enumi}{1}
\paragraph{\arabic{enumi}. Only $q$ can see $a$.}Examples of this case are depicted in Figures \ref{fi:longrule09}--\ref{fi:longrule13}. $o$ can ``hide" from $a$ on one of the sides of its non-visible terrain (such as in Figures \ref{fi:longrule09}--\ref{fi:longrule11}), or behind $q$ (such as in Figures \ref{fi:longrule12}--\ref{fi:longrule13}).  As in previous cases, we notice that in each triangulation either $aq$ or $bc$ must exist. We divide the rest of the analysis into the following subcases: \marfigsc{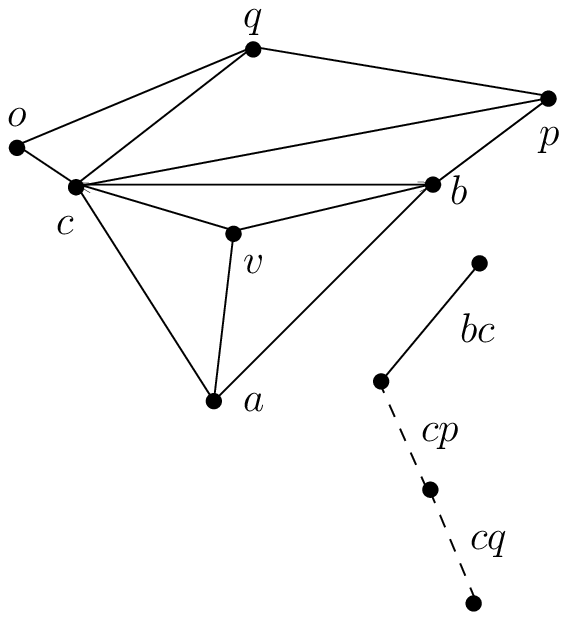}{1.4}

\medskip {\bf (i) \mbox{\boldmath $o$} is in its non-visible terrain,} as depicted in Figures \ref{fi:longrule09}, \ref{fi:longrule10}, and \ref{fi:longrule11}. In this case, $q$ is the only vertex which might be able to see vertices of added edges. For an extension vint, consider the set of the vertices of the added edges which are connected to $q$ in a specific triangulation. We can bound the number of triangulations of the extension vint, which do not contain a triangulation of the 6-vint, by counting the number of the possible different non-empty sets of this kind (as in the explanation of Rule \ref{rule:6-vint1}). Since the simple quadrilateral $qopb$ (which may not exist, as in Figure \ref{fi:longrule09}) might have two triangulations (as in the case of Figure \ref{fi:longrule10}), each set might correspond to two such triangulations (but not more than two). Let $x$ denote the support of the 6-vint;\marfigsc{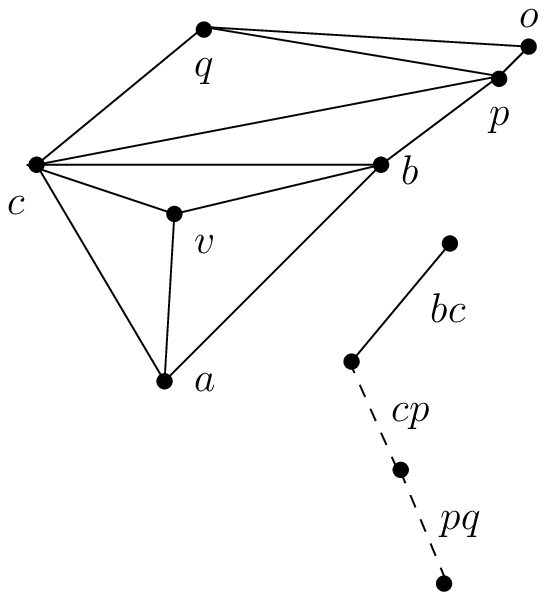}{1.4}the support of an extension vint is at most $x$ plus twice the number of non-empty sets (of $q$-connected vertices of the added RC edges). An 8-vint which extends the 6-vint with two level-1 edges, has at most three such sets. A standard 8-vint has at most two such sets. A 9-vint has at most five such sets (which is tight when there are two additional level-1 RC edges and one additional level-2 RC edge). That is, the absolute value of the overall negative charge from the three extension vints is at least $\frac{1}{x+4}+\frac{1}{x+6}+\frac{2}{x+10}$.

We notice that the inequality $\frac{1}{x}<\frac{1}{x+4}+\frac{2}{x+6}<\frac{1}{x+4}+\frac{1}{x+6}+\frac{2}{x+10}$ holds for every $x\geq3$ (the second inequality holds for every positive value of $x$). A 6-vint such as in Figure \ref{fi:longrule09} has a support of $tr(bc)+tr(aq)=5+1=6$. A 6-vint such as in Figure \ref{fi:longrule10} has a support of $tr(bc)+tr(aq)=5+2=7$.\marfigsc{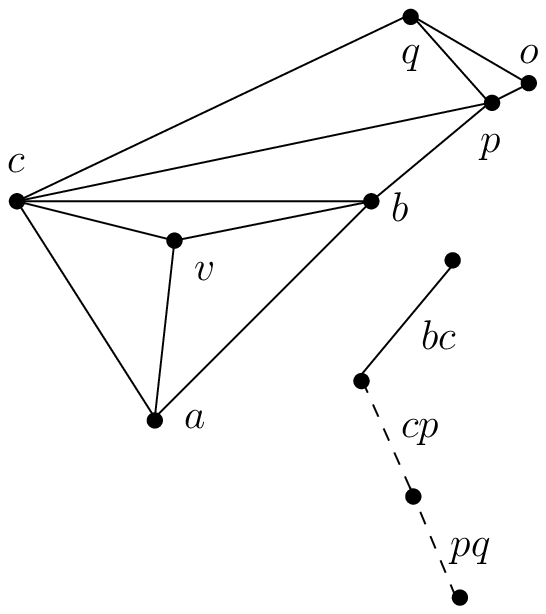}{1.4}A 6-vint such as in Figure \ref{fi:longrule11} has a support of $tr(bc)+tr(aq)=3+1=4$. Since these cases exhaust all possibilities, we always have $x\geq4$, which implies that a 9-vint and two 8-vints always (more than) neutralize the charge of the 6-vint.

\medskip {\bf (ii) \mbox{\boldmath $o$} is the vertex of \mbox{\boldmath $cq$} and is to the right of \mbox{\boldmath $\overrightarrow{aq}$},} as depicted in Figure \ref{fi:longrule12}. Since $o$ is hiding behind $q$, it also cannot see $b$ and $p$. Hence, the edge $cq$ must be present in every triangulation of the hole of the 6-vint, and thus, the 6-vint has a support of $C'_3=3$. For the larger vints, when $bc$ is present, there are two triangulations; when $aq$ is present, the number depends on the added edges (however, for any set of added edges, $bq$ must be present in the resulting triangulations; recall that we assume that $bp$ is not an added edge): \begin{list}{\labelitemi}{\leftmargin=1em}
 \item Without additional edges, each of the portions of the hole of the vint to the right of $aq$ and to its left has a single triangulation (which implies a single triangulation of the 6-vint when $aq$ is present). \marfigsc{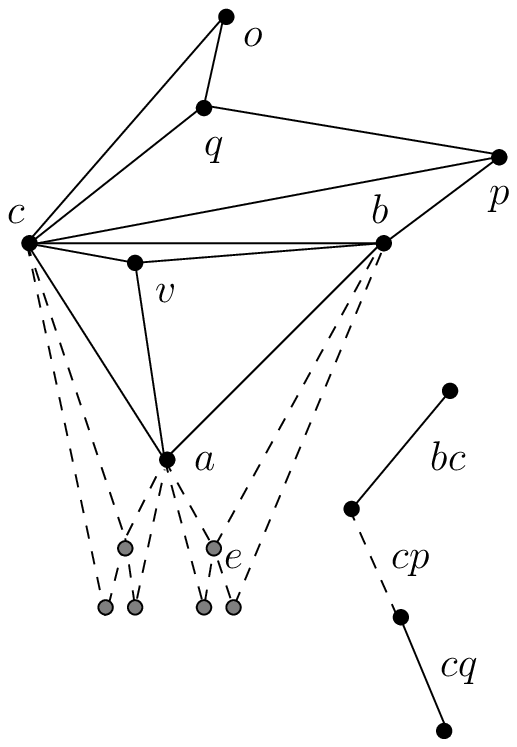}{1.4}
 \item Adding $ab$ results in at most two triangulations of the portion to the right of $aq$. Adding a child of $ab$ raises it up to three triangulations, and adding both children of $ab$ raises it up to five. This follows by noticing that each triangulation is uniquely determined by the set of vertices of added edges which are connected to $q$. In the latter case, for example, we have one triangulation when $ab$ is present; otherwise, $qe$ must be present (where, as in the figure, $e$ is the vertex of $ab$), and we have at most $C_2C_2=4$ ways to complete the triangulation.
 \item Adding $ac$ results in at most $C'_3=3$ triangulations of the portion to the left of $aq$. Adding a child of $ac$ raises it up to $C''_4=6$. Adding both children of $ac$ cannot cause this number to exceed $C''_5=19$ (this time, the analysis is not tight, since we allow several forbidden chords, such as the one connecting $a$ to the new lower left vertex).
\end{list}

\begin{table}[ht]
\caption{Possible supports of extension vints in case (ii) (Figure \ref{fi:longrule12})} \label{ta:comb1}
\centering

\begin{tabular}{c c c || c c c}
\hline\hline
\multicolumn{3}{c||}{8-vints} & \multicolumn{3}{c}{9-vints} \\

\hline\hline
Right of $aq$ & Left of $aq$ & Support & Right of $aq$ & Left of $aq$ & Support \\ [0.5ex]
\hline
 1 & 6 & 8 & 1 & $<19$ & $<21$\\
 2 & 3 & 8 & 2 & 6 & 14\\
 3 & 1 & 5 & 3 & 3 & 11\\
   &   &   & 5 & 1 & 7 \\ [1ex]
\hline\hline
\end{tabular}
\end{table}

\noindent The above analysis is summarized in Table \ref{ta:comb1}, which presents the maximal supports for the various extension vints (in the $i$-th row of the table, $i-1$ RC edges of the extension vint are added to the right of $aq$ and the rest are added to its left). Notice that the support of each vint is the number of triangulations of its right portion times the number of its left portion, plus 2.\marfigsc{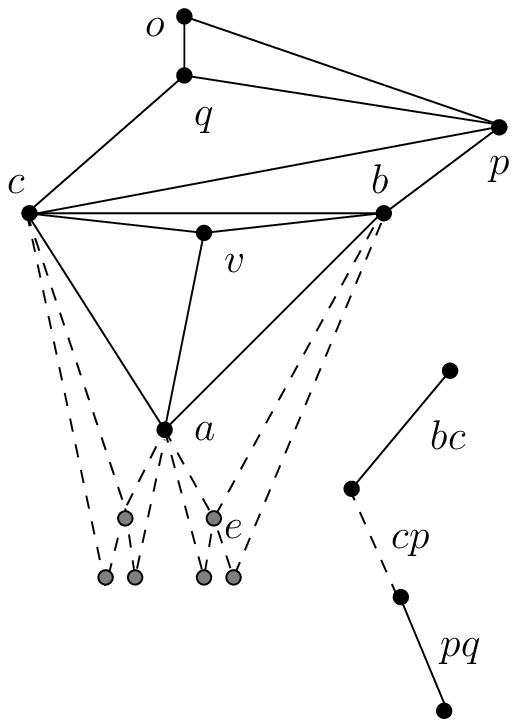}{1.4}We notice that an 8-vint has a support of at most 8, and a 9-vint has a support of at most 20. Two 8-vints and a 9-vint generate a negative charge of at least $2\cdot\frac{1}{8}+\frac{2}{20}>\frac{1}{3}$, so they neutralize the charge of the 6-vint.\medskip

{\bf (iii) \mbox{\boldmath $o$} is the vertex of \mbox{\boldmath $pq$} and is to the left of \mbox{\boldmath $\overrightarrow{aq}$},} as depicted in Figure \ref{fi:longrule13}. $o$ must be able to see $b$, since otherwise it will not be able to see $v$. It can easily be checked that each of the holes of the 6-vint and of the extension vints has three triangulations which contain $bc$ (again, by assumption, there are no added edges through $bp$). When $aq$ is present in the hole of the 6-vint, there are $C''_3=2$ triangulations of the portion to its right and one of the portion to its left. This implies that the support of the 6-vint is $3+2\cdot1=5$. We consider the number of triangulations of the holes of the larger vints, when $aq$ is present: \begin{list}{\labelitemi}{\leftmargin=1em}
 \item Without any additional edges, the portion of the hole of the vint to the left of $aq$ has a single triangulation, and the portion to its right has two triangulations (as in the case of the 6-vint).
 \item Adding $ac$ results in $C_2=2$ triangulations of the portion to the left of $aq$. Adding a child of $ac$ increases it up to $C'_3=3$. Adding both children of $ac$ increases it up to 5 (as in a preceding case, recalling that each triangulation is uniquely determined by the set of vertices which are connected to $q$).
 \item Adding $ab$ results in at most $tr(eo)+tr(bq)=1+C_2C_2=5$ triangulations to the right of $aq$. Adding a child of $ab$ increases it up to $tr(pq+bq)+tr(bo)=C'_3+C''_4=9$. Adding both children of $ab$ cannot cause it to exceed $tr(pq)+tr(bo)=(tr(ab)+tr(eq))+tr(bo)=(1+C_2C_2)+C''_5=24$ (the analysis is not tight for the case where $bo$ is present, since we admit some of the forbidden chords).
\end{list}

\begin{table}[ht]
\caption{Possible supports of extension vints in case (iii) (Figure \ref{fi:longrule13})} \label{ta:comb2}
\centering

\begin{tabular}{c c c || c c c}
\hline\hline
\multicolumn{3}{c||}{8-vints} & \multicolumn{3}{c}{9-vints} \\

\hline\hline
Right of $aq$ & Left of $aq$ & Support & Right of $aq$ & Left of $aq$ & Support \\ [0.5ex]
\hline
 9 & 1 & 12 & $<24$ & 1 & $<27$\\
 5 & 2 & 13 & 9 & 2 & 21\\
 2 & 3 & 9 & 5 & 3 & 18\\
   &   &   & 2 & 5 & 13 \\ [1ex]
\hline\hline
\end{tabular}
\end{table}

\noindent The above analysis is summarized in Table \ref{ta:comb2}, which presents the maximal supports for the various extension vints (with the same convention of display of the rows). The support of each vint is the number of triangulations of its right portion times the number of its left portion, plus 3. We notice that an 8-vint has a support of at most 13, and a 9-vint has a support of at most 26. Two 8-vints and a 9-vint generate a negative charge of at least $2\cdot\frac{1}{13}+\frac{2}{26}>\frac{1}{5}$, neutralizing the charge of the 6-vint. \marfigsc{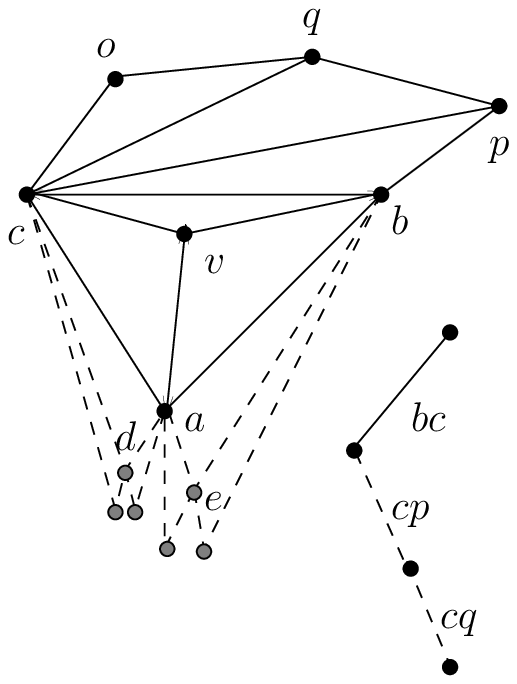}{1.4}

\addtocounter{enumi}{1}
\paragraph{\arabic{enumi}. Both $q$ and $o$ can see $a$.} Examples of this case are depicted in Figures \ref{fi:longrule14}--\ref{fi:longrule19}. Here we change the notation used in the preceding cases, and take $q$ and $o$ to be such that the clockwise order of the vertices on the boundary of the hole of the 6-vint is $a$, $c$, $o$, $q$, $p$, $b$. In the vertex group \{$a$,$b$,$c$,$p$\}, only $p$ and $a$ cannot see each other. Any additional visibility restrictions between the vertices of the 6-vint, have to involve either $o$ or $q$. If $o$ can see $p$, it can also see $b$ (since it is required to see $a$). Moreover, only the vertex of the level-3 edge of the 6-vint (which is either $o$ or $q$) might have visibility restrictions. This implies that there are only four possible cases: (i) $o$ and $q$ are unrestricted, and the 6-vint has a support of $C'_4=9$ (see Figure \ref{fi:longrule14}). (ii) $q$ is the level-3 vertex and it cannot see $c$ (it always sees $p$ since it precedes $p$ in the order along the hole); the 6-vint has a support of $C''_4=6$ (see Figure \ref{fi:longrule17}). (iii) $o$ is the level-3 vertex and it cannot see $p$ (but can see $b$); the 6-vint has a support of $C''_4=6$ (see Figure \ref{fi:longrule18}). (iv) $o$ cannot see $p$ and $b$, and the 6-vint has a support of $tr(ao)+tr(cq)=1+C'_3=4$ (see Figure \ref{fi:longrule19}). We analyze each of these subcases separately:

 \medskip {\bf (i) \mbox{\boldmath $o$} and \mbox{\boldmath $q$} are unrestricted} (see Figure \ref{fi:longrule14}, which depicts one of the two forms of such an ``almost convex" 6-vint; the analysis does not depend on the specific shape of the vint). Clearly, if a triangulation of the hole of the 6-vint (or of an extension vint) does not include $bc$, it must include either $ao$ or $aq$ (or both). Hence, the number of triangulations of the 6-vint (or of an extension vint) is $tr(bc)+tr(ao)+tr(aq)-tr(ao+aq)$. Notice that $tr(bc)=5$, both for the 6-vint and for any extension vint. We thus need to count the number of triangulations of two left portions --- the portion to the left of $ao$ and the portion to the left of $aq$ (which contains the chord $ao$ in some of its triangulations).\marfigsc{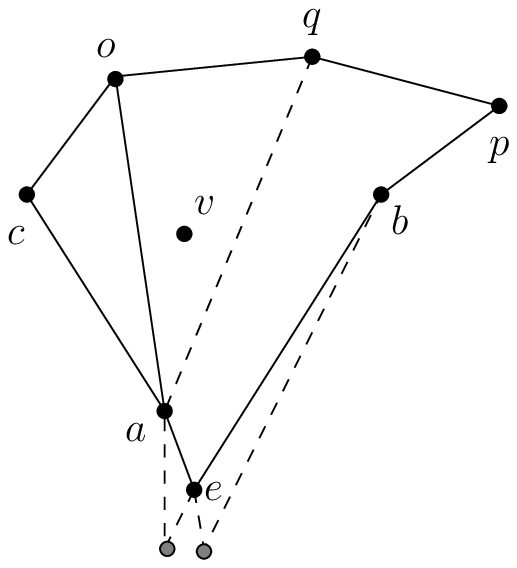}{1.4}Similarly, we need to count the triangulations of two respective right portions. In the following analysis, all of the results are upper bounds: \begin{list}{\labelitemi}{\leftmargin=1em}
 \item When there are no added edges, there is a single triangulation to the left of $ao$, a single triangulation to the right of $aq$, $C_2=2$ triangulations to the left of $aq$, and $C'_3=3$ triangulations to the right of $ao$. We can use these results to verify that the support of the 6-vint is $tr(bc)+tr(ao)+tr(aq)-tr(ao+aq)=5+1\cdot3+2\cdot1-1\cdot1=9$.
 \item After the addition of $ab$ (as depicted in Figure \ref{fi:longrule15}), there are $C_2=2$ triangulations to the right of $aq$ ($bq$ must be present), and $tr(op+bo)+tr(bq)=C_2+C_3=7$ to the right of $ao$. Adding a child edge of $ab$, results in $C'_3=3$ triangulations to the right of $aq$ (again, $bq$ must be present), and $tr(op+bo)+tr(bq)=C'_3+C'_4=12$ to the right of $ao$. Adding both child edges of $ab$ results in 5 triangulations to the right of $aq$ (each triangulation is uniquely defined by the set of vertices of added edges which are connected to $q$), and $tr(op+bo)+tr(bq)=5+[(tr(bq+eq)+tr(bq+eo)-tr(bq+eo+eq))+tr(bq+ab)]=5+[(C_3C_2+C_2C_3-C_2C_2)+C_2]=23$\marfigsc{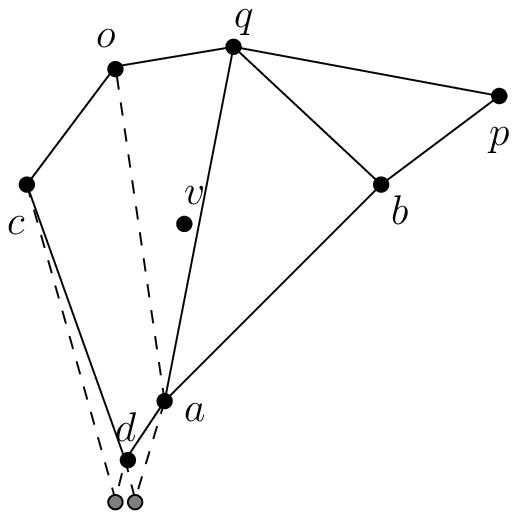}{1.4}triangulations to the right of $ao$.
 \item After the addition of $ac$ (as depicted in Figure \ref{fi:longrule16}), there are $C_2=2$ triangulations to the left of $ao$ and $C_3=5$ triangulations to the left of $aq$. Adding a child of $ac$ results in $C'_3=3$ to the left of $ao$, and $C'_4=9$ to the left of $aq$. Adding both child edges of $ac$ results in 5 triangulations to the left of $ao$ (each triangulation is uniquely defined by the set of vertices of added edges which are connected to $o$), and $[tr(do)+tr(dq)-tr(do+dq)]+tr(ac)=[C_2C_3+C_3C_2-C_2C_2]+C_2=18$ triangulations to the left of $aq$.
\end{list}

\noindent The above analysis is summarized in Table \ref{ta:comb3}, which presents the maximal supports for the various extension vints (each row in the table is indexed by the numbers of RC edges added through $ab$ and of RC edges added through $ac$).The support of each vint is given by the expression $tr(bc)+tr(aq)+tr(ao)-tr(aq+ao)$; by using the column names of Table \ref{ta:comb3}, and recalling that $tr(bc)=5$, the support can be written as $5+AB+CD-AD$.

Using the table, we notice that an 8-vint has a support of at most 25, and a 9-vint has a support of at most 38. Moreover, at most one extension 8-vint has a support of at most 25 (it is the 8-vint with three level-1 edges), and the other 8-vints have a support of at most 20.\marfigsc{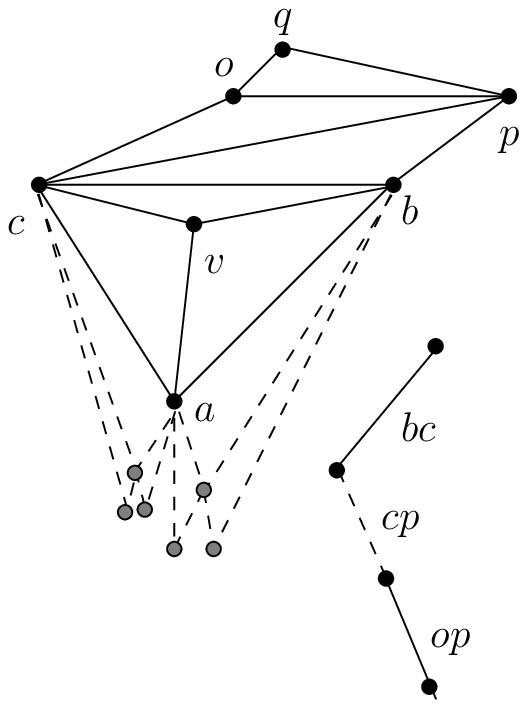}{1.4}The absolute value of the overall charge of two 8-vints and a 9-vint is at least $\frac{1}{20}+\frac{1}{25}+\frac{2}{38}=\frac{271}{1900}>\frac{1}{9}$, neutralizing the charge of the 6-vint.

\begin{table}[ht]
\caption{Upper bounds on the supports of the extension vints of a 6-vint with a support of 9.} \label{ta:comb3}
\centering

\begin{tabular}{c c c c c c c}
& & A & B & C & D & \\
\hline\hline
ab/ac & Type & Right   & Left    & Right   & Left    & Support \\
edges &      & of $aq$ & of $aq$ & of $ao$ & of $ao$ \\[0.5ex]
\hline\hline
2/0 & 8-vint & 3 & 2 & 12 & 1 & 20 \\
1/1 & 8-vint & 2 & 5 & 7 & 2 & 25 \\
0/2 & 8-vint & 1 & 9 & 3 & 3 & 20 \\
3/0 & 9-vint & 5 & 2 & 23 & 1 & 33 \\
2/1 & 9-vint & 3 & 5 & 12 & 2 & 38 \\
1/2 & 9-vint & 2 & 9 & 7 & 3 & 38 \\
0/3 & 9-vint & 1 & 18 & 3 & 5 & 33 \\ [1ex]
\hline\hline
\end{tabular}
\end{table}

\medskip {\bf (ii)/(iii) The 6-vint has a support of 6,} which implies that either $o$ or $q$ blocks the visibility between its two neighbors, as depicted in Figures \ref{fi:longrule17} and \ref{fi:longrule18}. We determine how many triangulations are lost in an 8-vint (or a 9-vint) when taking the previous case and adding such a restriction. This\marfigsc{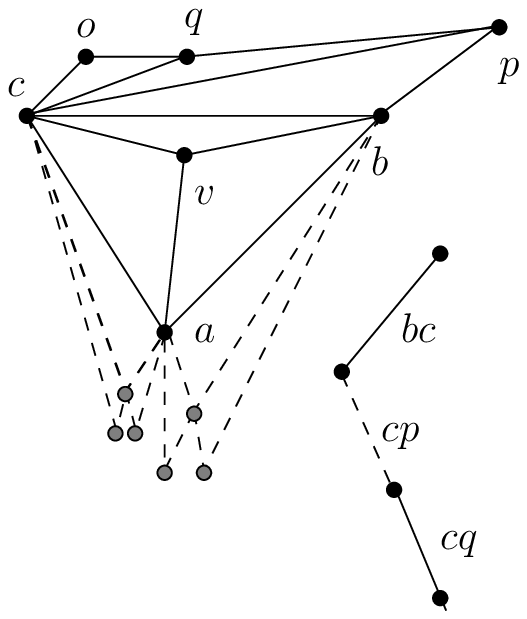}{1.4}is exactly the number of triangulations which contain the forbidden chord, which is also the number of triangulations after removing either $o$ or $q$ (without loss of generality, consider case (ii), where $o$ is removed). There are $C_2=2$ such triangulations when $bc$ is present.

When $aq$ is present, each triangulation holds a unique set of vertices of added edges which are connected to $q$; when there are at least two added edges, there are at least three such sets, and thus, there are $tr(bc)+tr(aq)\geq 2+3=5$ triangulations. Notice that even if some of these triangulations do not exist (since $q$ cannot see a vertex of an added edge), they were still counted in the analysis of case (i). This implies that the absolute value of the overall charge of two 8-vints and a 9-vint is more than $\frac{1}{20-5}+\frac{1}{25-5}+\frac{2}{38-5}=\frac{39}{220}>\frac{1}{6}$, neutralizing the charge of the 6-vint (a slightly higher charge is easily achieved using a separate analysis for each of the various extension vints. However, this simpler and weaker analysis is sufficient for our purpose).

\medskip {\bf (iv) The 6-vint has a support of 4,} as depicted in Figure \ref{fi:longrule19}. As in the previous case, we determine the number of triangulations of the 8-vints and 9-vints (from the first case), which use at least one of the forbidden chords. This time, these are the triangulations which contain $bo$, and an additional triangulation which contains $op$ but not $bo$ (both $cp$ and $bc$ must be present in this case). When $bc$ is present, there are $C_2=2$ triangulations which contain $bo$.

When $ao$ is present, each set of vertices of added edges which are connected to $o$ corresponds to two triangulations (since the quadrilateral $bpqo$ has two triangulations). In an 8-vint with three level-1 edges, there are at most four such sets, and in a standard 8-vint there are three.\marfigsc{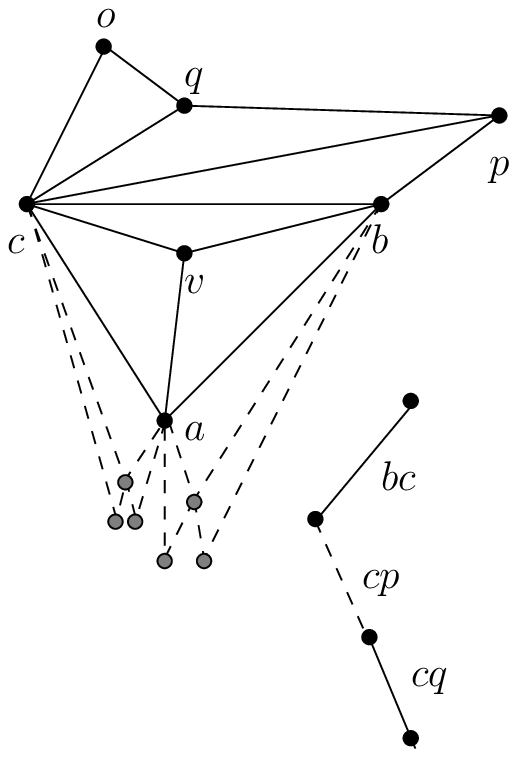}{1.4}In a 9-vint with three level-1 edges there are at most six such sets and in a 9-vint with two level-1 edges there are five. The absolute value of the overall charge of two 8-vints and a 9-vint is at least $\frac{1}{20-9}+\frac{1}{25-11}+\min(\frac{2}{38-15},\frac{2}{33-13})=\frac{883}{3542}>\frac{1}{4}-\frac{1}{1400}$, leaving the 6-vint with a charge smaller than $\frac{1}{1400}$ (note that this is the only case where we do not manage to completely neutralize the charge of the 6-vint). \hfill $\square$

\section{The analysis} \label{sec:analysis}
In the previous sections, we reviewed a variety of rules and methods for analyzing the charge of a 3-vint. In this section, we apply these rules to show that the charge of a 3-vint is always smaller than 30. This is achieved using a rather long case analysis, according to the possible rigid cores, which are grouped into subsections according to the number  $\lambda_1$ of their level-1 edges.

In order to bound the charge from any RC, with any non-rigid extensions, we first assume that its flip-tree is complete, up to level-3, and that each vint with a positive charge, which is not entirely in the RC, has a support of 2. This usually leads to a larger bound on the total charge (see the earlier version of the paper, \cite{ShWe06b}). To lower this bound, we remove edges from the flip-tree, by arguing that their presence can only lower the charge (for example, if they participate in 8-vints with a low support). Similarly, we argue that the worst-case charge is obtained when some vints have a higher support, either because a lower support would give a lower total charge, or by showing that it is impossible for them to have a low support.

\subsection{Analysis of $\lambda_1=0$} In this case, there are no edges in the rigid core; this implies that there are no vints with a support of 1, except for the 3-vint itself. There are three 5-vints and twelve 6-vints which contain two level-1 edges. All of these vints have a support of at least 3. The total charge is therefore at most \[4\cdot1+3\cdot\frac{1}{2}\cdot3+2\left(\frac{1}{2}\cdot6+\frac{1}{3}\cdot3\right)+1\left(\frac{1}{2}\cdot16+\frac{1}{3}\cdot12\right)=28\frac{1}{2}.\]

\subsection{Analysis of $\lambda_1=1$} This subsection analyzes the rigid cores with $\lambda_1=1$. We first analyze the {\em basic} RCs depicted in Figures \ref{fi:RC1a}, \ref{fi:RC1b}, and \ref{fi:RC1c}, and then deal with any other RC with $\lambda_1=1$, obtained by adding RC edges to one of the basic RCs. Any of these {\em extension} RCs is analyzed using a bound proved for the corresponding basic RC, and considering the changes in that bound caused by the rigidity of the new edges. In these flip-tree figures, the solid lines represent RC edges, and the dashed lines represent non-RC edges, which might, or might not be present in the flip-tree.

\vspace{5mm} \noindent {\bf RC 1a,} as depicted in Figure \ref{fi:RC1a}.

\noindent As already mentioned,\marfigsc{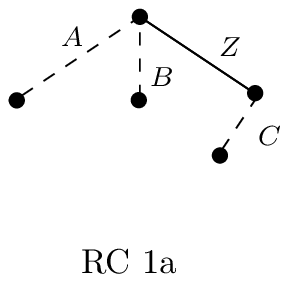}{1.2}here and later, we begin by assuming that the flip-tree is complete. This involves no loss of generality as long as we only consider vints with positive charges. We will drop this assumption and analyze the situation more carefully when we need to use vints with negative charges. \begin{list}{\labelitemi}{\leftmargin=1em}
 \item There are two non-rigid subtrees (as defined in Rule \ref{rule:subtree}; in the figure, these are the subtrees of $A$ and $B$). Each of those can contain five 6-vints. By Rule \ref{rule:subtree}, the charge from these ten 6-vints cannot exceed $2\cdot2=4$.
 \item The 6-vint which contains $C$ and $A$ has two non-adjacent flippable edges, which implies a support of at least 4.\marfigsc{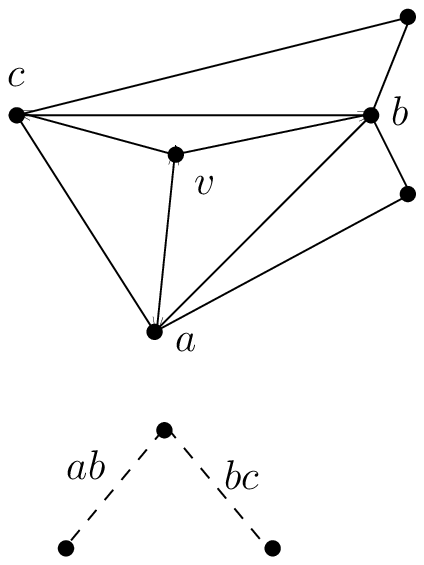}{1.2}There are four 6-vints of this sort, with a total charge of at most $4\cdot\frac{1}{4}=1$.
 \item The 5-vint which contains both $A$ and $B$ has a support of at least 3 (as depicted in Figure \ref{fi:RC1a1}). This also applies to the five 6-vints which extend this 5-vint. The overall charge from the above vints is at most $2\cdot\frac{1}{3}\cdot1+1\cdot\frac{1}{3}\cdot5=2\frac{1}{3}$.
\end{list}
In the above, we analyzed the supports of nineteen 6-vints. Since no 6-vint is fully contained in the RC, we assume that each of the other nine 6-vints has a support of 2. Using similar considerations for the other 5-vints and 4-vints, we conclude that the total charge cannot exceed \[4\cdot1+3\left( 1\cdot1+\frac{1}{2}\cdot2 \right)+2\cdot\frac{1}{2}\cdot8+1\cdot\frac{1}{2}\cdot9+4+1+2\frac{1}{3}=29\frac{5}{6}.\]

\noindent {\bf RC 1b,} as depicted in Figure \ref{fi:RC1b}. \begin{list}{\labelitemi}{\leftmargin=1em}
 \item As in RC 1a, there are five 6-vints and one 5-vint which contain both $A$ and $B$.\marfigsc{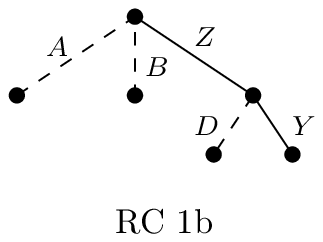}{1.2} Each of those vints has a support of at least 3, and their overall charge is, as above, at most $2\frac{1}{3}$.
 \item Similarly to RC 1a, the 6-vints which contain $D$ and either $A$ or $B$, have a support of at least 4. The charge from the two 6-vints cannot exceed $2\cdot\frac{1}{4}=\frac{1}{2}$.
 \item In the non-visible subtree of $Z$ and $Y$ (rooted at either $A$ or $B$), five 6-vints can be extended (using $Z$ and $Y$) into 8-vints with the same support (see Section \ref{sec:non-visib2}). We can ignore each of these 6-vints.
\end{list}
So far, we have accounted for twelve 6-vints, and one 5-vint. There is only one rigid 5-vint and no rigid 6-vints; so each of the other vints has a support of at least 2. We conclude that the total charge is at most\marfigsc{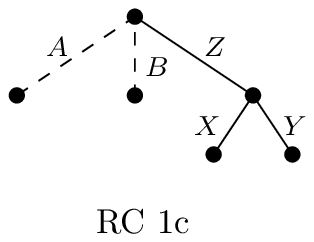}{1.2} \[4\cdot1+3\left(1\cdot1+\frac{1}{2}\cdot2\right)+2\left(1\cdot1+\frac{1}{2}\cdot7\right)+1\cdot\frac{1}{2}\cdot16+2\frac{1}{3}+\frac{1}{2}=29\frac{5}{6}.\]

\noindent {\bf RC 1c,} as depicted in Figure \ref{fi:RC1c}. \begin{list}{\labelitemi}{\leftmargin=1em}
 \item Assume, without loss of generality, that the non-visible subtree of $Z$, $Y$, and $X$ is the subtree of $A$ (see Section \ref{sec:non-visib2}). Using these RC edges, each of the five 6-vints lying entirely in the non-visible subtree can be extended into (more than) an 8-vint with the same support, neutralizing its charge. Similarly, each of the two 5-vints lying entirely in the non-visible subtree can be extended into an 8-vint with the same support, halving its charge. This implies that the overall charge from these seven vints is at most $2\cdot\frac{1}{2}\cdot\frac{1}{2}\cdot2=1$.
 \item As in the two preceding cases, there are five 6-vints and one 5-vint which contain both $A$ and $B$. Each of those vints has a support of at least 3, and their overall charge is, as above,  at most $2\frac{1}{3}$. \marfig{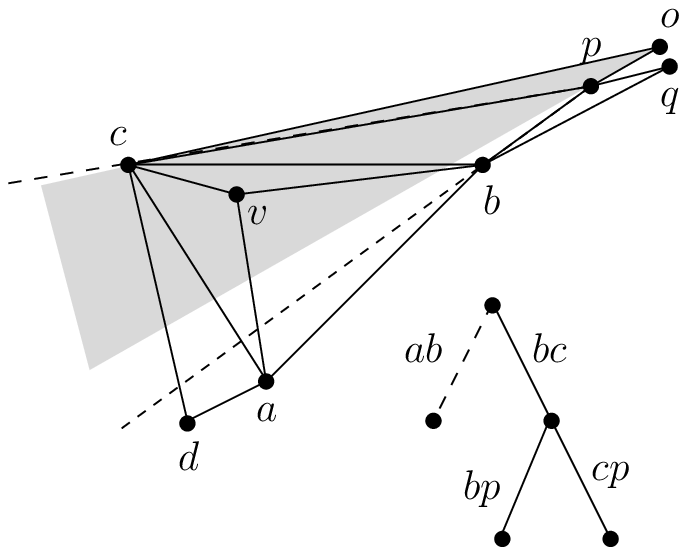}
 \item Consider the 5-vint using $B$ and $Z$. If this 5-vint has a support of 3, the four 6-vints extending it with either $X$, $Y$, or a child of $B$, also have a support of at least 3. This implies that, in this case, the charge is decreased by at least $(\frac{1}{2}-\frac{1}{3})(2\cdot1+1\cdot4)=1$. If the 5-vint has a support of 2, the vertex of $B$ cannot see the vertex of $Z$ (see Figure \ref{fi:RC1c1}, where $p$ is the vertex of $Z$ and $d$ is the vertex of $B$). Notice that any point that cannot be seen by $p$, cannot be seen neither by $o$ nor by $q$, which are the vertices of $X$ and $Y$, respectively (the line of sight of $o$ is shaded, and the line of sight of $p$ is bounded by the dashed lines). This implies that $o$ and $q$ also cannot see $d$. This, combined with the fact that the non-visible subtree of $Z$ is the subtree of $A$, imply that we can use $X$ and $Y$ to extend the 6-vint using $A$, $B$, and $Z$, into an 8-vint with the same support. We can also use $Z$ and $X$ (or $Y$) to extend the two 6-vints using $A$, $B$, and a child of $B$, into an 8-vint with the same support. Since each of these three 6-vints has a support of at least 3, the charge decreases by at least $\frac{1}{3}\cdot3=1$. We conclude that in either case the charge goes down by at least $1$.
\end{list}
The first two steps have taken care of ten 6-vints and three 5-vints. Using the default assumption that each of the remaining vints has a support of at least 2, with the exception of one 4-vint, two 5-vints, and one 6-vint which are rigid, and exploiting the charge reduction obtained in the first step, the overall charge is at most\marfigsc{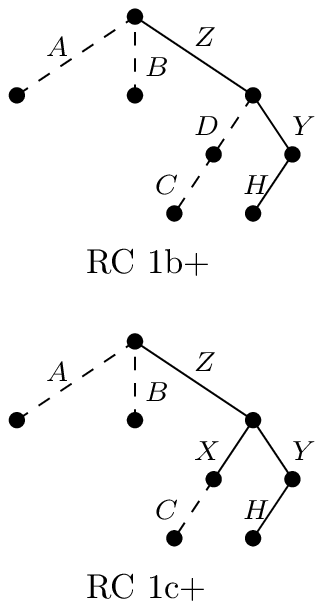}{1.2} \[4\cdot1+3\left(1\cdot1+\frac{1}{2}\cdot2\right)+2\left(1\cdot2+\frac{1}{2}\cdot4\right)+1\left(1\cdot1+\frac{1}{2}\cdot17\right)+1+2\frac{1}{3}-1=29\frac{5}{6}.\]

\paragraph{Extensions of the previous cases.} The only possible extension to the above RCs is the inclusion of additional level-3 RC edges (in cases RC 1b and RC 1c). Consider the addition of a \emph{single} level-3 edge, $H$, either to RC 1b or to RC 1c (without loss of generality, as a child edge of $Y$), as depicted in Figure \ref{fi:RC1ext}. There is only a single vint with a positive charge that uses $H$, which is a 6-vint previously considered as having a support of 2 (in the analysis of the bounds for these RCs). This increases the bound on the charge by $\frac{1}{2}$. In both cases, assume first that the edge $C$ exists, and consider the 6-vint which contains $C$. By Rule \ref{rule:RemainsRigid}, this 6-vint can be extended into an 8-vint with the same support, using $H$ and its parent (as depicted in Figure \ref{fi:RC1ext2}, where the 6-vint is shaded). This 6-vint, and the 6-vint obtained by replacing $C$ with its sibling edge (assuming that it exists), each added a charge of $\frac{1}{2}$ to the original total charge, which is now neutralized. Hence, the bound on the overall charge changes by $\frac{1}{2}(1\cdot1-1\cdot2)=-\frac{1}{2}$, implying that adding H to the RC can only lower the bound. If $C$ or $D$ (or both) do not exist, the overall charge decreases by at least $\frac{1}{2}$ (since we assumed each of the corresponding 6-vint has a support of at least 2; note that neither of these 6-vints participated in the special cases of charge reduction). This neutralizes the increase in the charge caused by $H$.

\marfig{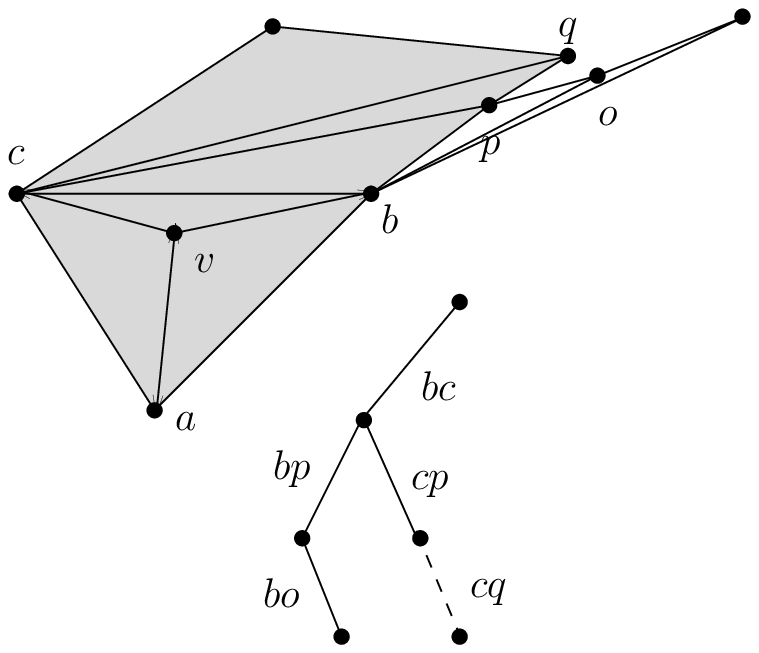}We now consider the addition of two level-3 RC edges. The subtree of the rigid level-1 edge can hold four 6-vints with a level-3 edge. Two of these 6-vints are entirely in the RC, which increases the bound on the charge by at most $\frac{1}{2}\cdot2=1$ (as in the previous paragraph, the bound on the charge of each 6-vint increases from $\frac{1}{2}$ to 1). For each of the two other 6-vints, either it is not present in the flip tree, or it can be extended into an 8-vint with the same support, as in the previous paragraph. In either case, the bound on the charge of the 6-vint decreases from $\frac{1}{2}$ to 0. This implies that the change in the charge cannot exceed $\frac{1}{2}(1\cdot2-1\cdot2)=0$.

In RC 1b, no more than two level-3 RC edges can be added. In RC 1c, after adding two such edges there are five RC edges, and by Rule \ref{rule:LargestRC}, additional level-3 RC edges cannot increase the charge.

\subsection{Analysis of $\lambda_1=3$} This subsection analyzes the rigid cores with $\lambda_1=3$. We first analyze the {\em basic} RCs depicted in Figures \ref{fi:RC3a}, \ref{fi:RC3b}, \ref{fi:RC3c}, and \ref{fi:RC3d}, and then deal with any other RC with $\lambda_1=3$, obtained by adding RC edges to one of the basic RCs. Any of these {\em extension} RCs is analyzed using a bound proved for the corresponding basic RC, and considering the changes in that bound caused by the rigidity of the new edges. In these flip-tree figures, the solid lines represent RC edges, and the dashed lines represent non-RC edges, which might, or might not be present in the flip-tree. \marfigsc{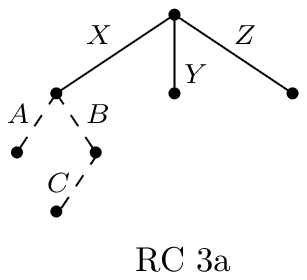}{1.3}

\vspace{5mm} \noindent {\bf RC 3a,} as depicted in Figure \ref{fi:RC3a}.

\noindent There are at most three 6-vints that use two level-2 edges, such as the one using $X$, $A$, and $B$. Each such 6-vint can be extended into an 8-vint by adding the two additional RC edges. The possible charges from such vints were analyzed in Section \ref{sec:2level2} (notice, though, that we cannot use Rule \ref{rule:6-vint1} from this section, since we do not have three additional RC edges). If the 6-vint has a support of at most 3, the 8-vint has the same support. If the 6-vint has a support of either 4 or 7, the 8-vint has a support of at most 7 or 13, respectively. In both cases, the combined charge from both vints cannot exceed $\frac{3}{28}$.

We first ignore every vint that is not entirely in the RC, except for the three 6-vints that were just considered. The charge from the remaining vints cannot exceed \[4\cdot1+3\cdot3+2\cdot3+1\cdot1+\frac{3}{28}\cdot3=20\frac{9}{28}.\]

\noindent We next analyze the possible charges coming from the ignored vints. First, there are six 5-vints that consist of a rigid level-1 edge and a level-2 edge (such as the one using $X$ and $A$). By Rule \ref{rule:SingleFlip}, three out of these six 5-vints are handicapped (if they are present in the flip-tree),\marfigsc{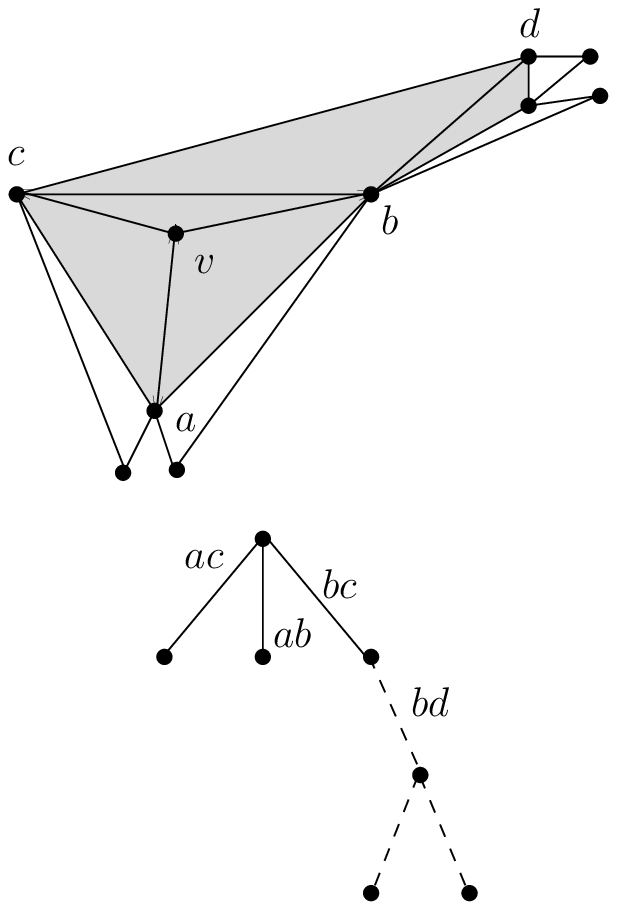}{1.4}and the other three can have a support of either 2 or 3. Moreover, each of the six 5-vints can be extended into two 6-vints by adding a level-3 edge (such as the one using $X$, $B$, and $C$), and into two more 6-vints by adding an additional level-1 edge (such as the one using $X$, $A$, and $Y$). Each 5-vint can be extended into a fifth 6-vint by an additional level-2 edge, but we already considered these 6-vints. We consider the possible cases for a 5-vint and its extensions, and bound the combined charge in each case: \begin{list}{\labelitemi}{\leftmargin=1em}

\item A handicapped 5-vint (as depicted in Figure \ref{fi:RC3a1}, where the hole of the 5-vint is shaded). Each of the two 6-vints that extend the 5-vint with a level-3 edge is entirely in its non-visible terrain. Therefore, by adding the two additional RC edges to such a 6-vint, we generate an 8-vint with the same support, neutralizing the charge of the 6-vint.\marfig{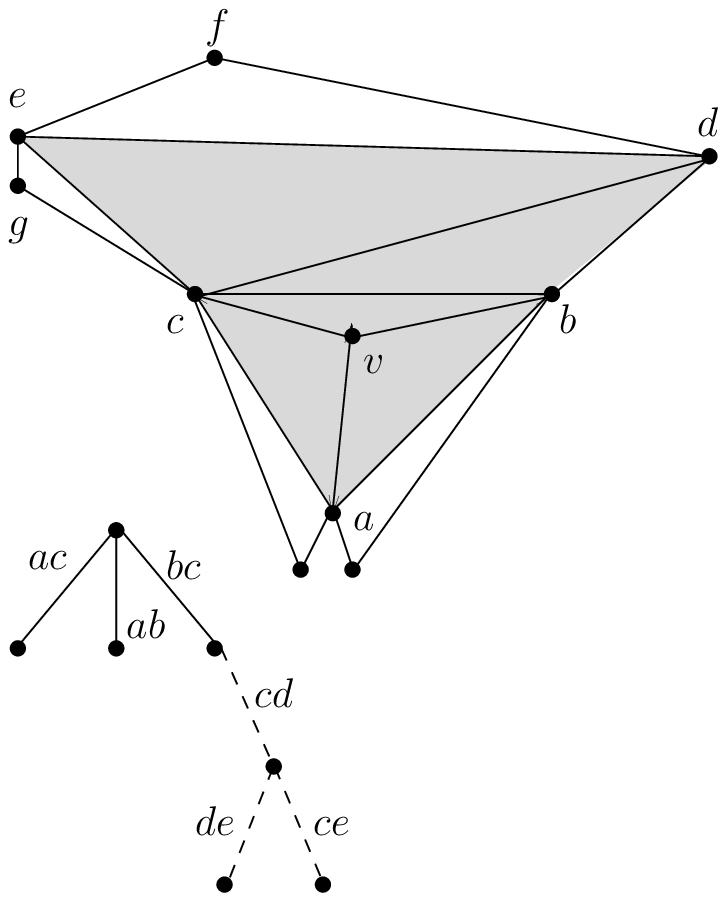}The charge from the 5-vint and the two remaining 6-vints is at most $2\cdot\frac{1}{2}\cdot1+1\cdot\frac{1}{2}\cdot2=2$.

\item A 5-vint which is not handicapped, but has a support of 2 (as depicted in Figure \ref{fi:RC3a2}, where the 5-vint is shaded). As in the previous case, each of the two 6-vints that extend the 5-vint with a level-3 edge (such as the 6-vint using $de$) can be extended into an 8-vint, by using the two additional RC edges. If the vertex of the level-3 edge of the 6-vint cannot see $a$ (such as the vertex $g$ in figure \ref{fi:RC3a2}), the 6-vint is entirely in its non-visible terrain, and thus, the 8-vint has the same support as the 6-vint. Otherwise, by Table \ref{ta:app} in Section \ref{sec:longPr} (the part where only $o$ --- f in Figure \ref{fi:RC3a2} --- can see $a$), the 6-vint has a support of either 4, 6, or 7, and the 8-vint has a support of at most 7, 9 or 13, respectively. Therefore, the overall charge from such a 6-vint and its extending 8-vint cannot exceed $\frac{1}{4}-\frac{1}{7}=\frac{3}{28}$. We conclude that the charge from the such a 5-vint and its four extension 6-vints is at most $2\cdot\frac{1}{2}\cdot1+1\cdot\frac{1}{2}\cdot2+\frac{3}{28}\cdot2=2\frac{3}{14}$.\marfigsc{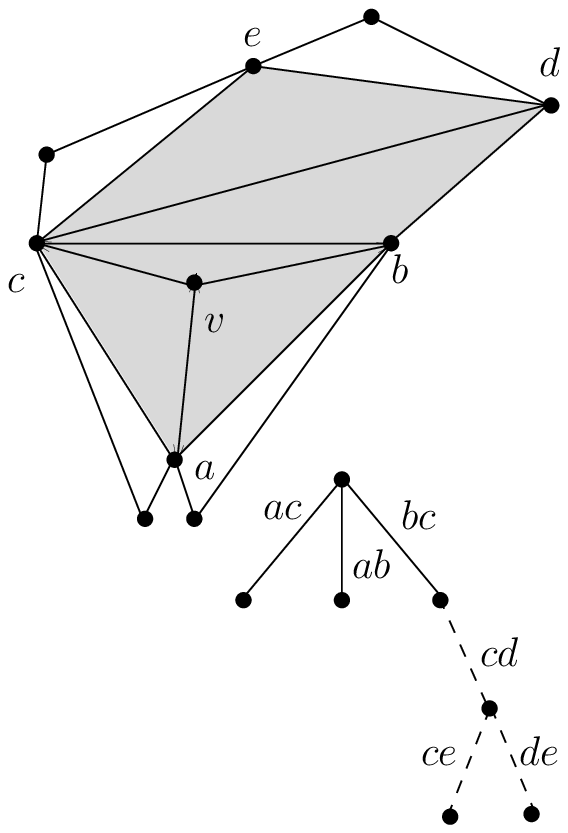}{1.4}

\item A 5-vint with a support of 3 (as depicted in Figure \ref{fi:RC3a3}, where the 5-vint is shaded). Once again, each of the two 6-vints that extend the 5-vint with a level-3 edge (such as the one using $de$) can be extended into an 8-vint, by using the two additional RC edges. The possible charges from such vints are listed in Table \ref{ta:app} in Section \ref{sec:longPr} (since the vertex of the level-2 edge can see $a$, we need to consider the part where only $q$ can see $a$, and the part where both $o$ and $q$ can see $a$). We notice that there is only a single case which generates a charge of more than $\frac{1}{5}$ --- when the 6-vint and the 8-vint have supports of 3 and 8, respectively. We conclude that the charge from the such a 5-vint and its four extension 6-vints is at most $2\cdot\frac{1}{3}\cdot1+1\cdot\frac{1}{3}\cdot2+\left(\frac{1}{3}-\frac{1}{8}\right)\cdot2=1\frac{3}{4}$.
\end{list}

\noindent So far we have accounted for all possible 5-vints and 6-vints. The overall charge depends on how many 5-vints with a level-2 edge are actually present (and what are their supports): \begin{list}{\labelitemi}{\leftmargin=1em}

\item  There are at most four 5-vints. By Rule \ref{rule:SingleFlip}, at least one of these 5-vints is handicapped. Therefore, the charge cannot exceed (the second term represents the handicapped 5-vint, and the third term represents the most pessimistic bound for the three other 5-vints)\marfigsc{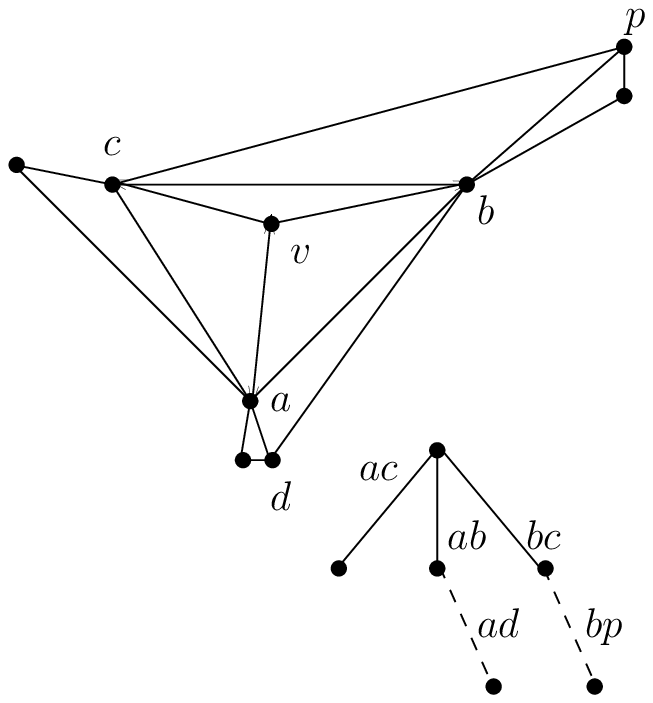}{1.5} \[20\frac{9}{28}+2\cdot1+2\frac{3}{14}\cdot3=28\frac{27}{28}.\]

\item There are five 5-vints, and at least two of them have a support of 3. There are at least two handicapped 5-vints, and by appending their edges together with the third RC edge, we form an 8-vint (as depicted in Figure \ref{fi:RC3a4}, where one handicapped 5-vint consists of $ab$ and $ad$, and the other one consists of $bc$ and $bp$). All the vertices of this 8-vint are in their non-visible terrains, and thus, its support is the product of the supports of the 5-vints, which is $2\cdot2=4$ (see Rule \ref{rule:non-visib}). Therefore, the total charge cannot exceed (the second term represents two 5-vints with a support of 3, the third term represents two handicapped 5-vints, the fourth term represents the remaining 5-vint, and the last term represents the 8-vint described above)\marfig{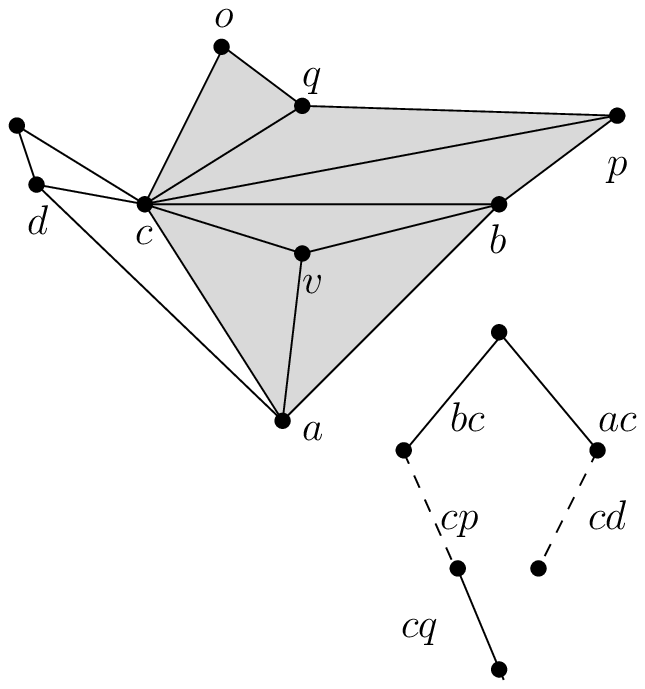} \[20\frac{9}{28}+1\frac{3}{4}\cdot2+2\cdot2+2\frac{3}{14}\cdot1-\frac{1}{4}\cdot1=29\frac{11}{14}.\]

\item There are five 5-vints, and at least four of them have a support of 2 (notice that this is the complement of the previous case for five 5-vints). We can use pairs of such 5-vints (from different subtrees) in order to create 8-vints such as the one described in the previous case. (For this, we only need to assume that the supports of each of the corresponding 5-vints is 2.) This time, there are at least four such 8-vints, each with a support of 4. Therefore, the total charge cannot exceed (the second term represents two handicapped 5-vints, the third term represents the three other 5-vints, and the last term represents the four 8-vints described above) \[20\frac{9}{28}+2\cdot2+2\frac{3}{14}\cdot3-\frac{1}{4}\cdot4=29\frac{27}{28}.\]

\item There are six 5-vints, and exactly three of them have a support of 2. These three 5-vints must be the handicapped 5-vints, and we can use the non-visible subtree method with each of them (see Section \ref{sec:non-visib2}). There are six 6-vints with a level-3 edge that were not yet ignored (those that extend a 5-vint with a support of 3), and we considered the charge generated by each of those as at most $\frac{1}{3}-\frac{1}{8}=\frac{5}{24}$ (in the case of ``A 5-vint with a support of 3"). Each non-visible subtree of a handicapped 5-vint contains two such 6-vints, and thus, by appending each of the 6-vints with the edges of the respective handicapped 5-vint, we generate an 8-vint with a double support (as depicted in Figure \ref{fi:RC3a5}, where the 6-vint is shaded and the handicapped 5-vint consists of $ac$ and $cd$). For simplicity, we will assume that each such 8-vint halves the net previous charge of its corresponding 6-vint, including the other extension 8-vint considered above, even though it actually gives a higher negative charge (for example, when the 6-vint has a support of 3, the charge should be $\left(\frac{1}{3}-\frac{1}{8}\right)-\frac{1}{3}\cdot\frac{1}{2}=\frac{1}{24}$ and not $\left(\frac{1}{3}-\frac{1}{8}\right)\cdot\frac{1}{2}=\frac{5}{48}$ ). Each of the three hadicapped 5-vints extends two such 6-vints, with an overall negative charge of at least $\frac{5}{24}\cdot\frac{1}{2}\cdot6=\frac{5}{8}$ (notice that such a 6-vint can be in at most two non-visible subtrees, which implies that the overall charge from a 6-vint and its extensions is non-negative).\marfig{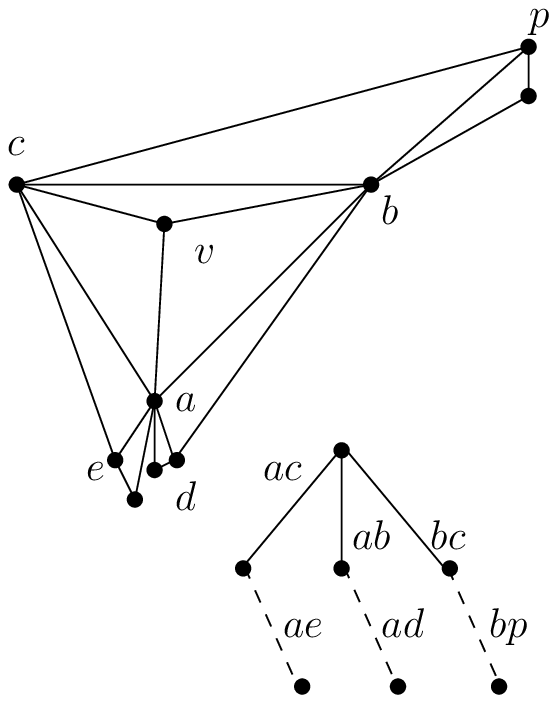} As in the previous cases, we can generate an 8-vint with a support of 4 from each pair of handicapped 5-vints. This time, there are three such 8-vints. Similarly, by appending the edges of all three handicapped 5-vints, we get a 9-vint with a support of $2\cdot2\cdot2=8$ (as depicted in Figure \ref{fi:RC3a6}). The total charge cannot exceed (the second term represents three handicapped 5-vints, the third term represents the three other 5-vints, the fourth term represents the negative charge from the three non-visible subtrees, the fifth term represents the three 8-vints described above, and the last term represents the 9-vint) \[20\frac{9}{28}+2\cdot3+1\frac{3}{4}\cdot3-\frac{5}{8}-\frac{1}{4}\cdot3-\frac{2}{8}\cdot1=29\frac{53}{56}.\]

\item There are six 5-vints, and at least four of them have a support of 2. We wish to show that the bound in this case cannot exceed the bound in the previous case. This is done by showing that in the previous case, changing the support of another 5-vint into 2 can only lower the overall charge. Such a change increases the bound on the charge of the 5-vint and its extensions by $(2\frac{3}{14}-1\frac{3}{4})=\frac{13}{28}$. Moreover, there might be a decrease in the negative charge attained from the use of non-visible subtrees (described in the previous case). Specifically, there are two 6-vints that extend the 5-vint with a level-3 edge, and each of them could be a 6-vint with a support of 3 contained in the non-visible subtree of at most two handicapped 5-vints. After the change, each such 6-vint generates a charge of at most $\frac{3}{28}$, instead of at most $\frac{5}{24}$. Since we assume that an 8-vint extending such a 6-vint neutralizes half of its charge, the negative charge from (at most) four 8-vints decreases by $\frac{1}{2}\cdot\left(\frac{5}{24}-\frac{3}{28}\right)\cdot4=\frac{17}{84}$. Nevertheless, we also get vints which decrease the total charge. That is, we can use the changed 5-vint to create additional extension vints, such as those described in the previous cases.\marfigsc{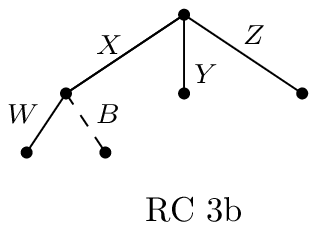}{1.3}We can create at least two new 8-vints with a support of 4 and one 9-vint with a support of 8. The above implies that the change can increase the total charge by at most $\frac{13}{28}+\frac{17}{84}-2\cdot\frac{1}{4}-\frac{2}{8}=-\frac{1}{12}$. We conclude that changing the support of more 5-vints into 2 cannot increase the total charge.
\end{list}

\noindent {\bf RC 3b,} as depicted in Figure \ref{fi:RC3b}. \begin{list}{\labelitemi}{\leftmargin=1em}
  \item Each of the ten 6-vints that consist of a rigid level-1 edge, a non-rigid level-2 edge, and a level-3 edge meets the conditions of Rule \ref{rule:6-vint2}. By the rule, the overall charge of these ten 6-vints cannot exceed $10\cdot\frac{1}{1400}=\frac{1}{140}$.
 \item Each of the two 6-vints which use two non-rigid level-2 edges meets the conditions of Rule \ref{rule:6-vint1}. By the rule, these 6-vints can be ignored.
 \item Each of the two 6-vints using $X$, $W$, and a level-3 edge can be extended into an 8-vint, by using $Y$ and $Z$. By Table \ref{ta:app} of Section \ref{sec:longPr} (the part where the level-2 edge is rigid), if the 6-vint has a support of at most 3, the 8-vint has the same support, and if it has a support of 4, the 8-vint has a support of at most 7. Therefore, the overall charge from these two 6-vints and their two extending 8-vints is at most $2(\frac{1}{4}-\frac{1}{7})=\frac{3}{14}$. \marfig{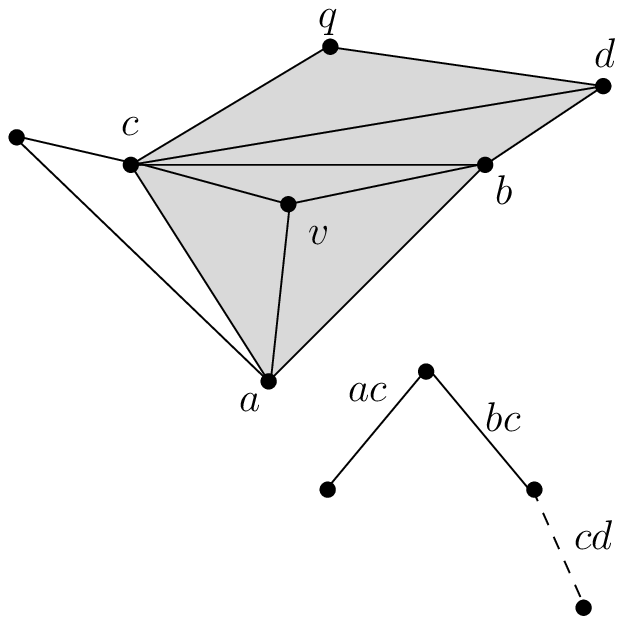}
\end{list}

\noindent We first ignore every vint not entirely in the RC, except for the fourteen 6-vints that were just considered, and the 6-vint that consists of $X$, $W$, and $B$. The charge from the remaining vints cannot exceed\[4\cdot1+3\cdot3+2\cdot4+1\left(1\cdot3+\frac{1}{2}\cdot1\right)+\frac{1}{140}+\frac{3}{14}=24\frac{101}{140}.\]

\noindent The possible additional charges come from the five 5-vints which use a rigid level-1 edge and a non-rigid level-2 edge, and from the ten 6-vints that extend such a 5-vint using an additional level-1 edge. At least two of these five 5-vints must be handicapped, and each of the other three can have a support of either 2 or 3. We analyze the possible charges from such 5-vints and their extension vints according to the following cases:\begin{list}{\labelitemi}{\leftmargin=1em}
 \item A 5-vint with a support of 2 (including the handicapped 5-vints). By appending additional RC edges to the 5-vint, we can generate two 6-vints and an 8-vint\footnote{Note that the extension 6-vints used here, and in the following cases, are indeed those not considered above. In the case of the 5-vint which uses $X$ and $B$, the extension 6-vints are those using $Y$ or $Z$, but not $W$}.\marfigsc{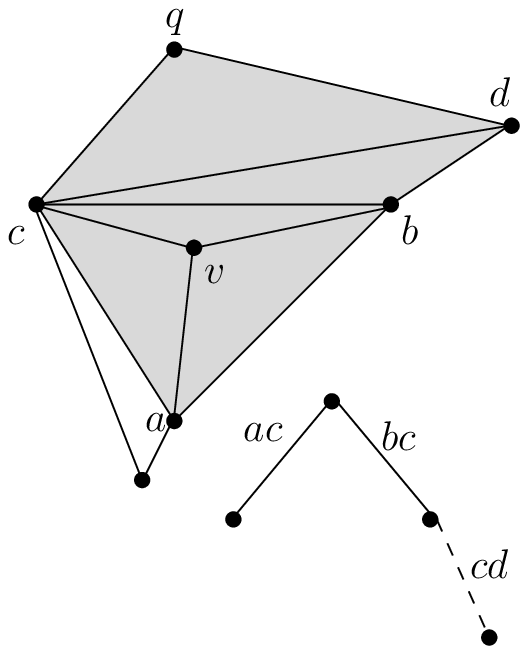}{1.4}Since all of the vertices of the 5-vint are in their non-visible terrain, adding additional RC edges cannot increase the support. Thus, all of the above vints have a support of 2. The overall charge of these vints is $\frac{1}{2}(2\cdot1+1\cdot2-1\cdot1)=1\frac{1}{2}$.
 \item A 5-vint with a support of 3 that does not use $X$ (as depicted in Figure \ref{fi:RC3b1}, where the 5-vint is shaded). By appending additional RC edges, the 5-vint can be extended into two 6-vints, each with a support of either 3 or 4 (as depicted in Figure \ref{fi:RC3b1} and in Figure \ref{fi:RC3b2}, respectively; in both figures, the 5-vint is shaded), and into one 8-vint with a support that can be bounded in terms of the supports of the 6-vints, as followd. (See Figure \ref{fi:RC3b3}, where the 5-vint is shaded and $ac$ represents $X$.) For each of these three extension vints, every triangulation of its hole must contain either $aq$ or $bc$ (in the notation of the figures), and $tr(bc)=C_2=2$. Let $i$ be a binary variable, which is 1 if and only if $q$ can see $d$ (in the notation of Figure \ref{fi:RC3b3}). (Notice that when $i=0$, $q$ cannot see $f$ either.) Let $j$ be a binary variable, which is 1 if and only if $q$ can see $e$. The support of the 6-vint that uses $X$ is $tr(aq)+tr(bc)=(1+i)+2=3+i$, and similarly, the support of the second 6-vint is $3+j$.\marfigsc{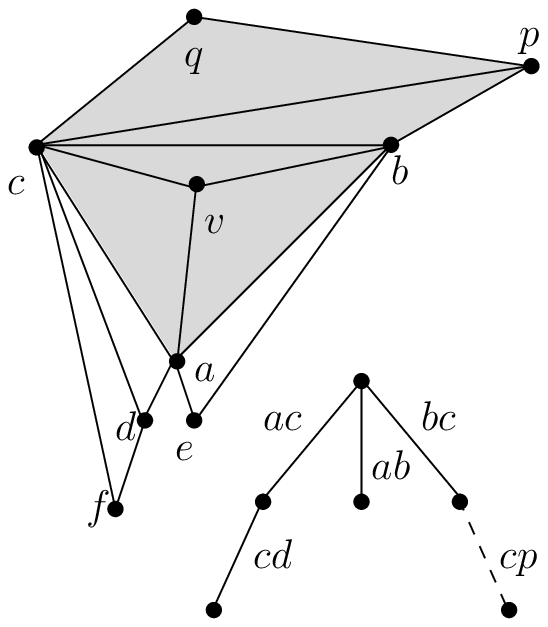}{1.4}We can bound the support of the 8-vint by the expression $tr(bc)+tr(aq)=2+(1+j)\cdot(1+2i)$ (where $tr(aq)$ is the maximal number of triangulations of the portion to the right of $aq$, times the maximal number of triangulations of the portion to its left). Table \ref{ta:RC3b} lists the possible cases for the supports of these vints. (The first term in the charge represents the charge of the 5-vint, which is always $2\cdot\frac{1}{3}$.) By examining the four possible cases, we conclude that the overall charge of these four vints cannot exceed $1\frac{1}{20}$.
\item A 5-vint with a support of 3 that contains $X$ (as depicted in Figure \ref{fi:RC3b4}, where the 5-vint is shaded, $bc$ represents $X$, and $cp$ represents $B$). By using the same analysis as in the previous case, we notice that the 6-vints have the same possible supports, and that the 8-vint has a smaller bound on its support --- $tr(bc)+tr(aq)=2+(1+j)\cdot(1+i)$ (since, by Rule \ref{rule:RemainsRigid}, adding $W$ cannot increase its charge). In this case, as is easily verified, the charge from the four vints cannot exceed $1$. \marfig{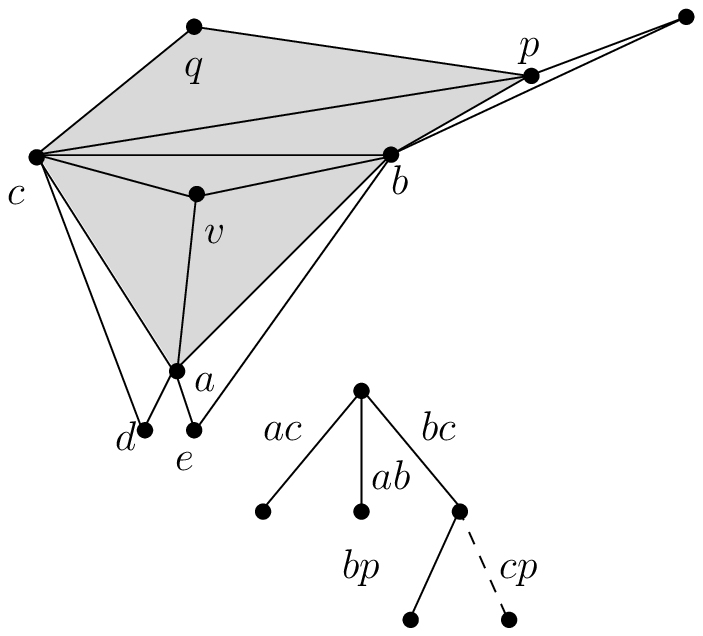}
\end{list}

\begin{table}[ht]
\caption{Possible charges from a 5-vint with a support of 3 and its extensions. The first 6-vint is the one containing $X$.} \label{ta:RC3b}
\centering
\begin{tabular}{c | c c c c}
\hline\hline
$i$/$j$ & 6-vint    & Second & Max support & Overall \\
        & using $X$ & 6-vint & of 8-vint   & charge  \\[0.5ex]

\hline
0/0 & 3 & 3 & 3 & $\frac{2}{3}+2\cdot\frac{1}{3}-\frac{1}{3}=1$\\
0/1 & 3 & 4 & 4 & $\frac{2}{3}+\frac{1}{3}+\frac{1}{4}-\frac{1}{4}=1$\\
1/0 & 4 & 3 & 5 & $\frac{2}{3}+\frac{1}{4}+\frac{1}{3}-\frac{1}{5}=1\frac{1}{20}$\\
1/1 & 4 & 4 & 8 & $\frac{2}{3}+2\cdot\frac{1}{4}-\frac{1}{8}=1\frac{1}{24}$\\

\hline
\end{tabular}
\end{table}

\noindent We divide the rest of the analysis according to the number of such 5-vints that are present in the flip-tree and their supports:
\begin{list}{\labelitemi}{\leftmargin=1em}
 \item At least two 5-vints are not present in the flip-tree. The charge is at most \[24\frac{101}{140}+1\frac{1}{2}\cdot3=29\frac{31}{140}.\]
 \item A single 5-vint is not present in the flip-tree, and at most two 5-vints have a support of 2. The charge is at most (the third term represents two 5-vints with a support of 3, and the second term represents the other two 5-vints) \[24\frac{101}{140}+1\frac{1}{2}\cdot2+1\frac{1}{20}\cdot2=29\frac{23}{28}.\]
 \item \marfig{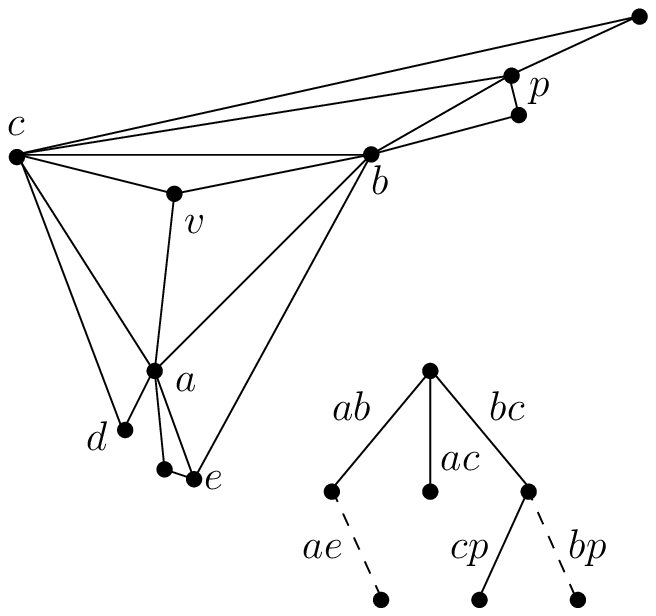}A single 5-vint is not present in the flip-tree, and at least three 5-vints have a support of 2. We consider the previous case and show that changing the support of another 5-vint into 2 cannot increase the total charge. The bound on the charge of the changed 5-vint increases by $1\frac{1}{2}-1\frac{1}{20}=\frac{9}{20}$. By appending the edges of the changed 5-vint with the edges of another 5-vint which has a support of 2 (and without a common level-1 edge), and with the additional RC edges, we form a 9-vint (as depicted in Figure \ref{fi:RC3b5}, where $bc$ represents $X$, the first 5-vint consists of $bc$ and $bp$, and the second 5-vint consists of $ab$ and $ae$). By Rule \ref{rule:non-visib}, the support of this 9-vint is the product of the supports of the two 5-vints, which is $2\cdot2=4$. Therefore, the change in the total charge is at most $\frac{9}{20}-\frac{2}{4}=-\frac{1}{20}$, which implies that giving a support of 2 to more 5-vints can only decrease the total charge. In conclusion, when a single 5-vint is missing, the charge cannot exceed $29\frac{23}{28}$.
 \item All five 5-vints are present, and only the two handicapped 5-vints have a support of 2. By appending the edges of the two handicapped 5-vints together with additional RC edges, we form an 8-vint and a 9-vint. As in the previous case, by Rule \ref{rule:non-visib}, both vints have a support of $2\cdot2=4$. The non-visible subtree of each of the handicapped 5-vints must contain a 5-vint with a support of 3.\marfig{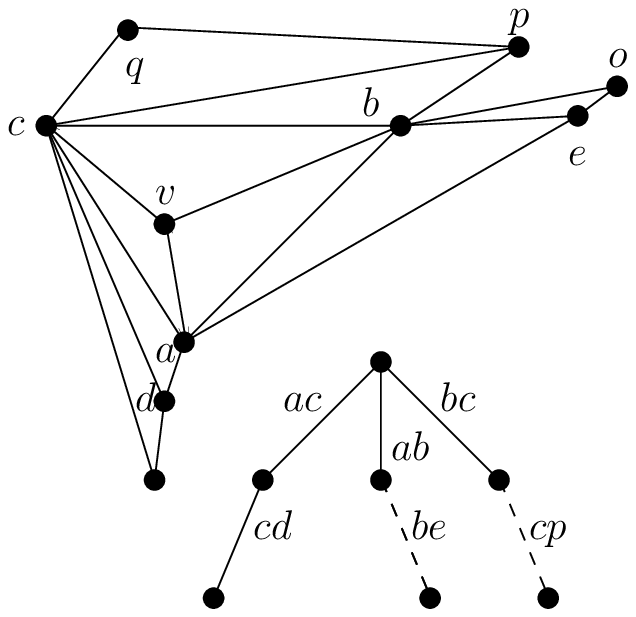}Appending the edges of a handicapped 5-vint together with the edges of the corresponding 5-vint with a support of 3, and with the rest of the RC, forms a 9-vint (as depicted in Figure \ref{fi:RC3b6}, where the handicapped 5-vint consists of $ab$ and $be$, and the corresponding 5-vint with a support of 3 consists of $bc$ and $cp$). If we ignore the edges of the handicapped 5-vint, we get a 7-vint with a support of at most $tr(bc)+tr(aq)=2+3=5$. Since $e$ and $o$ cannot see any vertices outside of the convex quadrilateral $aboe$, adding the edges of the handicapped 5-vint doubles the support of the 7-vint, to at most $5\cdot2=10$. There are two such 9-vints, since there are two handicapped 5-vints. Therefore, the total charge cannot exceed (the second term represents the two handicapped 5-vints, the third term represents the other three 5-vints, the fourth term represents the 8-vint and the 9-vint which have a support of 4, and the last term represents the two 9-vints that have a support of 10) \[24\frac{101}{140}+1\frac{1}{2}\cdot2+1\frac{1}{20}\cdot3-\frac{1}{4}(2\cdot1+1\cdot1)-\frac{2}{10}\cdot2=29\frac{101}{140}.\]
 \item All five 5-vints are present in the flip-tree, and there are at least three 5-vints with a support of 2. As before, we consider the previous case and change the support of another 5-vint into 2. The bound on the charge of the 5-vint and its extensions increases by at most $1\frac{1}{2}-1=\frac{1}{2}$. By appending the edges of the changed 5-vint together with the edges of a handicapped 5-vint that does not share a common level-1 edge with it, and with additional RC edges, we can create one additional 8-vint and one additional 9-vint, both with a support of 4 (as described in the previous case; notice that if the changed 5-vint is the one using $X$, we can use each of the handicapped 5-vints, and thus, have two 8-vints and two 9-vints). However, we might have already considered this 9-vint in the previous case as having a negative charge of at least $\frac{2}{10}=\frac{1}{5}$. Therefore, the bound increases by at most $\frac{1}{2}-\frac{1}{4}(2\cdot1+1\cdot1)+\frac{1}{5}=-\frac{1}{20}$ (the second term represents the additional 8-vint and 9-vint with a support of 4, and the third term represents the previous charge of the 9-vint). In conclusion, changing the supports of more 5-vints into 2 can only decrease the charge.
\end{list} \marfigsc{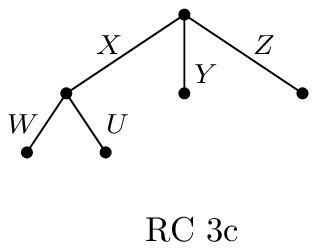}{1.3}

\noindent {\bf RC 3c,} as depicted in Figure \ref{fi:RC3c}. \begin{list}{\labelitemi}{\leftmargin=1em}
 \item There are five 5-vints, six 6-vints, and one 8-vint entirely in the RC. The overall charge of the vints entirely in the RC is $4\cdot1+3\cdot3+2\cdot5+1\cdot6-1\cdot1=28$.
 \item Each of the twelve 6-vints that use a level-3 edge meets the conditions of Rule \ref{rule:6-vint2}. By the rule, the overall charge of these twelve 6-vints cannot exceed $12\cdot\frac{1}{1400}<\frac{1}{100}$.
 \item Each of the two 6-vints that use two non-rigid level-2 edges meets the conditions of Rule \ref{rule:6-vint1}. By the rule, these two 6-vints can be ignored.
 \item Consider a 5-vint with a non-rigid level-2 edge and a support of 2. We already considered the two 6-vints that extend it with a level-3 edge and the 6-vint that extends it with an additional level-2 edge. By appending additional RC edges, the 5-vint can be extended into two additional 6-vints, three 8-vints and one 9-vint, all with a support of 2 (since the vertices of the 5-vint are in their non-visible terrain). The overall charge from the above vints is $\frac{1}{2}(2\cdot1+1\cdot2-1\cdot3-2\cdot1)=-\frac{1}{2}$. Therefore, the existence of a level-2 edge, which takes part in a 5-vint with a support of 2, can only decrease the total charge. By Rule \ref{rule:SingleFlip}, there can be at most two 5-vints with a level-2 edge and a support of 3, since the other two are handicapped 5-vints.
 \item \marfig{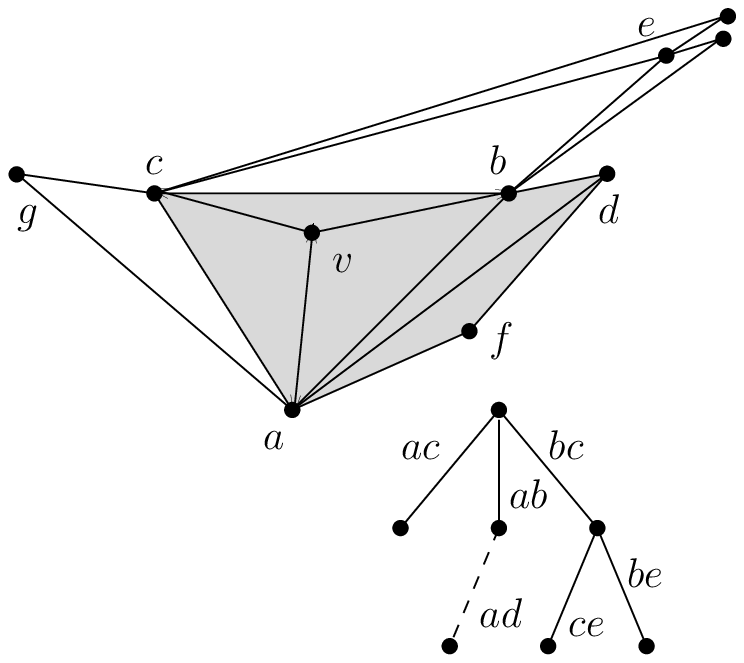}We still need to consider two possible 5-vints with a support of 3, and four 6-vints which extend them with an additional level-1 edge. One of these two 5-vints is in the non-visible subtree of $X$, $W$, and $U$ (as depicted in Figure \ref{fi:RC3c1}, where $X$, $W$, and $U$ correspond to $bc$, $be$, and $ce$, respectively, and the 5-vint is shaded). By using additional RC edges, we can extended the 5-vint into two 6-vints with a support of at least 3 (by adding either $bc$ or $ac$), one 8-vint with a support of exactly 3 (by adding $bc$, $be$, and $ce$), two 8-vints with a support of at most $tr(ab)+tr(cf)=2+2=4$ (by adding $ac$, $bc$, and a child-edge of $bc$), and one 9-vint with a support of at most 4 (by adding the entire RC). The overall charge from these seven vints is at most $\frac{2}{3}+\frac{1}{3}\cdot2-\frac{1}{3}-\frac{1}{4}\cdot2-\frac{2}{4}=0$ (the terms represent the vints in the order they were described above). We conclude that this 5-vint and its extensions cannot increase the total charge.
 \item We are left with a single 5-vint with a support of 3, and with the two 6-vints which extend it with an additional level-1 edge. Since these 6-vints also have a support of at least 3, the overall charge of these three vints cannot exceed $\frac{1}{3}(2\cdot1+1\cdot2)=1\frac{1}{3}$.
\end{list}

\noindent Hence, the total charge cannot exceed (the first term represents the vints that are entirely in the RC, the second term represents the twelve 6-vints with a level-3 edge, and the last term represents the only additional 5-vint that can generate positive charge with its extensions) \[28+\frac{1}{100}+1\frac{1}{3}=29\frac{103}{300}.\] \marfigsc{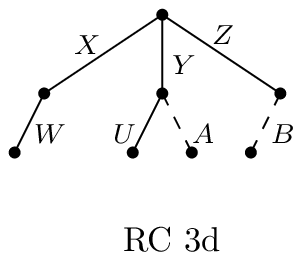}{1.3}

\noindent {\bf RC 3d,} as depicted in Figure \ref{fi:RC3d}. \begin{list}{\labelitemi}{\leftmargin=1em}
 \item There are five 5-vints, five 6-vints, and one 8-vint entirely in the RC. The overall charge of the vints entirely in the RC is $4\cdot1+3\cdot3+2\cdot5+1\cdot5-1\cdot1=27$.
 \item Each of the twelve 6-vints that use a level-3 edge meets the conditions of Rule \ref{rule:6-vint2}. By the rule, the overall charge of these twelve 6-vints cannot exceed $12\cdot\frac{1}{1400}<\frac{1}{100}$.
 \item The 6-vint using two non-rigid level-2 edges meets the conditions of Rule \ref{rule:6-vint1}. By the rule, this 6-vint can be ignored.
 \item Consider a 5-vint using $Z$ and a non-rigid level-2 edge (such as the one using $Z$ and $B$), and assume that it has a support of 2. The vertices of such a 5-vint are in their non-visible terrain, and thus, appending additional RC edges cannot increase its charge. By appending additional RC edges, we get two 6-vints (by appending either $X$ or $Y$), two 8-vints, and a 9-vint, all with a support of 2. There are three more vints that extend the 5-vint and may have a positive charge (the two 6-vints that extend the 5-vint with a level-3 edge, and the 6-vint that extends it with an additional level-2 edge), but we already considered them above. Thus, when the level-2 edge of such a 5-vint is present in the flip-tree, the bound on the charge can increase by at most $\frac{2}{2}\cdot1+\frac{1}{2}\cdot2-\frac{1}{2}\cdot2-\frac{2}{2}\cdot1=0$. We may therefore ignore the level-2 edge of the handicapped 5-vint in the subtree of $Z$, and assume that the other 5-vint in that subtree has a support of 3.
  \item Consider the 5-vint which contains $Z$ and has a support of 3. Out of the five 6-vints that extend it, we already considered the two 6-vints that extend the 5-vint with a level-3 edge and the 6-vint which extends the 5-vint with an additional level-2 edge. The overall charge from the 5-vint and the two remaining 6-vints cannot exceed $\frac{1}{3}(2\cdot1+1\cdot2)=1\frac{1}{3}$.
 \item Consider a 5-vint using either $X$ or $Y$ and a non-rigid level-2 edge (such as the one using $Y$ and $A$), and assume that it has a support of 2. The vertices of such a 5-vint are in their non-visible terrain, and thus, appending additional RC edges cannot increase its charge. By appending additional RC edges, we get three 6-vints (by appending either $X$, $Z$, or $U$), three 8-vints, and a 9-vint, all with a support of 2. There are two more vints that extend the 5-vint and may have a positive charge (the two 6-vints that extend the 5-vint with a level-3 edge), but we already considered them above. Thus, when the level-2 edge of such a 5-vint is present in the flip-tree, the bound on the charge can increase by at most $\frac{2}{2}\cdot1+\frac{1}{2}\cdot3-\frac{1}{2}\cdot3-\frac{2}{2}\cdot1=0$.\marfigsc{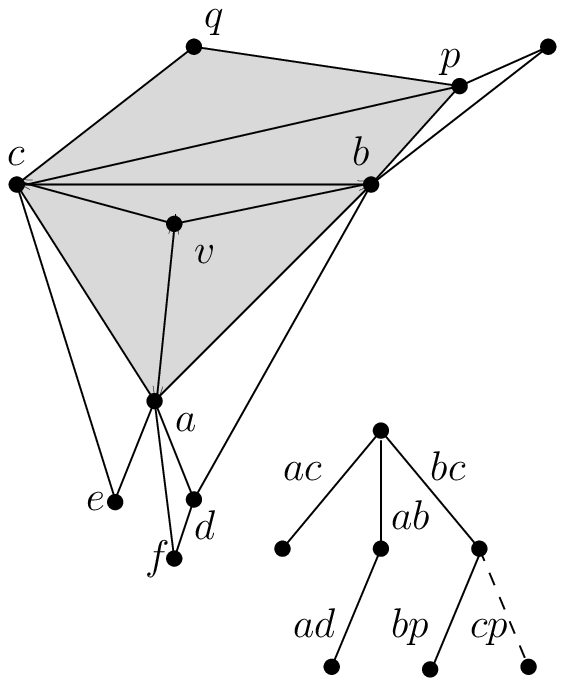}{1.4}We may therefore assume that these two 5-vints have a support of 3.
 \item Consider a 5-vint which contains either $X$ or $Y$, a non-rigid level-2 edge, and has a support of 3 (as depicted in Figure \ref{fi:RC3d1}, where $ac$ represents $Z$ and the 5-vint is shaded). Using RC edges, the 5-vint can be extended, as in the preceding case, into three additional 6-vints, three 8-vints, and one 9-vint. For each of these vints, every triangulation of its hole must contain either $bc$ or $aq$, and $tr(bc)=2$. There are two more vints that extend the 5-vint and might have a positive charge (the two 6-vints that extend the 5-vint with a level-3 edge), but they were already considered above. The 6-vint that extends the 5-vint with a rigid level-2 edge must have a support of 3, and we ignore it for now. Let $i$ be a binary variable, which is 1 if and only if $q$ can see $d$ (in the notation of the figure); notice that when $i=0$, $q$ cannot see $f$ either. Let $j$ be a binary variable, which is 1 if and only if $q$ can see $e$. The 6-vint using $ab$ has a support of $tr(bc)+tr(aq)=2+(1+i)=3+i$, and similarly, the 6-vint using $ac$ has a support of $3+j$. As in RC 3b, we use the supports of the 6-vints to bound the supports of the larger vints. The support of the 8-vint not using $ad$ is at most $tr(bc)+tr(aq)=2+(1+j)\cdot (1+i)$ (since $tr(aq)$ is the number of triangulation of the portion to the left of $aq$ times the number of triangulations of the portion to its right). Similarly, the support of the 8-vint not using $bp$ is $2+(1+j)\cdot (1+2i)$, the support of the 8-vint not using $ac$ is at most $2+1\cdot(1+2i)$, and the support of the 9-vint is at most $2+(1+j)\cdot (1+2i)$. Table \ref{ta:RC3d1} lists the possible cases of the overall charge of these six extension vints. Using the table, we notice that this charge cannot exceed $-\frac{29}{120}$, and thus, the overall charge from such a 5-vint and its seven extending vints (including the previously ignored 6-vint, with a support of 3) is at most $\frac{2}{3}+\frac{1}{3}-\frac{29}{120}=\frac{91}{120}$.

\begin{table}[ht]
\caption{The overall charge of six extension vints of a 5-vint using either $X$ or $Y$ and a non-rigid level-2 edge.} \label{ta:RC3d1}
\centering
\begin{tabular}{c | c c c c c c c}
\hline\hline
$i$/$j$ & 6-vint & 6-vint & first  & second & third  &  9-vint & Charge \\
        & using  & using  & 8-vint & 8-vint & 8-vint &         &        \\
        & $ab$   & $ac$   &        &        &        &         &        \\[0.5ex]
\hline
0/0 & 3 & 3 & 3 & 3 & 3 & 3 & $\frac{1}{3}+\frac{1}{3}-\frac{1}{3}-\frac{1}{3}-\frac{1}{3}-\frac{2}{3}=-1$\\
0/1 & 3 & 4 & 4 & 4 & 3 & 4 & $\frac{1}{3}+\frac{1}{4}-\frac{1}{4}-\frac{1}{4}-\frac{1}{3}-\frac{2}{4}=-\frac{3}{4}$\\
1/0 & 4 & 3 & 4 & 5 & 5 & 5 & $\frac{1}{4}+\frac{1}{3}-\frac{1}{4}-\frac{1}{5}-\frac{1}{5}-\frac{2}{5}=-\frac{7}{15}$\\
1/1 & 4 & 4 & 6 & 8 & 5 & 8 & $\frac{1}{4}+\frac{1}{4}-\frac{1}{6}-\frac{1}{8}-\frac{1}{5}-\frac{2}{8}=-\frac{29}{120}$\\

\hline
\end{tabular}
\end{table}
\end{list}

\noindent Hence, the total charge cannot exceed (the second term represents the 5-vint using $Z$, the third term represents the two other 5-vints that have a support of 3, and the last term represents the twelve 6-vints that use a level-3 edge) \[27+1\frac{1}{3}+\frac{91}{120}\cdot2+\frac{1}{100}=29\frac{43}{50}.\]

\paragraph{Extensions of the previous cases.} We start by considering additional level-3 RC edges. We cannot add any level-3 RC edges to RC 3a, since it does not have a level-2 RC edge. Each other RC with $\lambda_1=3$ has at least four RC edges, and thus, by Rule \ref{rule:LargestRC}, adding a level-3 RC edge cannot increase the bound on its charge. Therefore, we only need to consider additional level-2 RC edges. Adding a level-2 edge to RC3a results in RC 3b, and adding a level-2 edge to RC 3b results in RC 3c or RC 3d. Hence, it suffices to consider such extensions only of these two latter RCs.\marfigsc{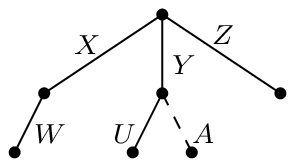}{1.3}

Consider an RC with a size of at least 5 (such as RC 3c, RC 3d, or an extension thereof), and assume that its total charge is bounded by $m$. We now show that adding a level-2 edge to the RC cannot increase the bound beyond $m$, if the sibling of the added edge is an RC edge. An example of such a case is depicted in Figure \ref{fi:RC3ext1}, where the changed edge is $A$. The change can increase the bound on the total charge by reducing the support of one 5-vint (using $Y$ and $A$), two 6-vints that extend the 5-vint with a level-3 edge, two 6-vints that extend the 5-vint with an additional level-1 edge (either $X$ or $Z$), and one 6-vint that extends the 5-vint with an additional level-2 edge ($U$). The charge from the two 6-vints that use a level-3 edge remains bounded by Rule \ref{rule:6-vint2}. (Notice that we have indeed used Rule \ref{rule:6-vint2} to bound the charge of these vints while analyzing the basic RCs, so the charge does not change.) Let $s$ be the support of the 5-vint before the change. Since after the change the 5-vint has a support of 1, the positive charge gained from the 5-vint by the change is $2\left(1-\frac{1}{s}\right)$. There is at least one 9-vint that consists of the edges of the 5-vint and additional RC edges. Before the change, this 9-vint had a support of at least $s$, and afterwards, it has a support of 1. The change in the charge of the 9-vint is at least $2\left(1-\frac{1}{s}\right)$, neutralizing the change in the charge of the 5-vint. (A similar case is described in Rule \ref{rule:LargestRC}.)\marfigsc{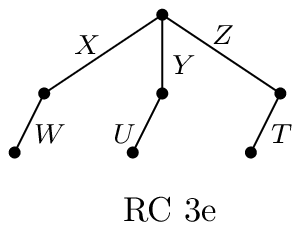}{1.3} Similarly, there are at least three 8-vints which consist of the edges of the 5-vint and additional RC edges, and the change in their charges neutralizes the change in the charges of the other three 6-vints. We conclude that the change did not increase the bound on the total charge.

Let RC 3e be the RC created by taking RC 3d and adding a child of $Z$ to the RC (as depicted in Figure \ref{fi:RC3ext2}, where the changed edge is $T$). Similarly to the previous paragraph, we show that the bound on the total charge of RC 3d applies also to RC 3e. Once again, we need to neutralize the positive charge generated by the reduction in the supports of one 5-vint (using $Z$ and $T$) and the five 6-vints that extend it. The charge of the two 6-vints using a level-3 edge is still bounded by Rule \ref{rule:6-vint2}. The charge of the 6-vint using $Z$ and two level-2 edges can still be ignored by Rule \ref{rule:6-vint1}. Finally, as in the previous paragraph, the change in the charge of the 5-vint and of the two 6-vints that extend it with an additional level-1 edge (either $X$ or $Y$)\marfigsc{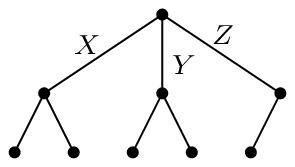}{1.3}is neutralized by the change in the charge of the 9-vint (using all the edges in Figure \ref{fi:RC3ext2}) and of the two 8-vints (removing either $U$ or $W$ from the 9-vint).

With all these observations, we can prove a bound of at most $29\frac{43}{50}$ on any extension RC with $\lambda_1=3$ and with at least two level-2 edges, by taking either RC 3d or RC 3e and using the above claim (that the addition of a level-2 RC edge with an already rigid sibling cannot increase the bound). For example, to show this bound on the RC in Figure \ref{fi:RC3ext3}, we start from RC 3e and add two level-2 edges to the RC, each with an already rigid sibling. We have already argued that the bound $29\frac{43}{50}$ holds for RC 3e, and the above claim implies that adding the two level-2 edges cannot increase the bound.

\subsection{Analysis of $\lambda_1=2$} This subsection analyzes the rigid cores with $\lambda_1=2$.\marfigsc{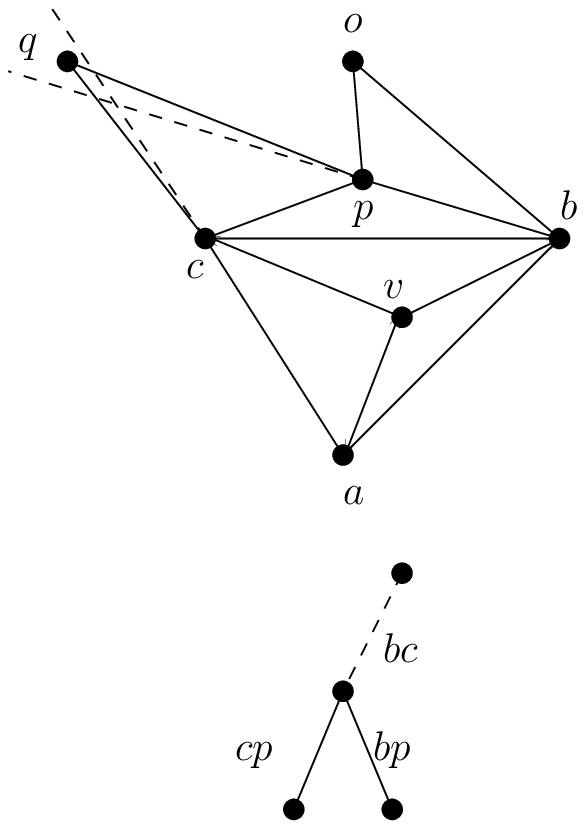}{1.4}We first analyze the {\em basic} RCs depicted in Figures \ref{fi:RC2a}, \ref{fi:RC2b}, \ref{fi:RC2c}, \ref{fi:RC2d}, and \ref{fi:RC2e}, and then deal with any other RC with $\lambda_1=2$, obtained by adding RC edges to one of the basic RCs. Any of these {\em extension} RCs is analyzed using a bound proved for the corresponding basic RC, and considering the changes in that bound caused by the rigidity of the new edges. In these flip-tree figures, the solid lines represent RC edges, and the dashed lines represent non-RC edges, which might, or might not be present in the flip-tree.

Before analyzing the basic RCs, we present two methods for analyzing RCs with $\lambda_1=2$. In these methods, we assume that there are three level-1 edges in the flip-tree, so there is exactly one non-rigid subtree (as defined in Rule \ref{rule:subtree}). In order to refer to specific subcases of these methods them while analyzing RCs, some of the following paragraphs are labeled with letters.

\paragraph{Method 1.} Assume that there are two 5-vints in the non-rigid subtree. In order for any of these 5-vints to have a support of 2, its level-2 edge must be rigid, so the vertex of the level-2 edge can see only its direct neighbors (along the boundry of the hole). In Figure \ref{fi:Meth1-1}, $o$ and $q$ are such vertices; in what follows, we focus on $o$, the vertex of $bp$.\marfigsc{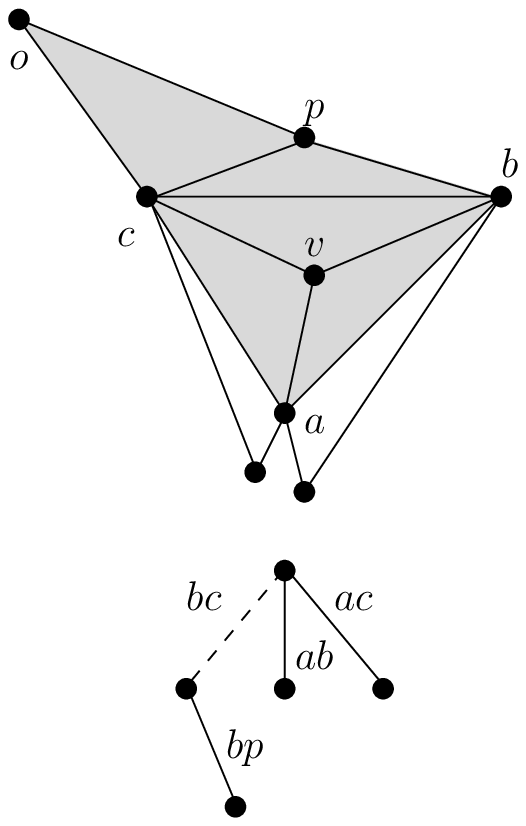}{1.4}In order not to see $a$, $o$ must ``hide" either to the right of the line supporting $\overrightarrow{ab}$, or to the left of the line supporting $\overrightarrow{ap}$ (the figure depicts the second situation). Similarly, in order not to see $c$, $o$ must hide either to the left of the line supporting $\overrightarrow{bc}$, or to the left of the line supporting $\overrightarrow{cp}$. However, the former of the last two cases is impossible, since in this case $o$ would not see $v$. This leaves two cases, which we refer to as a {\em type I 5-vint} and a {\em type II 5-vint}, respectively. In a type I vint, $o$ is to the right of the line supporting $\overrightarrow{ab}$ and to the left of the line supporting $\overrightarrow{cp}$. In a type II vint, $o$ is to the left of the line supporting $\overrightarrow{ap}$, in which case it is also to the left of the line supporting $\overrightarrow{cp}$. In Figure \ref{fi:Meth1-1}, the 5-vint which contains $o$ is a type II vint, and the 5-vint which contains $q$ is a type I vint (with the roles of $c$ and $b$ flipped in the definition). It cannot be the case that both 5-vints are of type II, since this would cause their level-2 triangles to overlap near $p$. We emphasize that these definitions only pertain to 5-vints with a support of 2 in the non-rigid subtree.

\mymark{A}In order to analyze a type I 5-vint (such as the shaded 5-vint in Figure \ref{fi:Meth1-2}), we use a similar argument to the one in Rule \ref{rule:6-vint1}. Consider a vint, $s$, which is created by adding RC edges to the 5-vint (such as the ones in the figure). As in Rule \ref{rule:6-vint1}, a triangulation of the hole of $s$ must contain exactly one of the edges $bc$, $ap$. It is easily checked that the hole of $s$ has a single triangulation that contains $bc$. Since $o$ is in its non-visible terrain, when $ap$ is present, only $p$ might be able to see vertices of added edges (along chords of the corresponding hole).\marfigsc{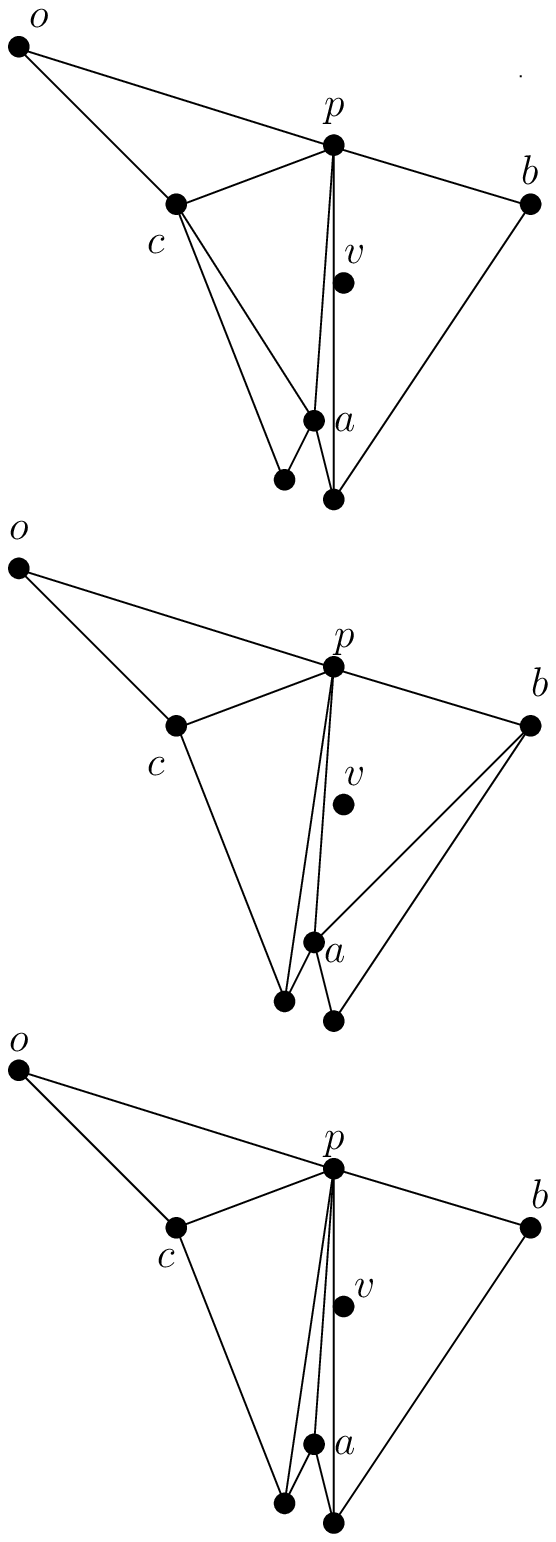}{1.25}This implies that every triangulation of the hole which contains $ap$ is uniquely determined by the subset of vertices of the added edges that are connected to $p$. The support of $s$ is two plus the number of these non-empty subsets (one additional triangulation that contains $bc$ and another that corresponds to the empty subset). For example, if there are two added edges, as depicted in Figure \ref{fi:Meth1-2}, there are two subsets with one vertex, and one subset with two vertices; the corresponding triangulations are depicted in Figure \ref{fi:Meth1-3}. This implies that the support of such a 7-vint is at most $2+3=5$. Although we have no use for a 7-vint, the support of larger vints can be analyzed by using this method.

\mymark{B}For each type I 5-vint, there might be two 6-vints which extend it with a level-3 edge. In one of these 6-vints, the vertex of the level-3 edge must be in its non-visible terrain; this is similar to the case of a handicapped 5-vint, which is why we refer to it as a {\em handicapped 6-vint}. We refer to the second 6-vint which extends a type I 5-vint with a level-3 edge as the {\em sibling} of the handicapped 6-vint. In Figure \ref{fi:Meth1-4}, the 6-vint using the edge $cq$ is a handicapped 6-vint, and the 6-vint using $pq$ is its sibling (the vertex of the level-3 edge of the sibling may or may not see $a$). Consider the vints that can be created by the addition of RC edges to a handicapped 6-vint. Since $p$ remains the only vertex in its visible terrain, we can analyze the supports of such vints in a manner similar to the analysis in the previous paragraph.

\mymark{C}If $r$, the level-3 vertex of the handicapped 6-vint, cannot see $p$, the handicapped 6-vint can be analyzed in the same way as the 5-vint. For example, consider the addition of two level-1 RC edges, as in the preceding example involving the 5-vint. In this case, there are still three non-empty subsets of vertices connected to $q$, which implies that the resulting 8-vint has a support of at most 3 plus the support of the 6-vint. If $r$ can see $p$, the quadrilateral $pqrc$ must be convex and have two triangulations; thus, each subset of vertices connected to $p$ in a triangulation corresponds to two additional triangulations. Notice that in this case the 6-vint has a support of at least 3.

\marfigsc{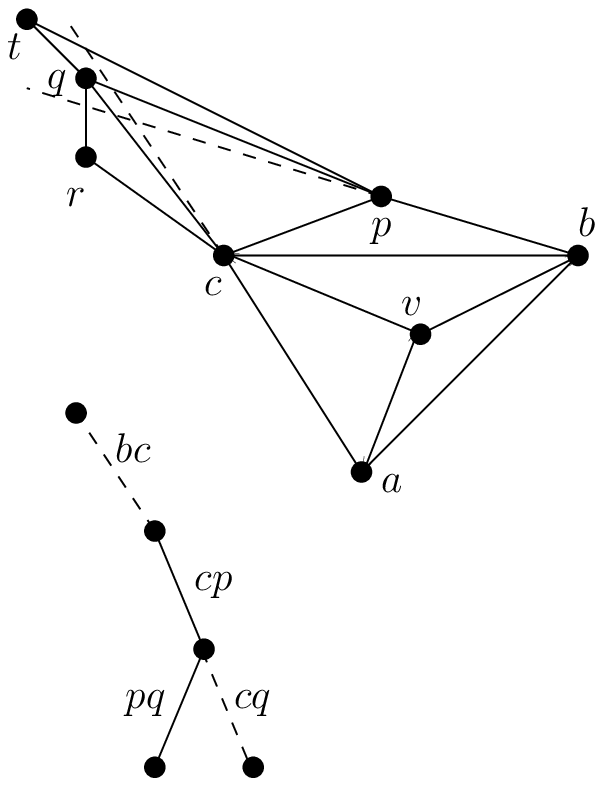}{1.4} \mymark{D}When there are three added edges (none of them a level-3 edge), the handicapped 6-vint can be extended into two 8-vints and one 9-vint. By examining the possible cases ($r$ can see $p$ and $b$, $r$ can see only $p$, and $r$ cannot see both $p$ and $b$; at most three non-empty subsets for an 8-vint, and at most four for a 9-vint), it is easily seen that these vints always neutralize the charge of the 6-vint.

\mymark{E}Consider a sibling of a handicapped 6-vint, such as the one using $t$ in Figure \ref{fi:Meth1-4}. If $t$ is in its visible terrain, the 6-vint has a support of $tr(cp)+tr(ta)=C_2C_2+C'_2C'_2=5$. This implies that if the 6-vint has a support of at most 4, the vertex of its level-3 edge must be in its non-visible terrain, and thus, it can be analyzed in the same way as the handicapped 6-vint. As in the previous case, if the sibling 6-vint has a support of 2 (such as the one depicted in Figure \ref{fi:Meth1-4}), each subset of added vertices connected to $p$ corresponds to one triangulation. If the 6-vint has a higher support (albeit still smaller than 5), the quadrilateral $ptqc$ must be convex, and thus, each subset corresponds to at most two triangulations.

\paragraph{Method 2.} This method applies to 5-vints with a rigid level-1 edge, a non-rigid level-1 edge, and a support of 2, so that the non-rigid edge is not in the non-visible subtree of the rigid edge.\marfigsc{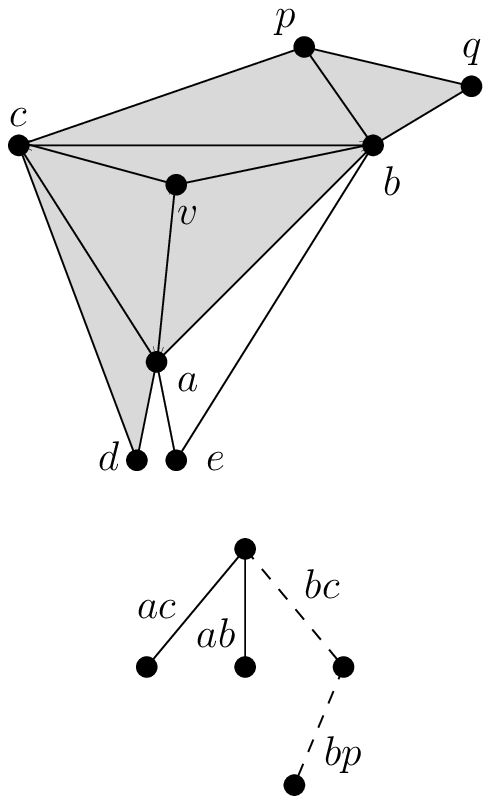}{1.4}Such a case is depicted in Figure \ref{fi:Meth2-1}, where the non-rigid edge is $bc$, the rigid edge is $ac$, and the 5-vint is shaded. Note that in this case the non-visible subtree of $ac$ is rooted at $ab$, and vice versa.

\mymark{A}Using the notations in the figure, we observe that for the 5-vint to have a support of 2, $p$ has to be either to the left of the line supporting $\overrightarrow{dc}$, or to the right of the line supporting $\overrightarrow{da}$. Assume, without loss of generality, that the latter occurs (as will follow, it will not matter that the two cases are not symmetric, since we will only use the fact that $p$ cannot see $d$). By construction, $p$ cannot see $ad$ (or $cd$ in the other situation) or any of its descendants. If $cd$ is rigid, then its vertex is to the right of the line supporting $\overrightarrow{da}$, and $p$ cannot see that vertex either. This implies that $p$ cannot see the vertices of the RC edges in the subtree of $ac$ (for our purposes, we do not need to consider level-3 RC edges, although this is also the case for them).

\mymark{B}After establishing which vertices of RC edges $p$ cannot see, we now consider the level-2 vertices of handicapped 5-vints (which might be present both in the subtree of $ab$ and in the subtree of $ac$). If $p$ can see the vertex of exactly one rigid level-1 edge (such as in Figure \ref{fi:Meth2-1}, where it can only see $e$),\marfigsc{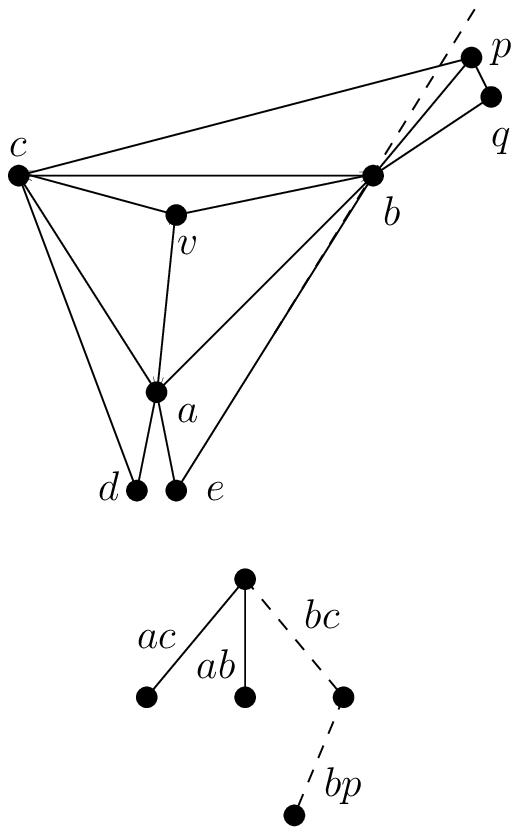}{1.4} it cannot see the level-2 vertex of the handicapped 5-vint using the other rigid level-1 edge (the vertex of $ad$ in the figure). If $p$ cannot see both vertices of the rigid level-1 edges (such as in Figure \ref{fi:Meth2-2}), it still cannot see the level-2 vertex of at least one handicapped 5-vint (again, the vertex of $ad$ in the figure). This can be easily proved using an analysis similar to the one in the previous paragraph.

\mymark{C}The edge $bc$ can have two child-edges; the vertex of one of those must be to the right of the line supporting $\overrightarrow{bp}$ ($q$ in Figure \ref{fi:Meth2-1}). This vertex cannot see $d$, which implies that the above analysis for $p$ also applies to it. Moreover, it also applies to the vertices of the child-edges of this level-2 edge, if they exist. Using the edges of these vertices, it is possible to form a 5-vint ($acpqb$ in Figure \ref{fi:Meth2-1}) and up to two 6-vints. Hence, extending these vints with RC edges from the subtree of $ac$ cannot increase their support.

\paragraph{A detailed example.} Since the use of the two methods is not trivial, we present a detailed analysis that uses both of them. Consider the flip-tree depicted in Figure \ref{fi:MethComb1}, where the non-rigid subtree is the subtree of $A$ and there are three RC edges --- $X$, $Y$, and $Z$ \marfigsc{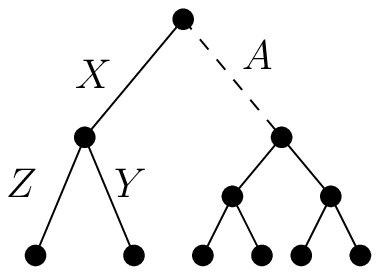}{1.1} (in order to keep this example simple, we ignore the third level-1 edge). We start with the naive approach of giving every (4-, 5-, or 6-) vint not entirely in the RC a support of 2. This implies that the total charge of the depicted flip-tree cannot exceed $4\cdot1+3\left(1\cdot1+\frac{1}{2}\cdot1\right)+2\left(1\cdot2+\frac{1}{2}\cdot3\right)+1\left(1\cdot1+\frac{1}{2}\cdot9\right)=21$.

We continue the analysis depending on whether the vertex of $X$ does or does not see the vertex of $A$. In the former case, after flipping $A$, $X$ becomes flippable (such a case is depicted in Figure \ref{fi:Meth2-1}, where $A$ and $X$ are $bc$ and $ab$, respectively), which implies that the 5-vint and four 6-vints which contain these two edges have a support of at least 3 (instead of a support of at least 2). This lowers the bound on the charge by at least $\left(\frac{1}{2}-\frac{1}{3}\right)\left(2\cdot1+1\cdot4\right)=1$. In the latter case (see Figure \ref{fi:Meth2-1}, with $A=bc$ and $X=ac$), we can use Method 2$(A,C)$) to extend a 5-vint and two 6-vints from the non-rigid subtree. The method states that extending these vints with RC edges cannot increase their support, and thus, each 6-vint can be extended into an 8-vint which neutralizes its charge\marfig{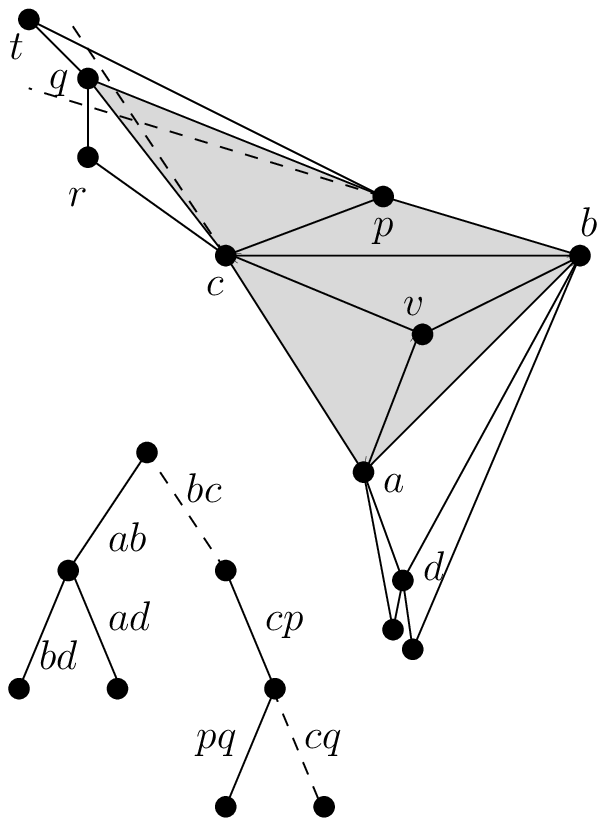}. (More precisely, a 6-vint with a support of $s$ can be extended into two 8-vints and a 9-vint, which have an overall charge of $\frac{1}{s}(1\cdot1-1\cdot2-2\cdot1)=-\frac{3}{s}$; however, the overall charge could be only 0 if we assume that the level-3 edge of the 6-vint is not present in the flip-tree. Thus, the overall charge from the 6-vint and its extensions is at most 0.) The 5-vint can be extended into an 8-vint which halves its charge. In this case, we lose a charge of at least $2\cdot\frac{1}{2}\cdot\frac{1}{2}+1\cdot\frac{1}{2}\cdot2=1\frac{1}{2}$. We conclude that the charge is decreased by at least $1$ in either case.

Our next observation is that either at least one of the two 5-vints in the non-rigid subtree has a support of at least 3, or there are two 5-vints with a support of 2, and then a Type I 5-vint must exist. In the former case, a 5-vint and the four 6-vints which extend it have a support of at least 3 (instead of a support at least 2), which lowers the charge by at least $\left(\frac{1}{2}-\frac{1}{3}\right)(2\cdot1+1\cdot4)=1$. In the latter case, we can use Method 1 to analyze the extensions of the Type I 5-vint, its handicapped 6-vint, and its sibling (such a case is depicted in Figure \ref{fi:MethComb2}, where the 5-vint is shaded and the subtree of $ab$ has three RC edges). According to Method 1$(B,C,D)$, three RC edges suffice to neutralize the charge of the handicapped 6-vint; this also applies to the sibling of this 6-vint, unless it has a support of 5 (Method 1$(E)$). The Type I 5-vint can be extended into an 8-vint with a support of at most 6 (there are four non-empty subsets of vertices of added edges connected to $p$).\marfigsc{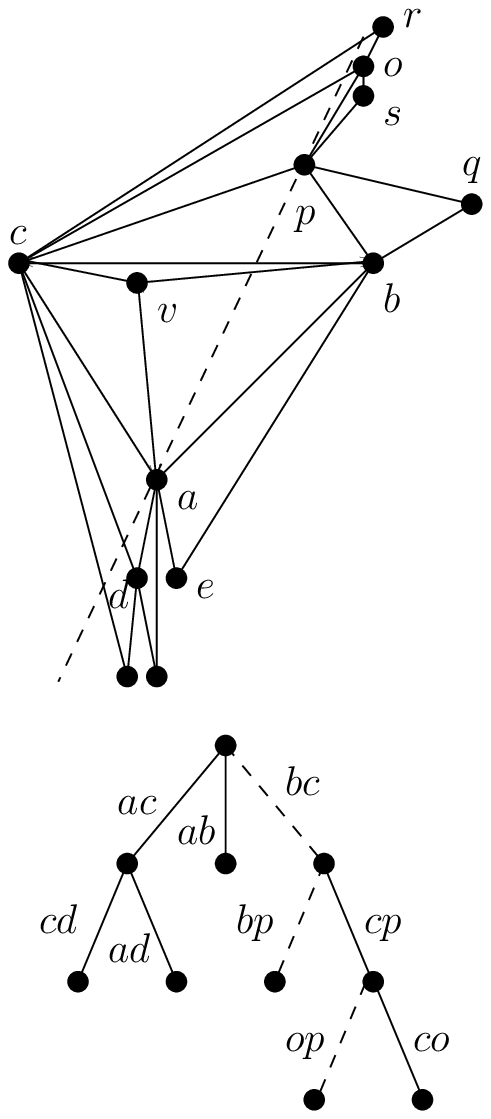}{1.25} These extensions imply that the charge is lowered by at least $\frac{1}{6}+\left(\frac{1}{2}-\frac{1}{5}\right)+\frac{1}{2}=\frac{29}{30}$. We conclude that the charge is decreased by at least $\frac{29}{30}$ in either case.

Variants of this example will show up while analyzing RCs with $\lambda_1=2$.

\paragraph{Combining the two methods.} Intuitively, we would like to continue the above example with the conclusion that the total charge cannot exceed $21-1-\frac{29}{30}=19\frac{1}{30}$. However, for this statement to be valid, we need to verify that the two methods can be combined without clashing with each other (i.e., multiply counting the same reduction in charge).

The first problem with combining the methods is that they both attempt to increase the bound on the support of the same 6-vint (from a bound of at least 2 to a bound of at least 3; in Figure \ref{fi:MethComb1}, a 6-vint of this kind is the one using $X$, $A$, and the level-2 edge of a Type I 5-vint). To prevent this problem, we ignore this 6-vint in the application of one of the methods. In the above example, it is better to ignore this 6-vint in the paragraph using Method 1, which lowers the decrease in the charge in its first case from 1 down to $\frac{5}{6}$.

The second, more complex problem, is that both methods might refer to the same 5-vint in the non-rigid subtree; Method 1 might extend it if it is a Type I 5-vint, while Method 2 might extend it for different reasons. Fortunately, this can be avoided, and to explain this we refer to Figures \ref{fi:MethComb3} and \ref{fi:MethComb4}. Figure \ref{fi:MethComb3} illustrates the case where a single 5-vint can be analyzed by Method 2 (the one using $bc$ and $ac$), and Figure \ref{fi:MethComb4} illustrates the case where there are two such 5-vints (the additional 5-vint is the one using $bc$ and $ab$).\marfig{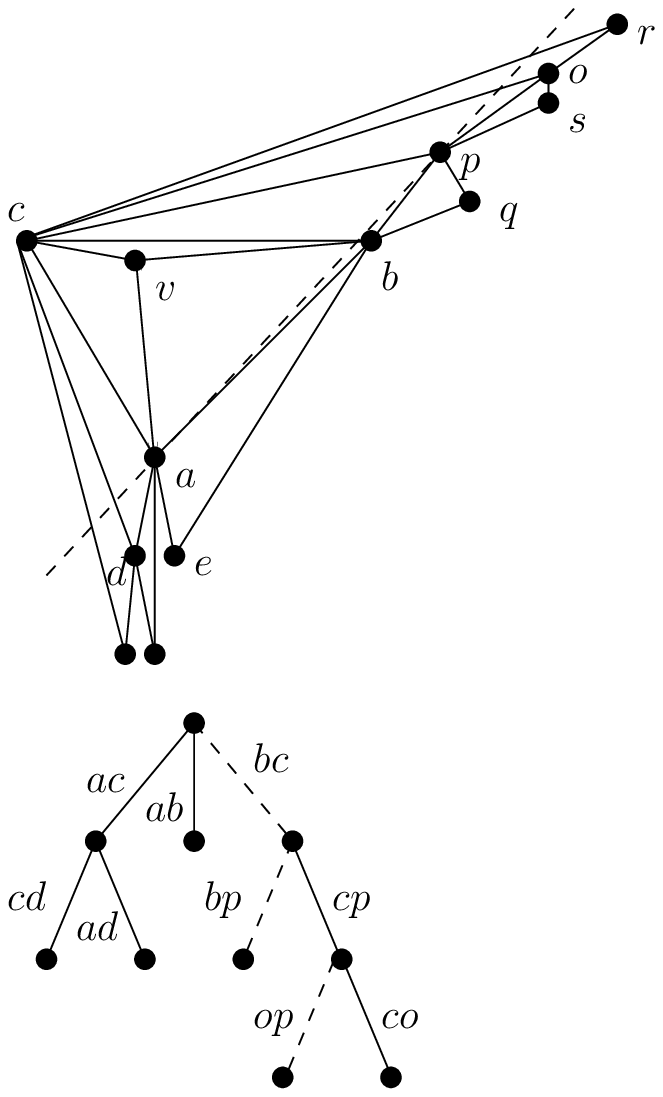} The following analysis is identical for both cases, but it may be easier to follow it when we keep both of them in mind. As before, we assume, as depicted in the figures, that the 5-vint in the non-rigid subtree which might be analyzed by Method 2 is the one using the level-2 edge $bp$. In the following paragraph, we prove that the 5-vint using the level-2 edge $cp$ can always be analyzed according to Method 1 (when it has a support of 2, even if it is a Type II 5-vint). This solves the second conflict between the methods, since it implies that we can apply each method to a different 5-vint of the non-rigid subtree.

If the 5-vint using $cp$ is a Type I vint, then, by definition, it can be analyzed by Method 1. We are left with the case of a Type II vint, as depicted in Figures \ref{fi:MethComb3} and \ref{fi:MethComb4}. In order not to see $a$, the vertex $o$ of $cp$ has to be to the right of the line supporting $\overrightarrow{ap}$ (the dashed line in the figures). This implies that $o$ cannot see any vertices of the subtree of $ab$. Moreover, since the 5-vint using $bc$ and $ac$ can be analyzed using Method 2, $d$ must be to the right of the line supporting $\overrightarrow{ap}$. This implies that $o$ cannot see $d$; the same applies to the vertices of the child edges of $ac$, if they are rigid. We conclude that $o$ cannot see the vertex of any RC edge, which is exactly the requirement for using the analysis of Method 1$(A)$. The same analysis applies to one of the child vertices of $o$ ($s$ in the illustration), and also for the other, if the 6-vint which contains it has a support of at most 4 ($r$ in the illustration; this follows from the same reasoning as given in Method 1$(B,C,E)$. This makes the two 6-vints, which extend the 5-vint with a level-3 edge, behave just as the handicapped 6-vint and its sibling.

We conclude that in order to combine the two methods without clashing, we only need to ignore a specific 6-vint in the method of our choise. When two 5-vints are analyzed according to method 2, we need to ignore two such 6-vints.

\vspace{5 mm} \marfigsc{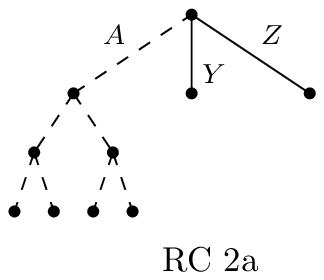}{1.2}\noindent {\bf RC 2a,} as depicted in Figure \ref{fi:RC2a} (the RC only consists of two level-1 edges).

We first ignore edges that are not in the RC and not in the non-rigid subtree (as defined in Rule \ref{rule:subtree}). The remaining edges are those depicted in Figure \ref{fi:RC2a} (although some of the non-RC edges might not be present). By applying a naive analysis, which gives a support of 2 to each vint not entirely in the RC, we get a charge of at most \[4\cdot1+3\left(2+\frac{1}{2}\right)+2\left(1+\frac{1}{2}\cdot4\right)+1\cdot\frac{1}{2}\cdot10=22\frac{1}{2}.\]

\noindent We observe that either at least one of the 5-vints of the non-rigid subtree has a support of 3, or a Type I 5-vint must exist (or one of the 5-vints does not exist, see below). In the former case, a 5-vint and three of the 6-vints which extend it have a support of at least 3 (we do not count the two additional 6-vints, using either $Y$ or $Z$, in order not to clash with Method 2); the charge is therefore reduced by at least $\left(\frac{1}{2}-\frac{1}{3}\right)(2\cdot1+1\cdot3)=\frac{5}{6}$. In the latter case, the handicapped 6-vint can be extended into an 8-vint with a support that can be bounded by Method 1$(B,C)$. Figure \ref{fi:RC2a1} depicts the resulting 8-vint, where the hole of the 6-vint is shaded. There are three non-empty subsets of vertices connected to $p$ (as defined in Method 1$(A)$)\marfigsc{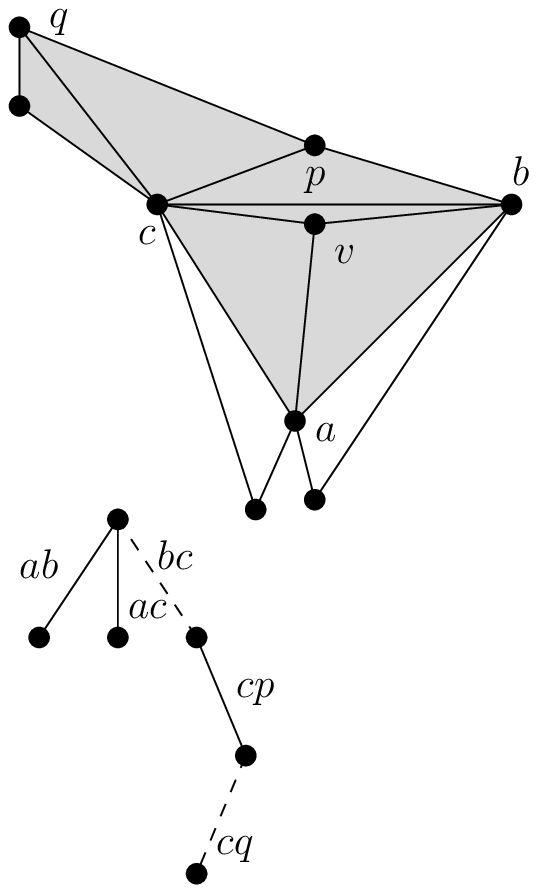}{1.4}. This implies that if the 6-vint has a support of 2, the 8-vint has a support of at most $2+3=5$, and if the 6-vint has a support of $m>2$, the 8-vint has a support of at most $m+6$ (we multiply the number of subsets of $p$-connected edges by 2). As is easily verified, the smallest decrease in the charge, i.e., $\frac{1}{5}$, occurs when the 6-vint has a support of 2. The same analysis remains valid for the sibling 6-vint, unless it has a support of 5 (see Method 1$(E)$; again, a sibling 6-vint with a support of 2 generates the smallest decrease in the charge). The reduction in the charge in this case is therefore at least $\frac{1}{5}+\frac{1}{5}=\frac{2}{5}$ (obtained from two 8-vints with a support of at most 5). In the third case, where a level-2 edge in the non-rigid subtree is missing, the 5-vint and its three extending 6-vints are missing, which obviously gives a lower charge than the case where each of them has a support of 3. We conclude that in either case the total charge goes down by at least $\frac{2}{5}$.

We next observe that either at least one 5-vint, with a rigid level-1 edge ($Y$ or $Z$) and a non-rigid level-1 edge ($A$), has a support of 3, or (recalling Figure \ref{fi:MethComb4}, which better captures this situation than Figure \ref{fi:Meth2-1}) we can use Method 2 to extend vints from the non-rigid subtree\marfigsc{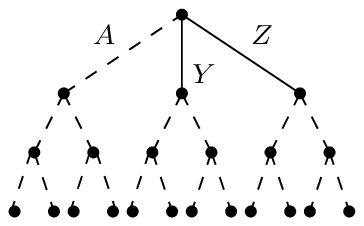}{1.2}(by using the RC edges). In the former case, the support of one 5-vint and the three 6-vints which extend it is at least 3, which lowers the bound by at least $\left(\frac{1}{2}-\frac{1}{3}\right)(2\cdot1+1\cdot3)=\frac{5}{6}$. In the latter case, we can extend two 6-vints from the non-rigid subtree (both obtained from the same 5-vint by adding a level-3 edge) into 8-vints with the same support by adding the two RC edges. This follows since the vertices of these 6-vints, as considered in Method 2$(C)$, cannot see vertices of RC edges (again, refer to Figure \ref{fi:MethComb4}); the charge is therefore reduced by at least $2\cdot\frac{1}{2}=1$. We conclude that in either case the bound on the total charge goes further down by at least $\frac{5}{6}$.

Hence, so far the total charge is at most \[22\frac{1}{2}-\frac{2}{5}-\frac{5}{6}=21\frac{4}{15}.\] \marfigsc{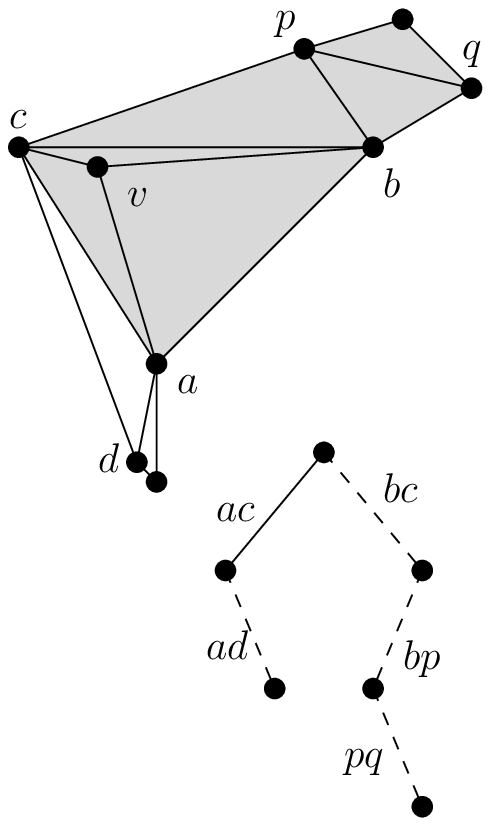}{1.3}

\noindent So far, we have ignored the portion of the flip-tree below the RC edges $Y$ and $Z$, but now we bring it into play (as depicted in Figure \ref{fi:RC2a2}). There might be additional charges from the four 5-vints with a rigid level-1 edge and a level-2 edge, and from their extensions into eighteen 6-vints (in all possible ways). Recall that, in this case, the four level-2 edges are assumed not to be rigid (see Figure \ref{fi:RC2a}). The two 6-vints which use two level-2 edges must therefore have a support of at least 3 (since both of the level-2 edges are flippable); the support of each of the other sixteen 6-vints depends on the support of its 5-vint: \begin{list}{\labelitemi}{\leftmargin=1em}
 \item A 5-vint with a support of 2 and its four other extensions into 6-vints generate a charge of at most $2\cdot\frac{1}{2}\cdot1+1\left(\frac{1}{2}\cdot3+\frac{1}{4}\cdot1\right)=2\frac{3}{4}$ (the 6-vint extending the 5-vint with $A$ has a support of at least 4, since it contains two non-adjacent flippable edges).
 \item A 5-vint with a support of 3 and its four other extensions into 6-vints generate a charge of at most $2\cdot\frac{1}{3}\cdot1+1\left(\frac{1}{3}\cdot3+\frac{1}{5}\cdot1\right)=1\frac{13}{15}$ (similarly to the previous case, the 6-vint extending the 5-vint with $A$ has a support of at least 5).
\end{list}

\noindent The above implies that when at least one of these four 5-vints is missing, the total charge cannot exceed $21\frac{4}{15}+3\cdot2\frac{3}{4}+\frac{1}{3}\cdot1=29\frac{17}{20}$ (the third term represents the single 6-vint which consists of a rigid level-1 edge and two level-2 edges; the second term represents the maximal charge from the other previously ignored vints that are present in this case). We may therefore assume that all four 5-vints are present, which implies that there are two handicapped 5-vints among them (see Section \ref{sec:foundations}). We can use these 5-vints to modify the above use of Method 2, and raise the negative charge attained by it from $\frac{5}{6}$ up to 1, as follows (see Method 2$(B)$).

Consider the two 5-vints, which use a rigid level-1 edge and a non-rigid level-1 edge. We already noticed that when both 5-vints have a support of 2, we gain a negative charge of at least 1 (when we used Method 2 earlier in the analysis of this RC).

If both 5-vints have a support of 3, there are five 6-vints which we previously considered to have a support of at least 2, and now have a support of at least 3\marfigsc{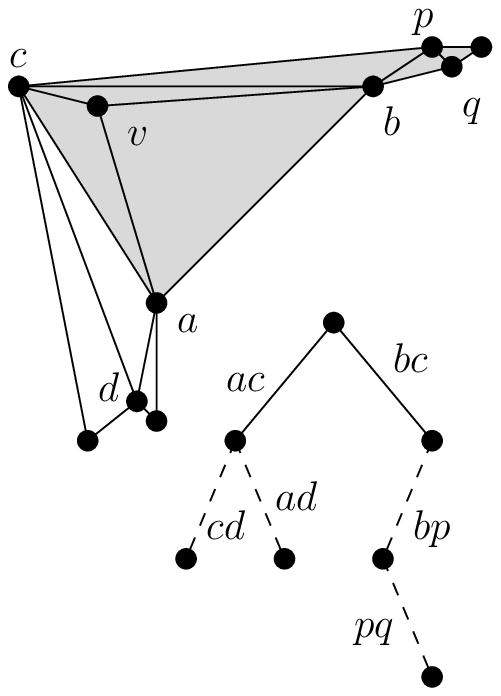}{1.4}(Namely, we have two 6-vints using $A$, $Y$, and a child of $A$, two using $A$, $Z$, and a child of $A$, and one using $A$, $Y$, and $Z$. However, we ignore one of the 6-vints using $Y$, $A$, and a child of $A$, in order not to clash with Method 1. Note that there are four additional 6-vints which extend these 5-vints, but they were already considered as having a higher support). Therefore, in this case, the charge goes down by at least $\left(\frac{1}{2}-\frac{1}{3}\right)(2\cdot2+1\cdot4)=1\frac{1}{3}$.

When exactly one 5-vint has a support of 2, we can use the edges of a handicapped 5-vint to extend two 6-vints from the non-rigid subtree (those considered in the first case, where the two 5-vints have a support of 2) into 8-vints with a double support (such an 8-vint is depicted in Figure \ref{fi:RC2a3}, where the 6-vint is shaded and the handicapped 5-vint is the one using $ac$ and $ad$); this raises the negative charge of this case up to $\frac{1}{4}\cdot2+\left(\frac{1}{2}-\frac{1}{3}\right)(2\cdot1+1\cdot2)=1\frac{1}{6}$ (the second term represents the change in the bound on the supports of the 5-vint and of two other 6-vints which extend it).

In either case, we lose an additional charge of at least 1, which implies that the total charge (not including the previously ignored vints) is at most \[22\frac{1}{2}-\frac{2}{5}-1=21\frac{1}{10}.\]

\noindent Moreover, a 6-vint which extends a handicapped 5-vint with a level-3 edge can be extended into an 8-vint with a double support, using the edges of the other handicapped 5-vint (since all of these edges are in their non-visible terrain). An example of such an 8-vint is depicted in Figure \ref{fi:RC2a4}, where the 6-vint is shaded and the other handicapped 5-vint is the one using $ac$ and $ad$. We will use this observation repeatedly in what follows.

The rest of the analysis is divided according to the supports of the two non-handicapped 5-vints which consist of a rigid level-1 edge and a level-2 edge: \begin{list}{\labelitemi}{\leftmargin=1em}
 \item Both 5-vints have support of 3. As explained above, each of the four 6-vints which extend a handicapped 5-vint with a level-3 edge, can be extended into an 8-vint with a double support.\marfigsc{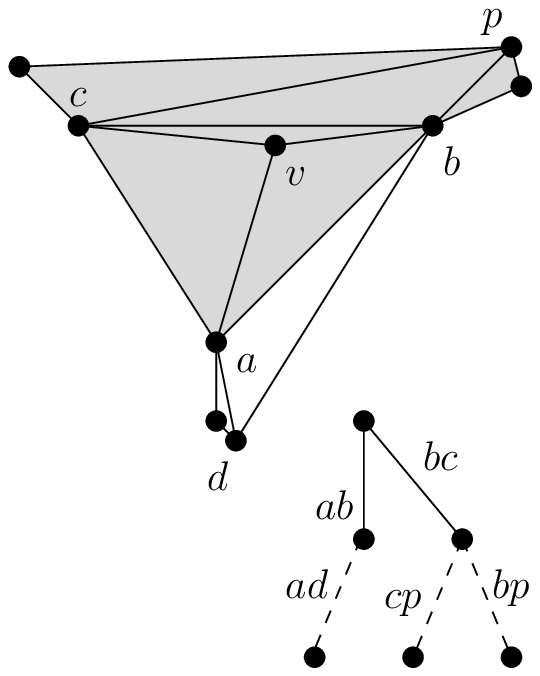}{1.4}There exists an additional 8-vint, consisting of the edges of the two 5-vints that have a support of 3, and of a level-2 edge of one of the handicapped 5-vints. This 8-vint has a support of at most $C_6=132$ (which is the maximal support for any 8-vint). Then the total charge cannot exceed \footnote{Without the additional 8-vint, the charge would be 30, and this is the only case in our analysis where the charge is not strictly smaller than 30. We use this vint to demonstrate that the maximum charge of a 3-vint is indeed smaller.} (the second and third terms account for the four addional 5-vints and for sixteen of their 6-vint extensions; the fourth term represents the remaining two 6-vints, which consist of a rigid level-1 edge and two level-2 edges; the fifth term represents four 8-vints, each with a support of 4, as just discussed; the last term represents an 8-vint with a support of at most 132) \[21\frac{1}{10}+2\cdot2\frac{3}{4}+2\cdot1\frac{13}{15}+\frac{1}{3}\cdot2-\frac{1}{4}\cdot4-\frac{1}{132}=29\frac{131}{132}.\]

\item Exactly one of the two 5-vints has a support of 3 (without loss of generality, we assume that it is in the subtree of $Z$). On top of the four 8-vints with a double support that were already mentioned, we show the existence of additional 8-vints, whose support is easily analyzed by noticing that all of their vertices are in their non-visible terrains.\marfigsc{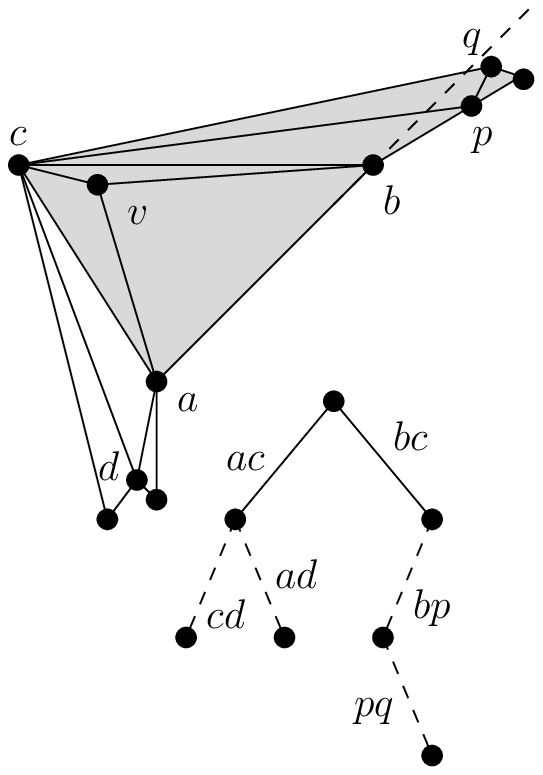}{1.4} First, each of the two 6-vints, which extend the handicapped 5-vint of $Z$ with a level-3 edge, can be extended into an 8-vint with a double support, using the edges of the non-handicapped 5-vint from the subtree of $Y$ (similarly and in addition to the reduction in charge yielded by the extension through the handicapped 5-vint; such an 8-vints is depicted in Figure \ref{fi:RC2a4}, where the 6-vint is shaded, and the additional 8-vint is using $ac$ and $cd$). Next, the 6-vint using $Y$ and two level-2 edges (which has a support of at least 3) can be extended into an 8-vint with a double support, using the edges of the handicapped 5-vint of $Z$ (as depicted in Figure \ref{fi:RC2a5}, where the 6-vint is shaded and the other handicapped 5-vint is using $ab$ and $ad$). Finally, One of the two 6-vints which extend the non-handicapped 5-vint of $Y$ with a level-3 edge, can be extended into an 8-vint with a double support, using the edges of the handicapped 5-vint of $Z$ (as depicted in Figure \ref{fi:RC2a6}, where the 6-vint is shaded and the extending edges are $ac$ and $ad$; the level-3 vertex of the 6-vint is in its non-visible terrain according to Rule \ref{rule:non-visib2}). The total charge cannot exceed (the first four terms of the sum have the same meaning as in the previous case; the last term accounts for eight 8-vints, four already encountered in the previous case, and the four new ones mentioned above; of those, one extends a 6-vint with a support of 3) \[21\frac{1}{10}+3\cdot2\frac{3}{4}+1\cdot1\frac{13}{15}+\frac{1}{3}\cdot2-\frac{1}{2}\left(\frac{1}{2}\cdot7+\frac{1}{3}\cdot1\right)=29\frac{29}{30}.\]

 \item Both 5-vints have a support of 2. As in the previous case, we show the existence of additional 8-vints, all of whose vertices are in their non-visible terrains. Each of the four 6-vints which extend a handicapped 5-vint with a level-3 edge can be extended into two 8-vints with a double support (instead of just one 8-vint), each using the edges of a different 5-vint from the other subtree (such as depicted in Figure \ref{fi:RC2a4}, where the 6-vint is shaded). This accounts for four additional 8-vints. Moreover, two of the four 6-vints, which extend with a level-3 edge a non-handicapped 5-vint with a rigid level-1 edge and a support of 2, can be extended into two 8-vints with a double support, each using the edges of a different 5-vint from the other subtree (as depicted in Figure \ref{fi:RC2a6}, where the 6-vint is shaded; the level-3 vertices of these 6-vints are in their non-visible terrains according to Rule \ref{rule:non-visib2}). There are four such 8-vints in this case, as opposed to only a single 8-vint in the previous one. The total charge cannot exceed (the first three terms of the sum have the same meaning as in the previous cases; the last term accounts for twelve 8-vints, as mentioned above)\marfigsc{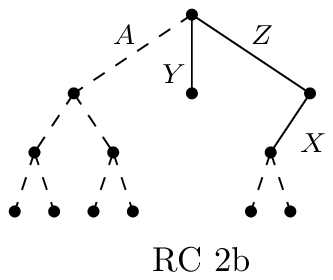}{1.2} \[21\frac{1}{10}+4\cdot2\frac{3}{4}+\frac{1}{3}\cdot2-\frac{1}{2}\cdot\frac{1}{2}\cdot12=29\frac{23}{30}.\]
\end{list}

\noindent {\bf RC 2b,} as depicted in Figure \ref{fi:RC2b}.

Similarly to the previous analysis, we start by considering only the edges from the RC, the edges from the non-rigid subtree, and the two child edges of $X$ (the edges depicted in Figure \ref{fi:RC2b}). By applying a naive analysis, which gives a support of 2 to each vint not entirely in the RC, we get a charge of at most \[4\cdot1+3\left(2+\frac{1}{2}\right)+2\left(2+\frac{1}{2}\cdot4\right)+1\left(1+\frac{1}{2}\cdot13\right)=27.\]

\noindent We observe that either at least one of the 5-vints of the non-rigid subtree has a support of 3, or a Type I 5-vint must exist.\marfigsc{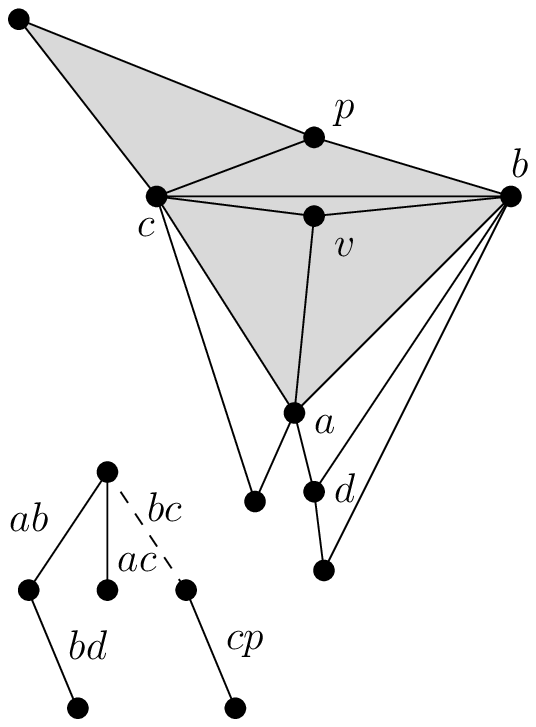}{1.2}In the former case, a 5-vint and four of the 6-vints which extend it have a support of at least 3 (we do not count the fifth 6-vint, in order not to clash with Method 2); the charge is therefore reduced by at least $\left(\frac{1}{2}-\frac{1}{3}\right)(2\cdot1+1\cdot4)=1$. In the latter case, the handicapped 6-vint can be extended into an 8-vint with a support that can be bounded by Method 1 --- as explained in Method 1$(D)$, we can ignore this 6-vint when there are at least three RC edges. This is also valid for the sibling of the handicapped 6-vint, unless it has a support of 5 (see Method 1$(E)$). Finally, according to Method 1$(A)$, we can extend the Type I 5-vint into an 8-vint, using the three RC edges (as depicted in Figure \ref{fi:RC2b1}, where the hole of the 5-vint is shaded). There are at most five non-empty sets of vertices connected to $p$, which implies that the support of this 8-vint is at most $2+5=7$. The reduction in the charge in this case is therefore at least $\frac{1}{2}+(\frac{1}{2}-\frac{1}{5})+\frac{1}{7}=\frac{33}{35}$. We conclude that in either case the total charge goes down by at least $\frac{33}{35}$.

We next observe that either the 5-vint using $A$ and $Z$ has a support of 3, or we can use Method 2 to extend vints from the non-rigid subtree (using $Z$ and $X$). In the former case, the support of one 5-vint and the four 6-vints which extend it is at least 3, which lowers the bound by at least $(\frac{1}{2}-\frac{1}{3})(2\cdot1+1\cdot4)=1$. In the latter case, there exist two 6-vints from the non-rigid subtree which can be extended into 8-vints with the same support, by appending $Z$ and $X$. This follows since the vertices of these 6-vints, as considered in Method 2$(C)$, cannot see vertices of RC edges (refer to Figures \ref{fi:MethComb3} and \ref{fi:MethComb4}); the charge is therefore reduced by at least $2\cdot\frac{1}{2}=1$. We conclude that in either case the bound on the total charge goes further down by at least $1$.

Hence, so far the total charge is at most \[27-\frac{33}{35}-1=25\frac{2}{35}.\] \marfigsc{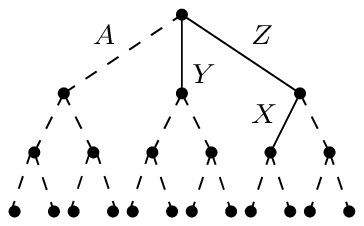}{1.4}

\noindent Similarly to the previous analysis, we now bring back the edges that were ignored up to now (resulting in the complete flip-tree depicted in Figure \ref{fi:RC2b2}; as before, the dashed edges are optional). There might be additional charges from three 5-vints with a rigid level-1 edge and a non-rigid level-2 edge, and from their extensions into fourteen 6-vints (in all possible ways). The 6-vint which contains $Y$ and two level-2 edges has a support of at least 3 (since both of the level-2 edges are flippable). The 6-vint which contains $Z$ and two level-2 edges has a support of at least 2. The support of each of the other twelve 6-vints depends on the support of its 5-vint: \begin{list}{\labelitemi}{\leftmargin=1em}
 \item A handicapped 5-vint. The two 6-vints which extend the 5-vint with a level-3 edge, use only vertices which are in their non-visible terrains (see Rule \ref{rule:non-visib2});\marfig{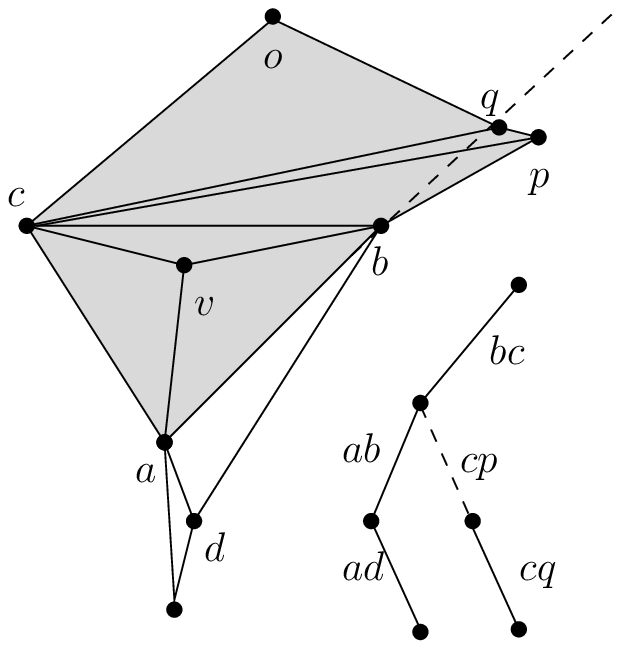} this implies that each of them can be extended into an 8-vint with the same support, using the additional RC edges. The 6-vint extending the 5-vint with $A$ has a support of at least 4, since it holds two non-adjacent flippable edges. This leaves one 6-vint with ``full" charge and another with a charge of $\frac{1}{4}$, so the overall charge from the 5-vint and its four extensions cannot exceed $\frac{1}{4}+\frac{1}{2}(2\cdot1+1\cdot1)=1\frac{3}{4}$.

 \item A non-handicapped 5-vint with a support of 2. One of the two 6-vints, which extend the 5-vint with a level-3 edge, must be entirely in its non-visible terrain (see Rule \ref{rule:non-visib2}); this 6-vint can be extended into an 8-vint with the same support, using RC edges. The second 6-vint and the 8-vint extending it with the other RC edges give an overall charge of at most $\frac{1}{4}-\frac{1}{6}=\frac{1}{12}$. (See Table \ref{ta:app} in Section \ref{sec:longPr}, where this is the worst possible charge for the case where only $o$ can see $a$, and the 8-vint is a standard 8-vint. This must be the case here, since the 5-vint has a support of 2; such a case is depicted in Figure \ref{fi:RC2b3}, where the 6-vint is shaded.) As in the previous case, the 6-vint extending the 5-vint with $A$ has a support of at least 4. The charge from 5-vint and its four extensions cannot exceed $\frac{1}{4}+\frac{1}{12}+\frac{1}{2}(2\cdot1+1\cdot1)=1\frac{5}{6}$.

 \item A 5-vint with a support of 3. The two 6-vints, which extend the 5-vint with a level-3 edge, can be extended into 8-vints, using RC edges. The appropriate parts of Section \ref{sec:longPr} (when only $q$ can see $a$, and when both $q$ and $o$ can see $a$) contain a single case where the 6-vint has a support of at most 3, and in this case, the overall charge of the two vints cannot exceed $\frac{1}{3}-\frac{1}{8}=\frac{5}{24}<\frac{1}{4}$. This implies that, no matter what the support of the 6-vint is, the overall charge of such a 6-vint and its extending 8-vint cannot exceed $\frac{1}{4}$. Similarly to the previous cases, the 6-vint extending the 5-vint with $A$ has a support of at least 5 (four triangulations as before, plus an additional triangulation where the level-1 edge of the 5-vint is flipped). The overall charge from the 5-vint and its four extensions cannot exceed $\frac{1}{4}\cdot2+\frac{1}{5}+\frac{1}{3}(2\cdot1+1\cdot1)=1\frac{7}{10}$. \marfigsc{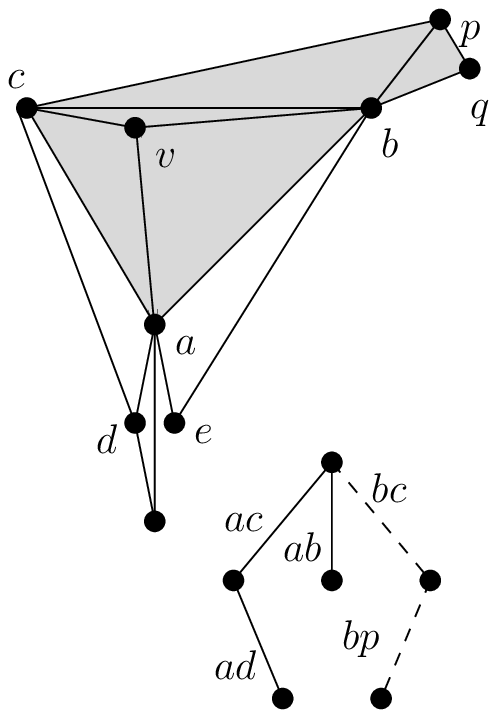}{1.2}
\end{list}

\noindent The above implies that when at least one of these three 5-vints is missing, the total charge cannot exceed $25\frac{2}{35}+\frac{1}{2}+2\cdot1\frac{5}{6}=29\frac{47}{210}$ (the second term represents the single 6-vint which consists of a rigid level-1 edge and two level-2 edges; the third term represents the maximal charge from the other previously ignored vints that are present in this case). We may therefore assume that all three 5-vints are present, which implies that there is at least one handicapped 5-vint (in the subtree of $Y$; see Section \ref{sec:foundations}). We can use this 5-vint to modify the above use of Method 2, and raise the negative charge attained by it from 1 up to $1\frac{1}{2}$, as follows (see Method 2$(B)$).

Consider the two 5-vints, which use a rigid level-1 edge and a non-rigid level-1 edge. When both 5-vints have a support of 2, we can extend a 5-vint and two 6-vints from the non-rigid subtree into three respective 8-vints with the same supports (according to Method 2; the extension of the 5-vint is depicted in Figure \ref{fi:RC2b4}, where the 5-vint is shaded). The negative charge in this case is at least $\frac{1}{2}\cdot3=1\frac{1}{2}$. \marfigsc{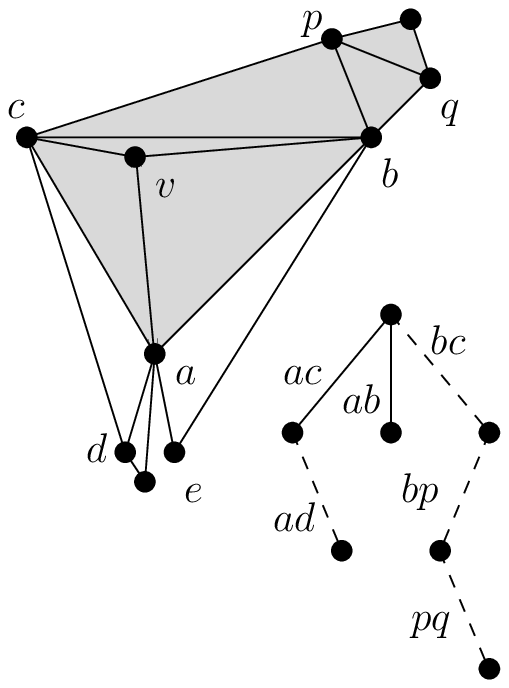}{1.3}

If both 5-vints have a support of 3, there are six 6-vints which we previously considered to have a support of at least 2, and now have a support of at least 3. Namely, we have two 6-vints using $A$, $Y$, and a child of $A$, two using $A$, $Z$, and a child of $A$, one using $A$, $Y$, and $Z$, and one using $A$, $Z$, and $X$. However, we ignore one of the 6-vints using $Y$, $A$, and a child of $A$, in order not to clash with Method 1; the other 6-vint clashing with Method 1 has been ignored in the application of Method 1, and we can safely use it here. (Note that there are three additional 6-vints which extend these 5-vints, but they were already considered as having a higher support.) Thus, in this case, the bound on the total charge decreases by at least $\left(\frac{1}{2}-\frac{1}{3}\right)(2\cdot2+1\cdot5)=1\frac{1}{2}$.

If only the 5-vint using $Y$ and $A$ has a support of 3, we can extend the two 6-vints from the non-rigid subtree (those considered in the first case, where the two 5-vints have a support of 2), but not necessarily the 5-vint. Therefore, in this case, the charge goes down by at least $\frac{1}{2}\cdot2+\left(\frac{1}{2}-\frac{1}{3}\right)(2\cdot1+1\cdot2)=1\frac{2}{3}$ (the second term represents the change in the bound on the supports of the 5-vint and of two other 6-vints which extend it). \marfig{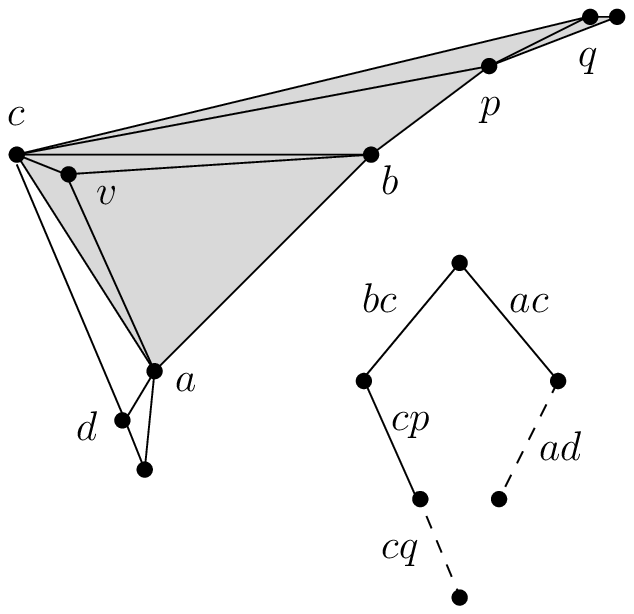}

If only the 5-vint using $Z$ and $A$ has a support of 3, we can extend two 6-vints from the non-rigid subtree into 8-vints with a double support, using the edges of the handicapped 5-vint (as depicted in Figure \ref{fi:RC2b5}, where the 6-vint is shaded and the extending edges are $ac$ and $ad$). Therefore, in this case, the charge goes down by at least $\frac{1}{4}\cdot2+\left(\frac{1}{2}-\frac{1}{3}\right)(2\cdot1+1\cdot4)=1\frac{1}{2}$ (the second term represents the change in the bound on the supports of the 5-vint and of four other 6-vints which extend it).

In either case, we lose a charge of at least $1\frac{1}{2}$, which implies that the total charge (not including the previously ignored vints) is at most \[27-\frac{33}{35}-1\frac{1}{2}=24\frac{39}{70}.\] \marfigsc{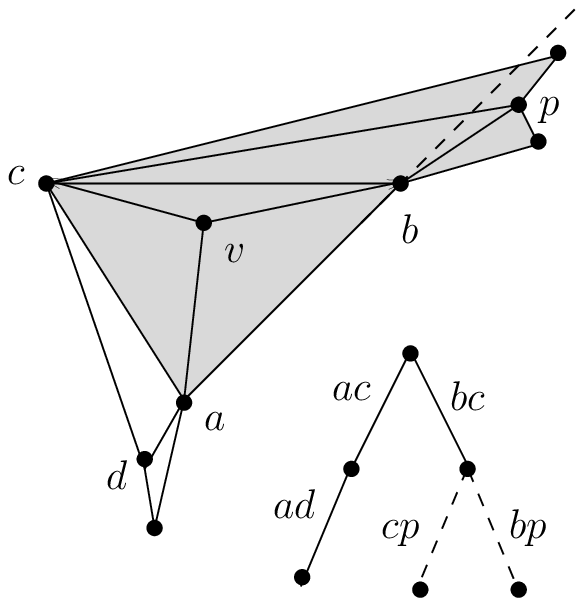}{1.4}

\noindent Moreover, one of the two 6-vints which contain $X$ and a level-3 edge is entirely in its non-visible terrain, according to Rule \ref{rule:non-visib2}. This 6-vint can be extended into an 8-vint with a double support, using the edges of the handicapped 5-vint in the subtree of $Y$ (as depicted in Figure \ref{fi:RC2b6}, where the 6-vint is shaded). We will use this observation repeatedly in what follows.

At least one of the three 5-vints that consist of a rigid level-1 edge and a non-rigid level-2 edge is handicapped, and thus, must have a support of 2 (it lies in the subtree of $Y$, as stated above). The rest of the analysis is divided according to the supports of the other two 5-vints:  \begin{list}{\labelitemi}{\leftmargin=1em}
 \item The (non-handicapped) 5-vint from the subtree of $Y$ has a support of 2 (with no restrictions on the support of the 5-vint in the subtree of $Z$). The above 6-vint using $X$ and a level-3 edge can be extended into two 8-vints with a double support (instead of just one), each using the edges of a different 5-vint from the subtree of $Y$. Moreover, the 6-vint using $Y$ and two level-2 edges is entirely in its non-visible terrain, and thus, can be extended into an 8-vint with the same support, using $Z$ and $X$, neutralizing its charge (as depicted in Figure \ref{fi:RC2b7}, where the 6-vint is shaded). Thus, the total charge cannot exceed (the fourth and fifth terms account for the three addional 5-vints and for twelve of their 6-vint extensions;\marfigsc{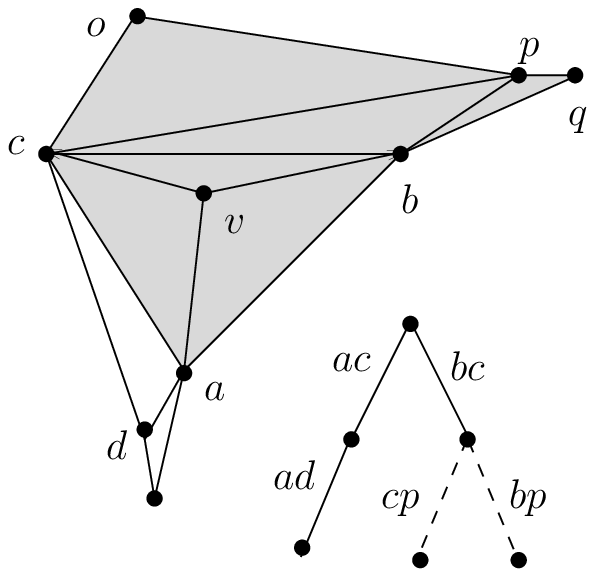}{1.4}the second and third terms represent the two remaining 6-vints, which consist of a rigid level-1 edge and two level-2 edges; the last term represents two 8-vints, each with a support of 4, as just discussed) \[24\frac{39}{70}+\frac{1}{2}+0+1\frac{3}{4}+2\cdot1\frac{5}{6}-\frac{1}{4}\cdot2=29\frac{409}{420}.\]

\item Only the 5-vint from the subtree of $Z$ has a support of 2. The 6-vint using $Z$ and two level-2 edges is entirely in its non-visible terrain, and thus, can be extended into an 8-vint with a double support, using the edges of the handicapped 5-vint. We next consider the 6-vint using $Y$ and two level-2 edges. We use Section \ref{sec:2level2} in order to bound the support of the 8-vint which extends it with $Z$ and $X$. If the two level-2 edges cannot see each other (as depicted in Figure \ref{fi:RC2b8}, where the 6-vint is shaded),\marfigsc{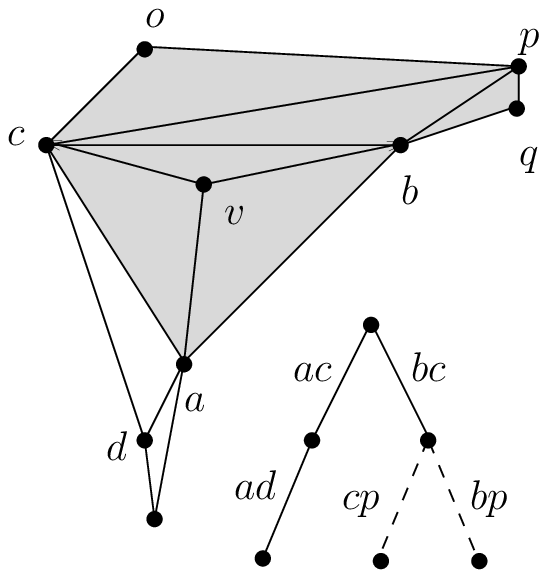}{1.4}the 6-vint has a support of $tr(bp)+tr(cq)=3+1=4$, and the 8-vint a support of at most $4+2=6$ (as explained in Section \ref{sec:2level2}; in our case, there are two non-empty sets of vertices connected to $o$). Otherwise, the 6-vint has a support of $tr(bc)+tr(ao)=5+2=7$, and the 8-vint has a support of at most $7+2\cdot2=11$ (as depicted in Figure \ref{fi:RC2b9}, where the 6-vint is shaded). In either case, the overall charge from the vints is at most $\frac{1}{4}-\frac{1}{6}=\frac{1}{12}$. Thus, the total charge cannot exceed (the fourth, fifth, and sixth terms account for the three addional 5-vints and for twelve of their 6-vint extensions; the second and third terms represent the two remaining 6-vints, which consist of a rigid level-1 edge and two level-2 edges; the last term represents two 8-vints, each with a support of 4, as just discussed) \[24\frac{39}{70}+\frac{1}{12}+\frac{1}{2}+1\frac{3}{4}+1\frac{5}{6}+1\frac{7}{10}-\frac{1}{4}\cdot2=29\frac{97}{105}.\]

 \item Both 5-vints have a support of 3. As in the previous case, the overall charge from the 6-vint using $Y$ and two level-2 edges, and from its extending 8-vint, cannot exceed $\frac{1}{12}$. The 6-vint using $Z$ and two level-2 edges has a support of 3. Thus, the total charge cannot exceed (the fourth and fifth terms account for the three addional 5-vints and for twelve of their 6-vint extensions; the second and third terms represent the two remaining 6-vints, which consist of a rigid level-1 edge and two level-2 edges; the last term represents an 8-vint with a support of 4, as discussed above) \[24\frac{39}{70}+\frac{1}{3}+\frac{1}{12}+1\frac{3}{4}+2\cdot1\frac{7}{10}-\frac{1}{4}\cdot1=29\frac{367}{420}.\] \marfigsc{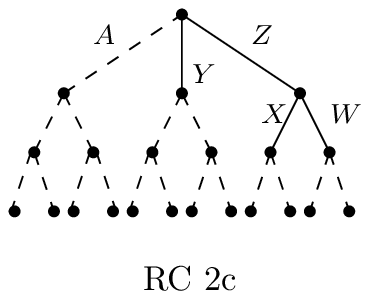}{1.3}
\end{list}

\noindent {\bf RC 2c,} as depicted in Figure \ref{fi:RC2c}. \begin{list}{\labelitemi}{\leftmargin=1em}
 \item The four 6-vints, which use $Y$ and a level-3 edge, meet the conditions of Rule \ref{rule:6-vint2}. By the rule, their overall charge cannot exceed $4\cdot\frac{1}{1400}=\frac{1}{350}$.
 \item The 6-vint, which consists of $Y$ and two level-2 edges, meets the conditions of Rule \ref{rule:6-vint1}. By the rule,  this 6-vint can be ignored.
 \item Each of the four 6-vints using $Z$ and a level-3 edge can be extended into an 8-vint, using the additional RC edges (as depicted in Figure \ref{fi:RC2c1}, where the hole of the 6-vint is shaded). When $o$ is in its non-visible terrain, the 6-vint is entirely in its non-visible terrain, and thus, the extending 8-vint has the same support (by Rule \ref{rule:RemainsRigid}, $o$ cannot see the vertex of $bp$), neutralizing its charge. By Rule \ref{rule:non-visib2}, three of the four 6-vints must be entirely in their non-visible terrains (the fourth 6-vint is the one using $cq$ as its level-3 edge, which is the one shaded in Figure \ref{fi:RC2c1}).\marfigsc{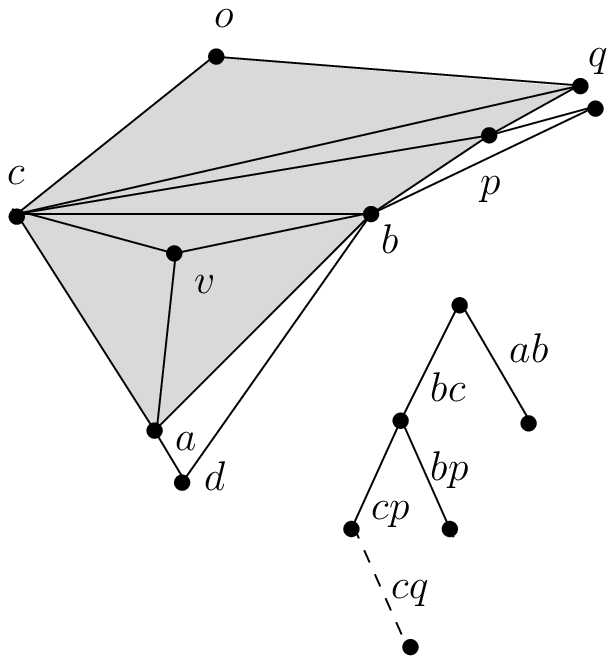}{1.5} For the fourth 6-vint, when $o$ is in its visible terrain, the 6-vint has a support of $tr(bc)+tr(ao)=3+1=4$, and its extending 8-vint has a support of at most $tr(bc)+tr(ao)=3+2=5$ (which occurs when $o$ can see the vertex of the RC edge $ab$, as in Figure \ref{fi:RC2c1}). This implies that the overall charge from the four 6-vints and their extension 8-vints cannot exceed $\frac{1}{4}-\frac{1}{5}=\frac{1}{20}$.
\end{list}

\noindent As in the previous cases, we first ignore some of the vints. This time, we only ignore the two 5-vints using $Y$ and one of its child edges, and the four 6-vints which extend one of these 5-vints with either $A$ or $Z$. We have already accounted for nine 6-vints (different from those ignored). Each of the other vints has a support of at least 2 (apart from the two rigid 4-vints, three rigid 5-vints, and three rigid 6-vints), and therefore, we get a charge of at most \[4\cdot1+3\left(2+\frac{1}{2}\right)+2\left(3+\frac{1}{2}\cdot4\right)+1\left(3+\frac{1}{2}\cdot12\right)+\frac{1}{20}+\frac{1}{350}=30\frac{387}{700}.\]

\noindent We observe that either at least one of the 5-vints of the non-rigid subtree has a support of 3, or a Type I 5-vint must exist (or one of these 5-vints is missing).\marfigsc{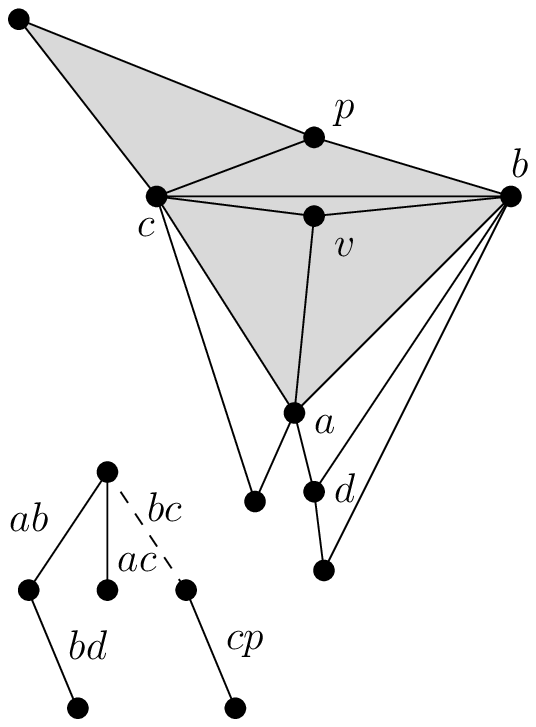}{1.3}In the former case, a 5-vint and four of the 6-vints which extend it have a support of at least 3 (we do not count the fifth 6-vint, in order not to clash with Method 2); the charge is  therefore reduced by at least $\left(\frac{1}{2}-\frac{1}{3}\right)(2\cdot1+1\cdot4)=1$. In the latter case, the handicapped 6-vint can be extended into an 8-vint with a support that can be bounded by Method 1 --- as explained in  Method 1$(D)$, we can ignore this 6-vint when there are at least three RC edges. This is also valid for the sibling of the handicapped 6-vint, unless it has a support of 5 (see Method 1$(E)$). Finally, according to Method 1$(A)$, we can extend the Type I 5-vint into two 8-vints, each using two level-1 RC edges and one level-2 RC edge (as depicted in Figure \ref{fi:RC2c2}, where the hole of the 5-vint is shaded; there are additional extension vints, but we do not consider them here). There are at most five non-empty sets of vertices connected to $p$, which implies that the support of either of these two 8-vints is at most $2+5=7$ (this analysis is identical to the one in RC 2b). The reduction in the charge in this case is therefore at least $\frac{1}{2}+\left(\frac{1}{2}-\frac{1}{5}\right)+\frac{1}{7}\cdot2=\frac{38}{35}$. We conclude that in either case the total charge goes down by at least $1$.

We next observe that either the 5-vint using $A$ and $Z$ has a support of 3, or we can use Method 2 to extend vints from the non-rigid subtree (using $Z$, $X$, and $W$). In the former case, the support of the 5-vint (the one using $A$ and $Z$) and the five 6-vints which extend it is at least 3, which lowers the bound by at least $\left(\frac{1}{2}-\frac{1}{3}\right)(2\cdot1+1\cdot5)=1\frac{1}{6}$. In the latter case, we can extend two 6-vints and a 5-vint from the non-rigid subtree into three respective 8-vints with the same supports, by appending $Z$ and $X$ (and also $W$ in the case of the 5-vint). This follows since the vertices of these 6-vints, as considered in Method 2$(C)$, cannot see vertices of RC edges. Therefore, in this case, the charge is reduced by at least $3\cdot\frac{1}{2}=1\frac{1}{2}$. We conclude that in either case the bound on the total charge goes further down by at least $1\frac{1}{6}$.

Hence, so far the total charge is at most \[30\frac{387}{700}-1-1\frac{1}{6}=28\frac{811}{2100}.\]

\noindent We next consider the possible charges of the two 5-vints and the four 6-vints that were previously ignored:\marfigsc{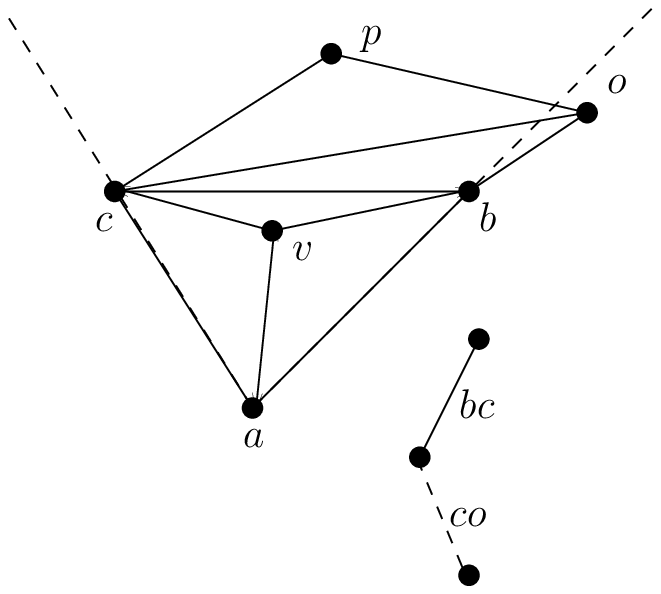}{1.5} \begin{list}{\labelitemi}{\leftmargin=1em}
 \item A 5-vint with a support of 2. The level-2 vertex of the 5-vint is in its non-visible terrain, since otherwise it would have a support of at least $tr(bc)+tr(ap)=2+1=3$ (as depicted in Figure \ref{fi:RC2c3}). This implies that the 5-vint can be extended into an 8-vint with the same support, using the additional RC edges. The 6-vint which extends the 5-vint using $A$ has a support of at least 4, since it holds two non-adjacent flippable edges. The overall charge from the 5-vint, its two extending 6-vints, and their extensions into larger vints, cannot exceed $\frac{1}{2}(2\cdot1+1\cdot1-1\cdot1)+\frac{1}{4}=1\frac{1}{4}$.
 \item A 5-vint with a support of 3. The 6-vint which extends the 5-vint using $A$ has a support of at least 5 (four triangulations as before, plus an additional triangulation where the level-1 edge of the 5-vint is flipped). The overall charge from the 5-vint, and its two extending 6-vints, cannot exceed $\frac{1}{3}(2\cdot1+1\cdot1)+\frac{1}{5}=1\frac{1}{5}$.
\end{list}

\noindent The above implies\marfigsc{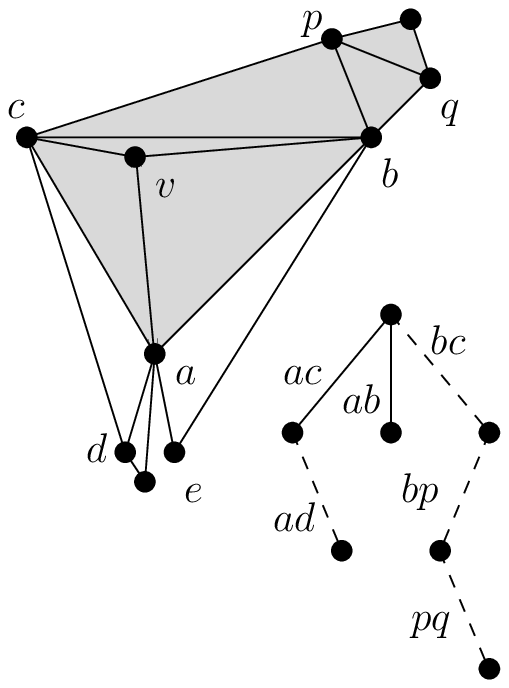}{1.4} that when at least one of these two 5-vints is missing, the total charge cannot exceed $28\frac{811}{2100}+1\frac{1}{4}=29\frac{334}{525}$. We may therefore assume that both 5-vints are present, which implies that there is at least one handicapped 5-vint (see Section \ref{sec:foundations}). We can use this 5-vint to modify the above use of Method 2, and raise the negative charge attained by it from $1\frac{1}{6}$ up to $1\frac{2}{3}$, as follows (see Method 2$(B)$).

Consider the two 5-vints, which use a rigid level-1 edge and a non-rigid level-1 edge. When both 5-vints have a support of 2, we can extend two 6-vints from the non-rigid subtre into 8-vints with the same support (according to Method 2); we can also extend a 5-vint from the non-rigid subtree into a 9-vint with the same support. The negative charge in this case is at least $1\cdot\frac{1}{2}\cdot2+2\cdot\frac{1}{2}=2$.

If both 5-vints have a support of 3, there are seven 6-vints which we previously considered to have a support of at least 2, and now have a support of at least 3. Namely, we have two 6-vints using $A$, $Y$, and a child of $A$, two using $A$, $Z$, and a child of $A$, one using $A$, $Y$, and $Z$, one using $A$, $Z$, and $X$, and one using $A$, $Z$, and $W$. However, we ignore one of the 6-vints using $Y$, $A$, and a child of $A$, in order not to clash with Method 1, and recall that, as above, another such 6-vint (using $Z$) was ignored in the application of Method 1. (Note that there are two additional 6-vints which extend these 5-vints, but they were already considered as having a higher support.)  Thus, in this case, the bound on the total charge decreases by $\left(\frac{1}{2}-\frac{1}{3}\right)(2\cdot2+1\cdot6)=1\frac{2}{3}$.

If only the 5-vint using $Y$ and $A$ has a support of 3, we can still extend the two 6-vints from the non-rigid subtree (those considered in the first case, where the two 5-vints have a support of 2) into 8-vints with the same support, but not necessarily the 5-vint. Therefore, in this case, the charge goes down by at least $\frac{1}{2}\cdot2+\left(\frac{1}{2}-\frac{1}{3}\right)(2\cdot1+1\cdot2)=1\frac{2}{3}$ (the second term represents the change in the bound on the supports of the 5-vint and of two other 6-vints that extend it).

If only the 5-vint using $Z$ and $A$ has a support of 3, we can extend two 6-vints from the non-rigid subtree into 8-vints with a double support, using the edges of the handicapped 5-vint in the subtree of $Y$ (as depicted in Figure \ref{fi:RC2c4}, where the 6-vint is shaded and the extending edges are $ac$ and $ad$), and thus, the charge goes down by at least $\frac{1}{4}\cdot2+\left(\frac{1}{2}-\frac{1}{3}\right)(2\cdot1+1\cdot5)=1\frac{2}{3}$ \marfigsc{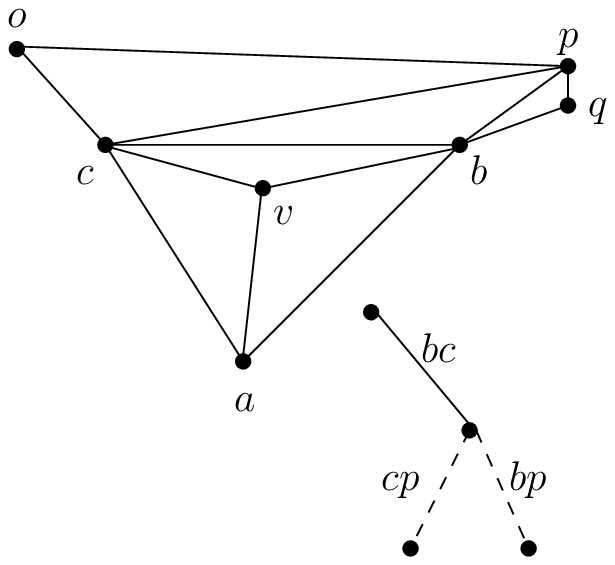}{1.4}(the second term represents the change in the bound on the supports of the 5-vint and of five other 6-vints that extend it).

In either case, we lose a charge of at least $1\frac{2}{3}$, which implies that the total charge (including the previously ignored vints) is at most \[28\frac{811}{2100}+\left(1\frac{2}{3}-1\frac{1}{6}\right)+1\frac{1}{4}\cdot2=30\frac{811}{2100}.\]

\noindent If the non-visible subtree of $Z,X,$ and $W$ is the non-rigid subtree, we can ignore our previous use of the two methods, and instead, extend each 5-vint and 6-vint from the non-rigid subtree into an 8-vint with the same support (adding RC edges from the subtree of $Z$ cannot increase the support). This increases the negative charge by $\frac{1}{2}\cdot(2+5)-\left(1+1\frac{2}{3}\right)=\frac{5}{6}$, and thus, the total charge cannot exceed $30\frac{811}{2100}-\frac{5}{6}=29\frac{387}{700}$. We may therefore assume that the non-visible subtree of $Z,X,$ and $W$ is the subtree of $Y$. \marfigsc{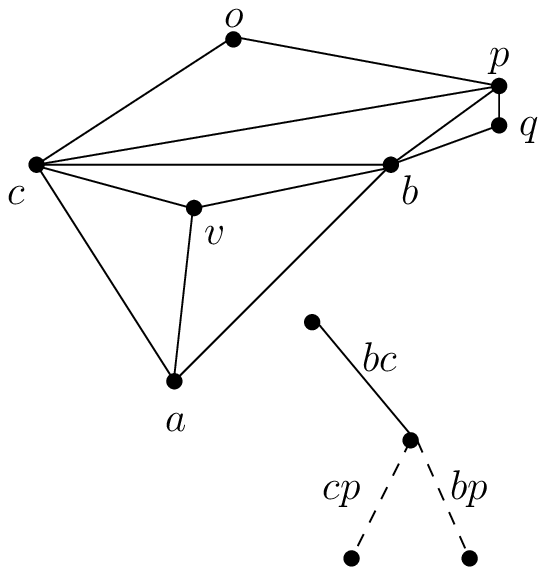}{1.4}

Consider the 6-vint using $Y$ and its two child edges, as depicted in Figures \ref{fi:RC2c5} and \ref{fi:RC2c6}. When both 5-vints from the subtree of $Y$ have a support of 2 (as depicted in Figure \ref{fi:RC2c5}), the chord $bc$ must be present in every triangulation of the hole of the 6-vint, and thus, the 6-vint has a support of at most $C_3=5$ (the number of triangulations of the pentagon $bqpoc$). When one of the 5-vints has a support of 3 (as depicted in Figure \ref{fi:RC2c6}; the other 5-vint is handicapped, and thus, must have a support of 2), the support is at most $tr(bc)+tr(ao+bo)=5+2=7$. We can ignore our previous use of Rule \ref{rule:6-vint1} (which implied that we can ignore the 6-vint) and extend the 6-vint into two 8-vints and one 9-vint with the same support, using RC edges (due to the assumption on the non-visible subtree of $Z$). The change in the charge is at most $\frac{1}{7}(1\cdot1-1\cdot2-2\cdot1)=-\frac{3}{7}$, and therefore, the charge cannot exceed \marfigsc{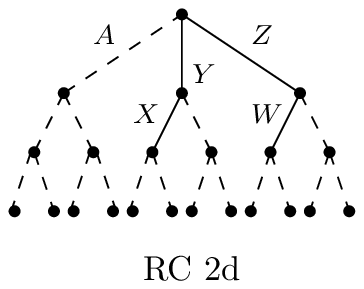}{1.3} \[30\frac{811}{2100}-\frac{3}{7}=29\frac{2011}{2100}.\]

\noindent {\bf RC 2d,} as depicted in Figure \ref{fi:RC2d}. \begin{list}{\labelitemi}{\leftmargin=1em}
 \item The four 6-vints which consist of a rigid level-1 edge, a non-rigid level-2 edge, and a level-3 edge, meet the conditions of Rule \ref{rule:6-vint2}. By the rule, their overall charge cannot exceed $4\cdot\frac{1}{1400}=\frac{1}{350}$.
 \item Each of the four 6-vints, which consist of a rigid level-1 edge, a rigid level-2 edge, and a level-3 edge, can be extended into an 8-vint, using the additional RC edges. By Rule \ref{rule:non-visib2}, at least two of these 6-vints are entirely in their non-visible terrains, and thus, each of their corresponding 8-vints has the same support as the 6-vint it extends. For each of the other two 6-vints, by Table \ref{ta:app} of Section \ref{sec:longPr} (the part where the level-2 edge is rigid), either the 6-vint has a support of at most 3 and the 8-vint has the same support, or the 6-vint has a support of 4 and the 8-vint has a support of at most 7. Thus, the overall charge from these vints cannot exceed $2\left(\frac{1}{4}-\frac{1}{7}\right)=\frac{3}{14}$.
\end{list}

\noindent As in the previous cases, we first ignore some of the vints. This time, we ignore the two 5-vints using a rigid level-1 edge and a non-rigid level-2 edge, and the six 6-vints which extend one of these 5-vints with either a level-1 edge or a level-2 edge. We have already accounted for eight 6-vints (different from those ignored). Each of the other vints has a support of at least 2 (apart from the two rigid 4-vints, three rigid 5-vints, and two rigid 6-vints), and therefore, we get a charge of at most \[4\cdot1+3\left(2+\frac{1}{2}\right)+2\left(3+\frac{1}{2}\cdot4\right)+1\left(2+\frac{1}{2}\cdot12\right)+\frac{3}{14}+\frac{1}{350}=29\frac{251}{350}.\]

\noindent We can use Method 1 in a manner completely identical (and essentially verbatim) to the one presented in RC 2c. This lowers the bound on the total charge by at least 1.

We next consider the two 5-vints which consist of a rigid level-1 edge and a non-rigid level-1 edge. When both 5-vints have a support of 2, according to Method 2$(A)$, we can extend a 5-vint from the non-rigid subtree into a 9-vint with the same support, by using RC edges. According to Method 1$(C)$, we can extend each of the two 6-vints, which extend this 5-vint with a level-3 edge, into an 8-vint with the same support. Therefore, in this case, the charge is reduced by at least $\frac{1}{2}(2\cdot1+1\cdot2)=2$.

When both 5-vints have a support of 3, there are seven 6-vints which we previously considered to have a support of at least 2, and now have a support of at least 3. Namely, we have two 6-vints using $A$, $Y$, and a child of $A$, two using $A$, $Z$, and a child of $A$, one using $A$, $Y$, and $Z$, one using $A$, $Y$ and $X$, and one using $A$, $Z$, and $W$. However, we ignore one of the 6-vints, say one of the two using $Y$, $A$, and a child of $A$, in order not to clash with Method 1, and recall that, as above, another such 6-vint (using $Z$) was ignored in the application of Method 1. (Note that there are two additional 6-vints which extend these 5-vints, but they were already considered as having a higher support.) Thus, in this case, the charge decreases by at least $\left(\frac{1}{2}-\frac{1}{3}\right)(2\cdot2+1\cdot6)=1\frac{2}{3}$.

When exactly one 5-vint has a support of 2, we can still extend the two 6-vints from the non-rigid subtree (those considered in the first case, where the two 5-vints have a support of 2), but not necessarily the 5-vint. Therefore, in this case, the charge is reduced by at least $\frac{1}{2}\cdot2+\left(\frac{1}{2}-\frac{1}{3}\right)(2\cdot1+1\cdot3)=1\frac{5}{6}$ (the second term represents the change in the bound on the supports of the 5-vint and of three other 6-vints that extend it).

In either case, the bound on the total charge goes further down by at least $1\frac{2}{3}$, which implies that the total charge is at most \marfig{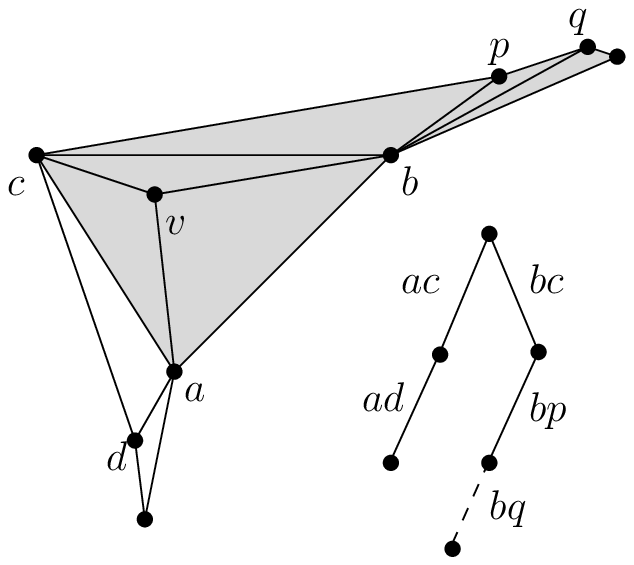} \[29\frac{251}{350}-1-1\frac{2}{3}=27\frac{53}{1050}.\]

\noindent We next consider the possible charges of the two 5-vints and the six 6-vints that were previously ignored: \begin{list}{\labelitemi}{\leftmargin=1em}
 \item A 5-vint with a support of 2. The 6-vint which extends the 5-vint using $A$ has a support of at least 4, since it holds two non-adjacent flippable edges. The 5-vint can be extended into an 8-vint with the same support, using RC edges. The overall charge from the 5-vint and the three 6-vints which extend it cannot exceed $\frac{1}{2}(2\cdot1+1\cdot2-1\cdot1)+\frac{1}{4}=1\frac{3}{4}$.
 \item A 5-vint with a support of 3. Without loss of generality, we refer to the 5-vint using $Y$ and its non-rigid child edge. The 6-vint which extends the 5-vint using $A$ has a support of at least 5 (four triangulations as before, plus an additional triangulation where the level-1 edge of the 5-vint is flipped). Moreover, by rule \ref{rule:non-visib2}(b),\marfig{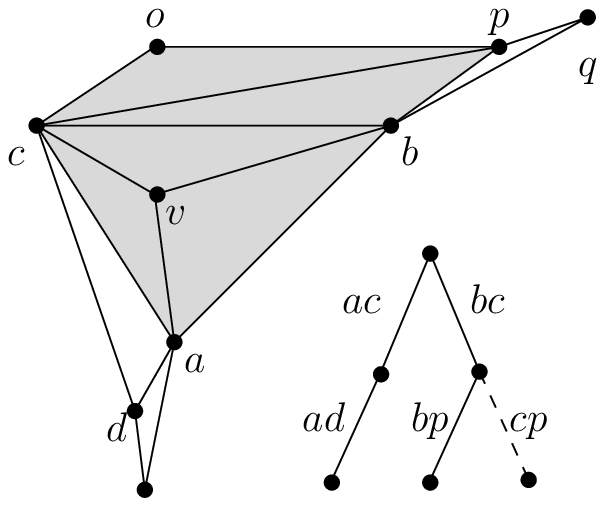}the two 6-vints using $Y$, $X$, and a level-3 edge are entirely in their non-visible terrain. This implies that each of these 6-vints can be extended into an 8-vint with the same support using $Z$ and $W$ (as depicted in Figure \ref{fi:RC2d2}, where $bc$ and $ac$ correspond to $Y$ and $Z$, respectively). This lowers the bound on the total charge by $\frac{1}{4}-\frac{1}{7}=\frac{3}{28}$ (which was the previous bound on the charge of those two 6-vints). Finally, consider the 8-vint which extends the 5-vint using the other RC edges, as depicted in Figure \ref{fi:RC2d3} (where $bc$ and $ac$ correspond to $Y$ and $Z$, respectively, and the 5-vint is shaded). Notice that $bp$ is present in every triangulation, which implies that the support of this 8-vint is at most $tr(bc)+tr(ao)=C_2+C'_3=2+3=5$. We conclude that the charge of this 5-vint and its extensions is at most $\frac{1}{3}(2\cdot1+2\cdot1)+\frac{1}{5}-\frac{3}{28}-\frac{1}{5}=1\frac{19}{84}$ (the second term represents the 6-vint using $A$, the third term represents the decrease in the two 6-vints that use a level-3 edge, and the last term represents the 8-vint that was discussed above).
\end{list}

\noindent The above implies that when at least one of these two 5-vints is missing,\marfig{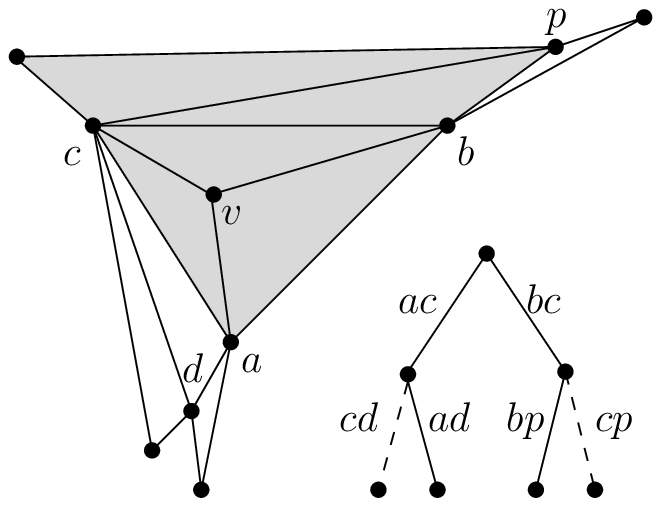}the total charge cannot exceed $27\frac{53}{1050}+1\frac{3}{4}=28\frac{1681}{2100}$. We may therefore assume that both 5-vints are present. We divide the rest of the analysis according to the supports of these two 5-vints: \begin{list}{\labelitemi}{\leftmargin=1em}
 \item Both 5-vints have a support of 3. The total charge cannot exceed \[27\frac{53}{1050}+1\frac{19}{84}\cdot2=29\frac{88}{175}.\]
 \item Both 5-vints have a support of 2. Consider the 9-vint which consists of the edges of both 5-vint and of the additional RC edges (as depicted in Figure \ref{fi:RC2d4}, where one of the 5-vints is shaded). Since all of these edges are in their non-visible terrains, the support of the 9-vint is the product of the supports of the two 5-vints, which is $2\cdot2=4$. By removing either $X$ or $W$, we generate an 8-vint with the same support. The total charge cannot exceed \marfig{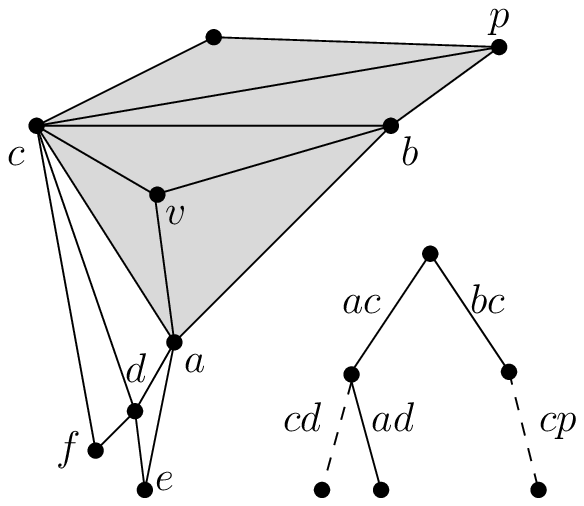} \[27\frac{53}{1050}+1\frac{3}{4}\cdot2-\frac{1}{4}(2\cdot1+1\cdot2)=29\frac{289}{525}.\]
 \item Exactly one 5-vint has a support of 2. We once again consider the two 8-vints which were analyzed in the previous case. Such an 8-vint is depicted in Figure \ref{fi:RC2d5}, where the 5-vint with a support of 3 is shaded and the RC edge that is not used is the one adjacent to this 5-vint. Since $p$ cannot see $a$, and $e$ cannot see $f$, the support of such an 8-vint is smaller than $C''_6=C_6-C_5\cdot2+C_4=132-42\cdot2+14=62$. The second 8-vint can be analyzed in a similar manner (albeit not symmetric to the previous case). Therefore, the total charge cannot exceed \[27\frac{53}{1050}+1\frac{3}{4}+1\frac{19}{84}-\frac{1}{62}\cdot2=29\frac{2312}{2325}.\]
\end{list}

\noindent {\bf RC 2e,} as depicted in Figure \ref{fi:RC2e}.\marfigsc{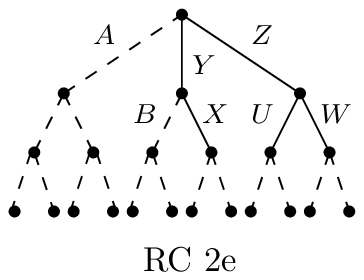}{1.3} \begin{list}{\labelitemi}{\leftmargin=1em}
 \item Each of the eight 6-vints, which use a rigid level-1 edge and a level-3 edge, meets the conditions of Rule \ref{rule:6-vint2}. By the rule, their overall charge 6-vints cannot exceed $8\cdot\frac{1}{1400}=\frac{1}{175}$.
 \item The 6-vint using $A$ and $B$ has a support of at least 4, since it holds two non-adjacent flippable edges.
\end{list}
So far, we have accounted for nine 6-vints. Each of the other vints has a support of at least 2 (apart from the two rigid 4-vints, four rigid 5-vints, four rigid 6-vints, and one rigid 8-vint), and therefore, we get a charge of at most \[4\cdot1+3\left(2+\frac{1}{2}\right)+2\left(4+\frac{1}{2}\cdot5\right)+1\left(4+\frac{1}{2}\cdot15\right)-1\cdot1+\frac{1}{175}+\frac{1}{4}=35\frac{179}{700}.\] \marfig{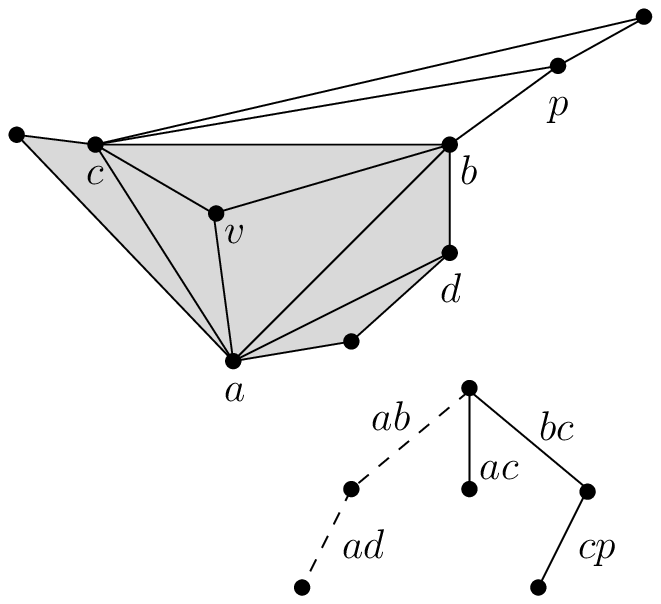}

\noindent If the non-rigid subtree is the non-visible subtree of $Z$, we can extend each of the five 6-vints and two 5-vints from the non-rigid subtree into an 8-vint with the same support. We can also extend the 5-vint which consists of $A$ and $Y$, the 6-vint which consists of $A$, $Y$, and $X$, and the two 6-vints which consist of $A$, $Y$, and a child-edge of $A$, each into an 8-vint with the same support (a case of a 6-vint which consists of $A$, $Y$, and a child-edge of $A$ is depicted in Figure \ref{fi:RC2e1}, where $ab$, $bc$, and $ac$ correspond to $A$, $Z$, and $Y$, respectively, and the 6-vint is shaded). Thus, the total charge cannot exceed $35\frac{179}{700}-\frac{1}{2}\cdot11=29\frac{529}{700}$. (Note that in this analysis, we assume that all these eleven 5-vints and 6-vints are present. If any of them is missing, we lose at least as much positive charge as negative charge, so the total charge can only decrease.) We may therefore assume that the non-visible subtree of $Z$ is the subtree of $Y$.

If $B$ is present in the flip-tree, it generates a positive charge by participating in a 5-vint and three 6-vints (we ignore two additional 6-vints which use a level-3 edge, since we consider, somewhat loosely, their charge as part of the term $\frac{1}{175}$,\marfig{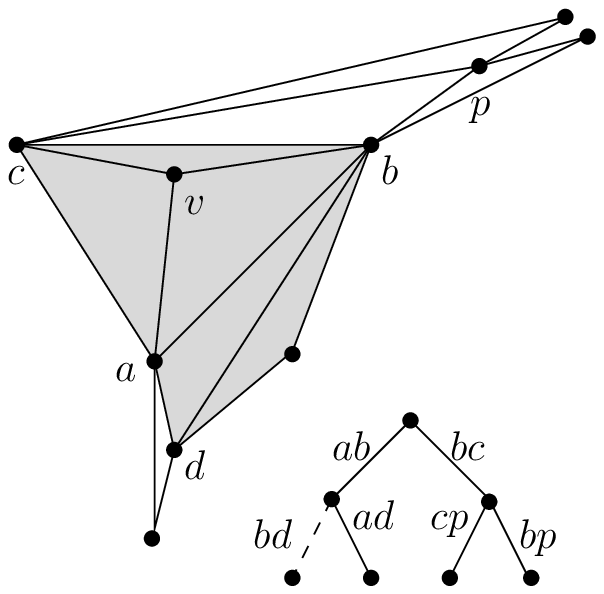} even when $B$ is not present). We can extend the 5-vint into a 9-vint with the same support, using $X$, $Z$, $W$, and $U$ (by Rule \ref{rule:RemainsRigid}, appending $X$ cannot increase the support of the vint; such a 9-vint is depicted in Figure \ref{fi:RC2e2}, where $ab$ and $bc$ correspond to $Y$ and $Z$, respectively, and the 5-vint is shaded). Similarly, we can extend the 5-vint into three 8-vints (each obtained by removing a single RC edge from the 9-vint), which more than neutralize the charge of the three 6-vints (each of the three 8-vints has the same support as the 5-vint, and at least one 6-vint has a higher support, since it is using the flippable edge $A$). We conclude that adding $B$ to the flip-tree can only decrease the bound on the total charge, and we may therefore assume that $B$ is not present in the flip-tree.

Hence, we now consider the flip-tree depicted in Figure \ref{fi:RC2e3}, and so far, the charge cannot exceed \[4\cdot1+3\left(2+\frac{1}{2}\right)+2\left(4+\frac{1}{2}\cdot4\right)+1\left(4+\frac{1}{2}\cdot13\right)-1\cdot1+\frac{1}{175}=33\frac{1}{175}.\]

\noindent Once again,\marfigsc{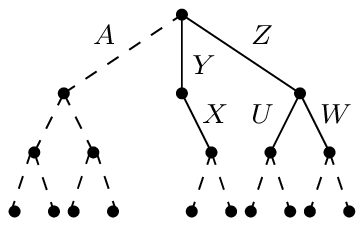}{1.3}we can use Method 1 in a manner completely identical (and essentially verbatim) to the one presented in RC 2c. This lowers the bound on the total charge by at least 1.

We next consider the two 5-vints which consist of a rigid level-1  edge and a non-rigid level-1 edge. When both 5-vints have a support of 2, according to Method 2$(A)$, we can extend a 5-vint from the non-rigid subtree into a 9-vint with the same support, by using RC edges. According to Method 1$(C)$, we can extend each of the two 6-vints, which extend this 5-vint with a level-3 edge, into an 8-vint with the same support. Therefore, in this case, the charge is reduced by at least $\frac{1}{2}(2\cdot1+1\cdot2)=2$.

When both 5-vints have a support of 3, there are eight 6-vints which we previously considered to have a support of at least 2, and now have a support of at least 3. Namely, we have two 6-vints using $A$, $Y$, and a child of $A$, two using $A$, $Z$, and a child of $A$, one using $A$, $Y$, and $Z$, one using $A$, $Y$, and $X$, one using $A$, $Z$, and $U$, and one using $A$, $Z$, and $W$. However, we ignore one of the 6-vints using $Y$, $A$, and a child of $A$, in order not to clash with Method 1, and recall that, as above, another such 6-vint (using $Z$) was ignored in the application of Method 1. (Note that there is an additional 6-vint which extends these 5-vints, but it was already considered as having a higher support.) Thus, the charge decreases by at least $\left(\frac{1}{2}-\frac{1}{3}\right)(2\cdot2+1\cdot7)=1\frac{5}{6}$.

When only the 5-vint using $A$ and $Z$ has a support of 2, we can still use $Z$, $U$, and $W$ to extend the 5-vint and two 6-vints from the non-rigid subtree (those considered in the first case, where the two 5-vints have a support of 2) into respective 8-vints with the same support. Therefore, in this case, the charge is reduced by at least $\frac{1}{2}\cdot3+\left(\frac{1}{2}-\frac{1}{3}\right)(2\cdot1+1\cdot4)=2\frac{1}{2}$ (the second term represents the change in the bound on the supports of the 5-vint that uses $A$ and $Y$ and of four other 6-vints that extend it).

When only the 5-vint using $A$ and $Y$ has a support of 2, we can still use $Y$ and $X$ to extend the two 6-vints from the non-rigid subtree, but not necessarily the 5-vint. Therefore, in this case, the charge is reduced by at least $\frac{1}{2}\cdot2+\left(\frac{1}{2}-\frac{1}{3}\right)(2\cdot1+1\cdot4)=2$ (the second term represents the change in the bound on the supports of the 5-vint that uses $A$ and $Z$ and of four other 6-vints that extend it). \marfigsc{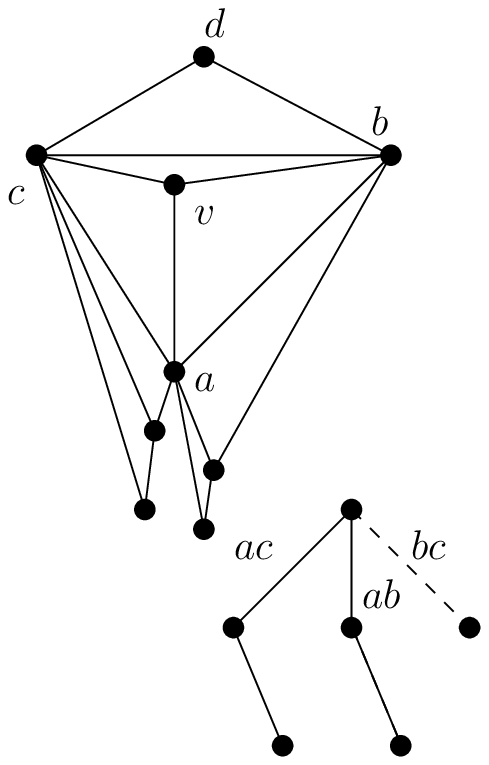}{1.4}

In either case, the bound on the total charge goes further down by at least $1\frac{5}{6}$, which implies that the total charge is at most \[33\frac{1}{175}-1-1\frac{5}{6}=30\frac{181}{1050}.\]

\noindent Consider the two 8-vints which consist of $A$, $X$, $Y$, $Z$, and a child-edge of $Z$ (as depicted in Figure \ref{fi:RC2e4}, where $bc$, $ab$, and $ac$ correspond to $A$, $Y$, and $Z$, respectively). The number of  triangulations of the hole of such an 8-vint is at most $tr(bc)+tr(ad)=1+C'_3\cdot C'_3=1+3\cdot3=10$. Therefore, the total charge cannot exceed \[30\frac{181}{1050}-\frac{1}{10}\cdot2=29\frac{1021}{1050}.\]

\paragraph{Extensions of the previous cases.} We start by treating additional level-2 RC edges. There remains a single case that has not yet been handled --- an RC with four level-2 edges (it is depicted in Figure \ref{fi:RC2ext1} and we refer to it as RC 2e+).\marfigsc{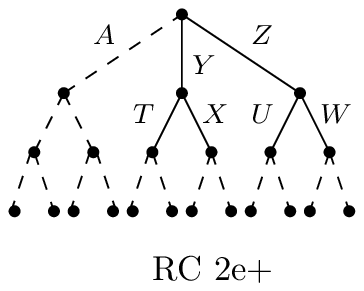}{1.3}We analyze this case by showing that adding a level-2 edge to RC 2e cannot increase the bound on its charge (without loss of generality, we assume that this level-2 edge is $T$). The following proof is very similar to the one in the extensions part of $\lambda_1=3$. The positive charge gained from the change in the RC comes from the 5-vint using $Y$ and $T$, the two 6-vints which extend this 5-vint with a level-3 edge, the two 6-vints which extend it with a level-1 edge, and the 6-vint which extends it with $X$. In the analysis of RC 2e, by Rule \ref{rule:6-vint2}, the charge from the two 6-vints using a level-3 edge was bounded by $\frac{1}{1400}\cdot2=\frac{1}{700}$, and this remains valid after the change. Thus, we only need to consider the change in the charges coming from the 5-vint and the three other 6-vints. By extending the 5-vint with the additional RC edges, we create a 9-vint (as depicted in Figure \ref{fi:RC2ext2}, where the 5-vint is shaded). If before the change the 5-vint had a support of $m\geq 2$,\marfigsc{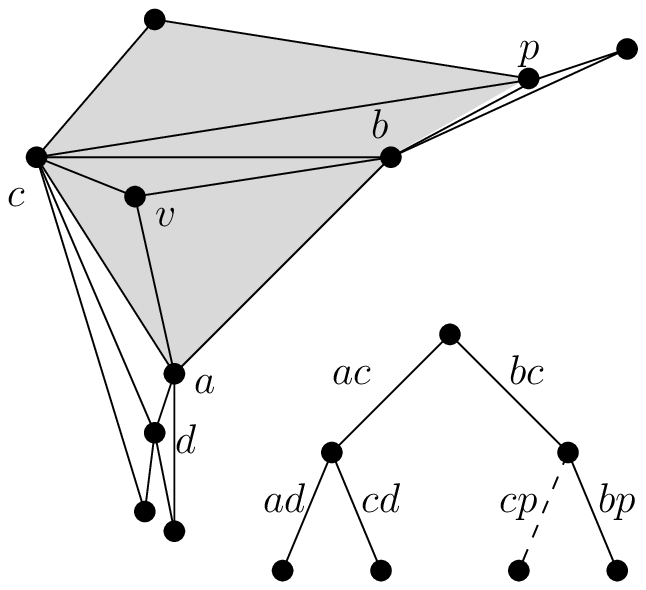}{1.4} this 9-vint had a support of at least $m$. The change in the overall charge coming from  these two vints is at most $2\left(1-\frac{1}{m}\right)-2\left(1-\frac{1}{m}\right)=0$ (the first term represents the change in the 5-vint, and the second term represent the change in the 9-vint). This implies that the 9-vint neutralizes the change in the charge coming from the 5-vint. We can use a similar argument to neutralize the change in the charges coming from the 6-vint using $Y$, $T$, and $X$, and from the 6-vint using $Y$, $T$, and $Z$ (by using two out of the three 8-vints that extend the 5-vint with additional RC edges). We are left with the 6-vint using $T$, $Y$, and $A$, and with the third 8-vint that extends the 5-vint with RC edges. After the change, the 6-vint has a support of at least 2, which implies that the change raised its charge by less than $\frac{1}{2}$. Similarly, after the change, the 8-vint has a support of 1, which implies that the change raised its (negative) charge by at least $\frac{1}{2}$, more than neutralizing the change in the charge coming from the 6-vint. In conclusion, adding another level-2 edge to RC 2e cannot increase the bound on its charge.

We next deal with the addition of level-3 edges to the basic RCs. By Rule \ref{rule:LargestRC}, adding a level-3 RC edge to RC 2c, RC 2d, or RC 2e (or RC 2e+), cannot increase the bound on their charge. Moreover, we cannot add a level-3 RC edge to RC 2a, since it does not have any level-2 RC edges.\marfigsc{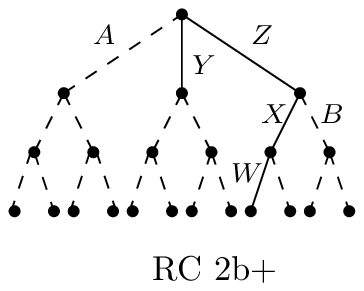}{1.4} This implies that we only need to treat the addition of a single level-3 RC edge to RC 2b (by Rule \ref{rule:LargestRC}, adding a second level-3 edge to this RC cannot increase the charge). This RC, denoted by RC 2b+, is depicted in Figure \ref{fi:RC2ext3}, where $W$ is the new RC edge. We modify our previous analysis of RC 2b so that it applies to RC 2b+, as follows.

Making $W$ rigid can only increase the positive charge by reducing the support of a single 6-vint (the one using $Z$, $X$, and $W$); we will refer to this 6-vint as $u$. In the original analysis of RC 2b, we  considered $W$ to be part of the flip-tree only if the overall charge from $u$ and from the various extensions thereof was positive; otherwise, we would have assumed that $W$ is missing, in order to get a larger charge. Thus, it suffices to show that, after the change, the overall charge from $u$ and from its various extensions is non-positive; this would imply that making $W$ rigid can only decrease the overall charge.

In the original analysis of RC 2b, we show that if at least one of the three 5-vints using a rigid level-1 edge and a non-rigid level-2 edge is not present in the flip-tree, the charge cannot exceed $29\frac{47}{211}$. While proving that claim, we consider $u$ as having a support of at least 2, and use the worst-case value 2 in the calculations. Moreover, we do not use $W$ to generate any extension vints. Thus, after the change, if at least one of these 5-vints is not present in the flip-tree, the total charge cannot exceed $29\frac{47}{211}+(1-\frac{1}{2})=29\frac{305}{422}$ (the second term represents the change in the charge coming from $u$). We may therefore assume that all three 5-vints are present in the flip-tree. \marfigsc{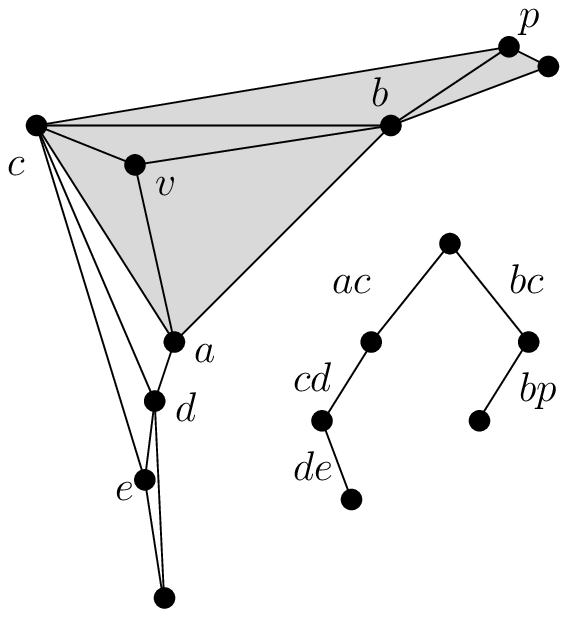}{1.4}

We next append the edges of the handicapped 5-vint in the subtree of $Y$ to the edges $Z$, $X$, and $W$. Since all of these edges are in their non-visible terrain, we get an 8-vint with a support of 2 (as depicted in Figure \ref{fi:RC2ext4}, where the 5-vint is shaded). This leaves a charge of $1-\frac{1}{2}=\frac{1}{2}$ to neutralize. We divide the rest of the analysis into the following two cases: \begin{list}{\labelitemi}{\leftmargin=1em}
 \item The 5-vint using $Z$ and $B$ has a support of 2. We can extend this 5-vint with $Y$, $X$, and $W$, which will generate another 8-vint with a support of 2 (since all of the edges of the 8-vint are in their non-visible terrains; see Section \ref{sec:non-visib}), neutralizing the rest of the charge.
 \item The 5-vint using $Z$ and $B$ has a support of 3. In the analysis of RC 2b, we argued that each of the 6-vints extending the 5-vint with a level-3 edge, and its possible extensions, generate a positive charge of at most $\frac{1}{4}$ (see the case of ``A 5-vint with a support of 3" in RC 2b). By Rule \ref{rule:RemainsRigid}, appending $X$ and $W$ to such a 6-vint (where now $W$ is rigid) cannot increase its support. Therefore, we can create two 8-vints that neutralize the charge of these two 6-vints, which decreases the bound on the charge by at least $\frac{1}{4}\cdot2=\frac{1}{2}$.
\end{list}

\changetext{}{1.23in}{}{}{}

\section{Conclusion}
By a rather meticulous case analysis, we have shown that every set of $n$ points in the plane admits at most $30^n$ different triangulations. We have also noted that our proof technique cannot decrease the base below $28\frac{17}{28}$, so we are very close to the best base that this approach can yield. Nevertheless, we strongly believe that the true upper bound is much smaller. A major weakness of our machinery is that it caters to the worst possible charge that a 3-vint can receive, as opposed to the average charge. Obtaining a sharper upper bound on the average charge requires a totally different approach, which we leave open for future research. For example, it might be possible to prove that for every 3-vint with $\lambda_1=3$, there must exist a 3-vint with $\lambda_1=1$ (with bijective correspondence).

Still, we note that all the cases in our analysis actually yielded charges that were strictly smaller than 30, and each of them could be further improved if we were to consider further expansions of the flip-tree. It might therefore be interesting to find the best base that this approach can yield, which might well be the ``lower bound" $28\frac{17}{28}$. For this, an exhaustive search by computer is probably the way to proceed. This in turn requires the generation of all possible combinatorially distinct configurations of up to 25 points and the flip trees that they generate. The existing databases of \emph{order types} (see, e.g., \cite{AK06}) are not yet powerful enough to provide the data we need. However, it might be possible to drastically decrease the number of relevant configurations by using heuristics.

\subsection*{Acknowledgements.}
We extend our warmest thanks to Emo Welzl, for his invaluable contributions to this research, and for his kind permission to include in this paper the basic infrastructure that he has laid down in the preceding paper \cite{ShWe06b}. The present work would not have materialized without his insights and creativity.

\end{document}